\documentclass[showpacs,prd,amsfonts]{revtex4}

\usepackage{graphicx}
\usepackage{bm}

\begin{document}

\title{Relativistic  gravitation theory for the MOND paradigm}

\author{Jacob D. Bekenstein}\email{bekenste@vms.huji.ac.il}
\homepage{http://www.phys.huji.ac.il/~bekenste/}
\affiliation{Racah Institute of Physics, Hebrew University of
Jerusalem\\ Givat Ram, Jerusalem 91904 ISRAEL}

\date{\today}
\begin{abstract} 
The modified newtonian dynamics (MOND)  paradigm of
Milgrom can boast of a number of successful predictions regarding
galactic dynamics; these are made without the assumption that dark
matter plays a significant role.  MOND requires gravitation to
depart from Newtonian theory in the extragalactic regime where
dynamical accelerations are small.  So far relativistic gravitation
theories proposed to underpin MOND have either clashed with the
post-Newtonian tests of general relativity, or failed to provide
significant gravitational lensing, or violated hallowed principles by
exhibiting superluminal scalar waves or an \textit{a priori} vector
field.  We develop a relativistic MOND inspired theory which
resolves these problems.  In it gravitation is mediated by metric, a
scalar field and a 4-vector field, all three dynamical.  For a simple
choice of its free function,  the theory has a Newtonian limit for
nonrelativistic dynamics with significant acceleration, but a MOND
limit when accelerations are small. We calculate the
$\beta$ and $\gamma$ PPN coefficients showing them to agree with
solar system measurements.   The gravitational light deflection by
nonrelativistic systems is governed by the same potential responsible
for dynamics of particles.  To the extent that MOND successfully
describes dynamics of a system, the new theory's predictions for
lensing by that system's visible matter will agree as well with
observations as general relativity's predictions made with a
dynamically successful dark halo model.  Cosmological models based on
the theory are quite similar to those based on general relativity;
they predict slow evolution of the scalar field.  For a range of
initial conditions, this last result makes it easy to rule out 
superluminal propagation of metric, scalar and vector waves.
 
\end{abstract}


\maketitle 

\section{Introduction}\label{intro}

In the extragalactic regime, where Newtonian gravitational theory
would have been  expected to be an excellent description,
accelerations of stars and gas, as estimated from  Doppler 
velocities and geometric considerations, are as a rule much larger 
than those  due to the Newtonian field generated by the visible 
matter in  the system~\cite{Oort,Zwicky}.  This is the  ``missing
mass'' problem~\cite{DMU} or ``acceleration 
discrepancy''~\cite{Can2}.  It is fashionable to infer from it the
existence of much dark matter in systems ranging from dwarf
spheroidal galaxies with masses $\sim 10^6 M_\odot$ to great
clusters of galaxies in the $10^{13} M_\odot$
regime~\cite{DMU,Sweden}.  And again, galaxies and clusters of
galaxies are found to gravitationally lense  background sources. 
When interpreted within general relativity (GR), this lensing is
anomalously large unless one assumes the presence of dark matter in
quantities and with distribution similar to those required to
explain the accelerations of stars and gas.  Thus extragalactic
lensing has naturally been regarded as confirming the presence of
the dark matter suggested by the dynamics.  

But the putative dark matter has never been  identified despite much
experimental and observational effort~\cite{TurnerEllis}.  This 
raises the possibility that the acceleration discrepancy as well as
the gravitational lensing anomaly may reflect departures from
Newtonian gravity and GR on galactic and larger scales.  Now
alternatives to GR are traditionally required to  possess a Newtonian
limit for small velocities and potentials; thus the acceleration
discrepancy also raises the possibility that the correct relativistic
gravitational theory may be of a kind not generally considered
hitherto.

In the last two decades Milgrom's modified Newtonian dynamics (MOND)
paradigm~\cite{M1,M2,M3} has gained recognition as a successful
scheme for unifying much of extragalactic dynamics phenomenology
without invoking ``dark matter''.  In contrast with earlier
suggested modifications of Newton's law of universal
gravitation~\cite{Berendzen,Jeans,Morpho,Finzi}, MOND is
characterized by an acceleration scale
$\mathfrak{a}_0$, not a distance scale, and its departure from
Newtonian predictions is acceleration dependent:  
\begin{equation}
\tilde\mu(|\mathbf{a}|/\mathfrak{a}_0)\mathbf{a} =
-\bm{\nabla}\Phi_{\rm N}. 
\label{MOND}
\end{equation} Here $\Phi_{\rm N}$ is the usual Newtonian potential
of the visible matter, while $\tilde\mu(x)\approx x$ for
$x\ll 1$ and 
$\tilde\mu(x)\rightarrow 1$ for
$x\gg 1$.  Milgrom estimated  $\mathfrak{a}_0\approx 1\times
10^{-8}$ cm s$^{-2}$ from the empirical data. In the laboratory and
the solar system where accelerations are strong compared to
$\mathfrak{a}_0$, formula (\ref{MOND}) goes over to the Newtonian law
$\mathbf{a} =-\bm{\nabla}\Phi_{\rm N}  $.

Milgrom constructed formula (\ref{MOND}) to agree with the fact
that  rotation curves of disk galaxies become flat outside their
central parts.  That far out a galaxy of mass $M$ exhibits an
approximately spherical Newtonian potential.  The scales are such
that
$|\bm{\nabla}\Phi_{\rm N}|\approx GM r^{-2} \ll
\mathfrak{a}_0$ in this region, and so Eq.~(\ref{MOND}) with
$\tilde\mu(x)\approx x$ gives
$|\mathbf{a}|\approx (GM\mathfrak{a}_0)^{1/2} r^{-1}$ which has the
$r$ dependence appropriate for the centripetal acceleration
$v_c^2/r$ of a radius independent rotational velocity
$v_c$---an asymptotically  flat rotation curve.   In fact one
obtains the relation
$M=(G\mathfrak{a}_0)^{-1}v_c{}^4$ which leads to the
\textit{prediction} that for any class of galaxies with a constant
mass to luminosity ratio $\Upsilon$ in a specified spectral band, the
luminosity in that band should scale as
$v_c{}^4$.    And indeed, there exists an empirical law of just this
form: the Tully--Fisher law~\cite{TullyFisher} (TFL) relating near infrared
(H--band) luminosity $L_H$ of a spiral disk galaxy to its rotation
velocity,
$L_H\propto v_c{}^4$, with the proportionality factor being constant
within each galactic morphology  class.   

This version of the TFL was established only after MOND
was enunciated~\cite{Aaronson}.  It is in harmony with the MOND
prediction in two ways.   First, the infrared light of a galaxy comes
mostly from cool dwarf  stars which make up most of its mass (hence
giving a tight correlation between
$M$ of the predicted relation and $L_H$ of the empirical law). 
Second, the proportionality coefficient varies from class to class as
would be expected from the observed correlation between
$\Upsilon$ of a galaxy and its morphology.  

In the alternative dark matter paradigm (which casts no doubt on
standard gravity theory), flat rotation curves are explained by
assuming  that every disk galaxy is nested inside a roundish
spherical halo of dark matter~\cite{OstrikerPeeblesYahil} whose
mass density drops approximately like
$r^{-2}$.  The halo is supposed to dominate the gravitational field
in the outer parts of the galaxy.  This makes the Newtonian potential
approximately logarithmic with radius in those  regions, thereby
leading to an asymptotically flat rotation curve.   In practice the
dark halo resolution works  only after fine tuning.   It is an
observational fact that for bright spiral galaxies the rotation
curve in the optically bright region is well explained in Newtonian
gravity  by the observed matter~\cite{Kalnajs}.  But, as mentioned,
in the outer regions the visible matter's contribution must be
dwarfed by the halo's.  So fine tuning is needed between the dark
halo parameters (velocity  dispersion and core radius) and the
visible disk ones~\cite{BahcallCassertano,Sancisi}.  

This fine  tuning problem is exacerbated by the TFL
$L_H\propto v_c{}^4$.   Because the infrared luminosity comes from
the visible matter in the  galaxy, but  the rotation velocity is
mostly set by the halo, the TFL  also requires fine
tuning between halo and disk parameters. The standard dark matter
explanation of the
$r^{-2}$ profile of an halo is that it arises naturally from
primordial cosmological perturbations~\cite{QuinnSalmonZurek}.  The
visible galaxy is regarded as forming by dissipational collapse of
gas into the potential well of  the halo.  The fine tuning mentioned
is then viewed as resulting  from the adjustment of the halo to the
gravitation of the incipient disk~\cite{Sancisi,Barnes}.  But the
TFL is observationally a  very sharp correlation; in
fact, it is the basis for one of the most reliable methods for
gauging distances to spiral galaxies.  Such sharpness is hardly to
be expected from statistical  processes of the kind envisaged in
galaxy formation, a point emphasized by Sanders~\cite{SReview}. So
in the dark matter picture the TFL is something of a
mystery.

There are other MOND successes.  Milgrom predicted early that in
galaxies with surface mass density well below $\mathfrak{a}_0
G^{-1}$, the acceleration discrepancy should be especially
large~\cite{M2}.   In dwarf spirals this property was established
empirically years later~\cite{JobinCarignan}, and it is now known to
be exhibited by a large number of low surface brightness
galaxies~\cite{McGaugh}.  Another example: MOND successfully
predicts the detailed shape of a rotation curve from the observed
matter (stars and gas) distribution on the basis of a single free
parameter, $\Upsilon$, down to correlating  features in the velocity
field with those seen in the light distribution~\cite{Kent,McKent,BBS,Gentile}.    This is especially true in the case of low surface mass density disk galaxies for which MOND's predictions are independent of the specific choice of
$\tilde\mu(x)$~\cite{MLSD}, and these MOND theoretical rotation
curves fit the observed curves of a number of low surface brightness
dwarf galaxies~\cite{BBS,MCGdB,Begum} very well.   By contrast, the dark halo paradigm requires one or two free parameters apart from
$\Upsilon$  to approximate the success of the MOND
predictions~\cite{SandersMcGaugh}.   In fact, even when the empirical data is analyzed within the dark halo paradigm, it displays the preferred acceleration scale $\mathfrak{a}_0$ of MOND~\cite{McGaugh2}.

Occasionally doubt has been cast on MOND's ability to describe clusters of galaxies properly~\cite{Aguirre}.  Many of these exhibit accelerations not small on scale $\mathfrak{a}_0$, yet conventional analysis suggests they contain much dark matter in opposition to what MOND would suggest.  Sanders has recently reanalyzed the problem~\cite{Sanders} with the conclusion that these clusters may contain much as yet undiscovered baryonic matter in the core which should be classed as ``visible'' in connection with MOND. Other MOND successes, outside the province of disk galaxies, have been reviewed elsewhere~\cite{SReview,SandersMcGaugh,more}.

So the simple MOND formula~(\ref{MOND}) is very successful.  But it
is not a theory.  Literally taken the MOND recipe for acceleration
violates the conservation of momentum (and of energy and of angular
momentum)~\cite{M1}.   And MOND entails a paradox: why does the
center of mass of a star orbit in its galaxy with anomalously large
acceleration given by Eq.~(\ref{MOND}) with $\tilde\mu\ll 1$, while
each parcel of gas composing it is subject to such high acceleration
that is should, by the same formula, be accelerated Newtonially
?~\cite{M1}.   In short, the MOND formula is not a consistent
theoretical scheme.  Neither is MOND, as initially stated,
complete.  For example, it does not specify how to calculate
gravitational lensing by galaxies and clusters of galaxies.  As is
well known, in standard gravity theory light deflection is well
described only by relativistic theory (GR).  And whereas Newtonian
cosmological models work well for part of the cosmological
evolution, MOND cosmological models built in analogy with their
Newtonian counterparts, though sometimes agreeing with
phenomenology~\cite{Aguirre},  can  yield peculiar
predictions~\cite{Felten} (but see Ref.~\onlinecite{SandersC}).  In
short, a complete, consistent theoretical underpinning of the MOND
paradigm which accords with observed facts, and is also relativistic,
has been lacking.

This lack is being resolved in measured steps.   A first step was the
lagrangian reformulation of MOND~\cite{BekMilg}  called AQUAL (see
Sec.~\ref{sec:nonrelat}).   AQUAL cures the nonconservation
problems and resolves the paradox of the galactic motion of an object
whose parts accelerate strongly relative to one another; it does so
in accordance with a conjecture of Milgrom~\cite{M1}.  And for
systems with high symmetry AQUAL reduces exactly to the MOND formula
(\ref{MOND}).

A relativistic generalization of AQUAL is easy to construct with help
of a scalar field which together with the metric describes
gravity~\cite{BekMilg} (see Sec.~\ref{sec:rAQUAL} below).  It
reduces to MOND approximately in the weak acceleration regime, to
Newtonian gravity for strong accelerations, and can be made
consistent with the post-Newtonian solar system tests for GR.  But
relativistic AQUAL is acausal: waves of the scalar field can
propagate superluminally in the MOND regime (see the appendix of
Ref.~\onlinecite{BekMilg} or Appendix~\ref{sec:A} here).  The
problem can be traced to the aquadratic kinetic part of the
lagrangian of the theory which mimics that in the original AQUAL.  A
theory involving a second scalar field, PCG, was thus developed to
bypass the problem~\cite{Can2,phase,SPCG} (see Sec.~\ref{PCG}
below).   PCG may be better behaved causally  than relativistic
AQUAL~\cite{Rosen}, but it brings woes of its own.  It is marginally
in conflict with the observed perihelion precession of
Mercury~\cite{Can2}, and in common with relativistic AQUAL, PCG
predicts extragalactic gravitational lensing which is too weak if
there is indeed no dark matter.   This last problem is  traceable to
a feature common to PCG and relativistic AQUAL: the physical metric
is conformal to the metric appearing in the Einstein-Hilbert
action~\cite{Kyoto}.

One way to sidestep this problem without discarding the MOND
features is to exploit the direction defined by the gradient of the
first scalar field to relate the physical metric to the Einstein
metric by a disformal transformation  (see Ref.~\cite{Kyoto} or
Sec.~\ref{disf} below).  But  it turns out that with this relation
the requirement of causal propagation acts to depress gravitational
lensing~\cite{BekSan}, rather than enhancing it as is
observationally required.  The persistence of the lensing problem in
modified gravitational theories has engendered a folk theorem to the
effect that it is impossible for a relativistic theory to
simultaneously incorporate the MOND dynamics, observed
gravitational lensing and correct post-Newtonian behavior without
calling on dark matter~\cite{Nester,Edery,Soussa1,Soussa2}.  

Needless to say, this theorem cannot be proved~\cite{Soussa3}. 
Indeed, by the simple device of relating the physical and Einstein
metrics via a disformal transformation based on a \textit{constant}
time directed 4-vector, Sanders~\cite{Sandersstratified} has
constructed an AQUAL like ``stratified''  relativistic theory which
gives the correct lensing while ostensibly retaining the MOND
phenomenology and consistency with the post-Newtonian tests.
Admittedly Sanders' stratified theory is a preferred frame theory, 
and as such outside the traditional framework for gravitational
theories.  But it does point out a trail to further progress.   

The present paper introduces T$e$V$e\,$S, a new relativistic
gravitational theory devoid of \textit{a priori} fields, whose
nonrelativistic weak acceleration limit accords with MOND while its
nonrelativistic strong acceleration regime is Newtonian. 
T$e$V$e\,$S is based on a metric, and dynamic scalar and 4-vector
fields (one each); it naturally involves one free function, a length
scale, and two positive dimensionless parameters,
$k$ and $K$.   T$e$V$e\,$S passes the usual solar system tests of GR,
predicts gravitational lensing in agreement with the observations
(without requiring dark matter), does not exhibit superluminal
propagation, and provides a specific formalism for constructing
cosmological models.

In Sec.~\ref{sec:Found} we summarize the foundations on which a
workable relativistic formulation of MOND must stand.  We follow
this with a brief critical review of relativistic AQUAL, PCG and
disformal metric theories, some of whose elements we borrow. 
Sec.~\ref{sec:actions} builds the action for  T$e$V$e\,$S while
Sec.~\ref{sec:eqns} derives the equations for the metric, scalar and
vector fields.  In Sec.~\ref{sec:GRlimit} we demonstrate that
T$e$V$e\,$S has a GR limit for a range of small $k$ and $K$.  This is
shown explicitly for cosmology (Sec.~\ref{sec:cosmology}) and for
quasistatic situations like galaxies
(Sec.~\ref{sec:quasistationary}).   All the above applies for any
choice of the free function; in Sec.~\ref{sec:choice} we make a
simple choice for it which facilitates further elaboration.   For
spherically symmetric systems the nonrelativistic MOND limit is
derived in Sec.~\ref{sec:Mondish}, while the Newtonian limit is
recovered for modestly small $k$ in Sec.~\ref{sec:Newtonian}.  The
above conclusions are extended to nonspherical systems in 
Sec.~\ref{sec:nonspherical}.  Sec.~\ref{sec:postN} shows that the
theory passes the usual post-Newtonian solar system tests.  
Sec.~\ref{sec:lensing} demonstrates
that for given dynamics, T$e$V$e\,$S gives the same gravitational
lensing as does a dynamically successful dark halo model within GR. 
In Sec.~\ref{sec:models} we discuss T$e$V$e\,$S cosmological models
with flat spaces showing that they are very similar to the
corresponding GR models (apart from the question of cosmological
dark matter which is left open), and demonstrating that the scalar
field evolves little, and so can be taken to be small and
positive.   As discussed next in Sec.~\ref{sec:causality}, this
last conclusion serves to rule out superluminal propagation in
T$e$V$e\,$S.

\section{Theoretical Foundations for the MOND
paradigm\label{sec:Found}}

\subsection{\label{sec:nonrelat}AQUAL: nonrelativistic field
reformulation of MOND}

However successful empirically when describing motions of test
particles e.g. stars in the collective field of a galaxy,  formula
(\ref{MOND}) is not fully correct.  It is easily checked that a pair
of particles accelerating one in the field of the other according to
(\ref{MOND}) does
\textit{not} conserve momentum.   Thus the MOND formula by itself is
not a theory.  It is, however, a simple matter to construct a fully
satisfactory
\textit{nonrelativistic} theory for MOND~(~\cite{BekMilg}). 
Suppose we retain  the Galilean and rotational invariance of the
Lagrangian density which gives Poisson's equation, but drop the
requirement of linearity of the equation.  Then we come up with  
\begin{equation} 
\mathcal{L} = -{\mathfrak{a}_0^2\over 8\pi G}\,
f\Big({|\bm{\nabla}\Phi|^2\over
\mathfrak{a}_0^2}\Big)-\rho\Phi. \label{AQUAL}
\end{equation} Here $\rho$ is the mass density, $\mathfrak{a}_0$ is
a scale of acceleration introduced for dimensional consistency, and
$f$ is some function.   Newtonian theory (Poisson's equation)
corresponds to the choice 
$f(y)=y$.   From Eq.~(\ref{AQUAL}) follows  the gravitational field
equation
\begin{equation}
\bm{\nabla}\cdot\left[\tilde\mu(|\bm{\nabla}\Phi|/\mathfrak{a}_0)\,\bm{\nabla}\Phi\right]=4\pi
G\rho,
\label{AQUALeq} 
\end{equation} where $\tilde\mu(\surd y)\equiv df(y)/dy$.  Because
of its AQUAdratic Lagrangian,  the theory has been called
AQUAL~\cite{Can2}.  The form of $f$ and the value of
$\mathfrak{a}_0$ must be supplied by phenomenology.  We assume
\begin{equation} f(y)\longrightarrow\cases {y &$y\gg 1$;\cr 
{\scriptstyle 2\over\scriptstyle 3}y^{3/2} &$y\ll 1$.\cr}
\label{f}
\label{cases}
\end{equation}

For systems with spherical, cylindrical or planar geometry,
Eq.~(\ref{AQUALeq}) can be integrated once immediately.  With the
usual prescription for the acceleration,  
\begin{equation}
\textbf{a}=-\bm{\nabla}\Phi, \label{Newton}
\end{equation} the solution corresponds precisely to the MOND
formula (\ref{MOND}).   This is no longer true for lower symmetry. 
However, numerical integration reveals that (\ref{MOND}) is
approximately true, in most cases to respectable
accuracy~\cite{Mnumer}.  

The mentioned  inexactness of Eq.~(\ref{MOND})  for systems such as
a discrete collection of particles is at the root of the mentioned
violation of the conservation laws. Because AQUAL starts from a
Lagrangian, it respects all the usual conservation laws (energy,
momentum and angular momentum), as can be checked
directly~\cite{BekMilg}.  This supplies the appropriate perspective
for the mentioned failings of MOND.  AQUAL also supplies the tools
for showing that Newtonian behavior of the constituents of a large
body, e.g. a star, is consistent with non-Newtonian dynamics of the
latter's center of mass in the weak collective field of a larger
system, e.g. a galaxy.

To summarize, whenever parts of a system devoid of high symmetry
move with accelerations weak on scale $\mathfrak{a}_0$, the field
$\bm{\nabla}\Phi$ which defines their accelerations is to be
calculated by solving the AQUAL equation (\ref{AQUALeq}).  AQUAL
then becomes the nonrelativistic field theory on which to model the
relativistic formulation of the MOND paradigm.  

\subsection{Principles for relativistic MOND}\label{principles}

A relativistic MOND theory seems essential if gravitational lensing
by extragalactic systems and  cosmology are to be understood without
reliance on dark matter.  What principles should the relativistic
embodiment of the MOND paradigm adhere to ?  The following list is
culled from those suggested by Bekenstein~\cite{Can2,Kyoto},
Sanders~\cite{Smn86} and Romatka~\cite{Romatka}. 

\subsubsection{\label{sec:principles}Principles}

$\hskip -5pt\bullet $\quad \textit{Action principle} \quad The
theory must be derivable from an action principle. This is the only
way  known to guarantee that the necessary conservation laws of
energy, linear and angular momentum are incorporated automatically. 
It is simplest to take the action as an integral over a local
lagrangian density.  A nonlocal action has been
tried~\cite{Soussa1}, but the resulting theory fails on account of
gravitational lensing.

$\hskip -5pt\bullet $\quad \textit{Relativistic invariance}\quad 
Innumerable elementary particle experiments  provide direct evidence
for the universal validity of special relativity.  The action should
thus be a relativistic scalar so that all equations of the  theory
are relativistically invariant.  Implied in this is the
correspondence of the theory with special relativity when
gravitation is negligible.   This proviso rules out preferred frame
theories.

$\hskip -5pt\bullet $\quad \textit{Equivalence principle}\quad As
demonstrated with great accuracy (1 part in $10^{12}$) by the 
E\"otv\"os--Dicke experiments~\cite{Will}, free particles with
negligible self--gravity  fall in a gravitational field along
universal trajectories  (weak equivalence principle).  For slow
motion (the case tested by the experiments), the equation
$\mathbf{a}=-\bm{\nabla}\Phi$, which encapsulates the universality,
is equivalent to the geodesic equation in a (curved) metric
$\tilde g_{\alpha\beta}$ with $\tilde g_{tt}\approx -1-2\Phi$.   For
light propagating in a static gravitational field, such a metric
would predict that all frequencies as measured with respect to
(w.r.t.) observers at rest in the field undergo a redshift measured
by $\Phi$.  This is experimentally verified~\cite{VessotLevine} to 1
part in
$10^4$.  It thus appears that a curved metric $\tilde
g_{\alpha\beta}$ describes those properties of spacetime in the
presence of gravitation that are sensed by material objects. 
According to Schiff's conjecture~\cite{Schiff,Will}, this implies
that the theory must be a metric theory, i.e., that in order to
account sfor the effects of gravitation, all
\textit{nongravitational} laws of  physics, e.g. electromagnetism,
weak interactions,
 etc. must be expressed in their usual laboratory forms but with the 
metric
$\tilde g_{\alpha\beta}$ replacing the Lorentz metric.  This is the
Einstein equivalence principle~\cite{Will}.

$\hskip -5pt\bullet $\quad \textit{Causality}\quad  So as not to
violate causality and thereby  compromise the logical consistency of
the theory,  the equations deriving from the action should not permit
superluminal  propagation of any measurable field or of energy and
linear and angular momenta.   Superluminal here means exceeding the
speed which is invariant under the Lorentz transformations.  By
Lorentz invariance of Maxwell's equations this is also the speed of
light.   In curved space, where curvature can cause waves to develop
tails, the maximal speed is that of wavefronts, typically that of the
high frequency components.  

$\hskip -5pt\bullet $\quad \textit{Positivity of energy}\quad
 Fields in the  theory should never carry negative energy. From the
quantum point of view this is a precaution against instability of the
vacuum.  This principle is usually taken to mean that the energy
density of each field should be nonnegative at each event (local
positivity). The fact that the gravitational field itself cannot be
generically assigned an energy density shows that this popular
conception is overly stringent.  A more useful statement of
positivity of energy is that any bounded system must have positive
energy (global positivity instead of the stronger local
positivity).  For example, the gravitational field can carry
negative energy density locally (at least in the Newtonian
conception), yet for pure gravity and in some cases in the presence
of matter, a complete gravitating system is subject to the positive
energy theorems~\cite{Witten}.  Also, there are examples of scalar
fields whose local energy density is of indefinite sign, yet a
complete stationary system of such fields with sources has positive
mass~\cite{Bmass}.   Of course, local positivity implies global
positivity.

$\hskip -5pt\bullet $\quad \textit{Departures from Newtonian
gravity}\quad The theory should  exhibit a preferred scale of
\textit{acceleration} below which departures from Newtonian gravity
should set in, even at low velocities.  

\subsubsection{Requirements}

The relativistic embodiment of MOND should  predict a number of well
established phenomena.  For example, we expect the following.

$\hskip -5pt\star $\quad \textit{Agreement with the extragalactic
phenomenology}: The nonrelativistic limit of the theory should make
predictions in agreement with those of the AQUAL equation, which is
known to subsume much extragalactic phenomenology.  This is checked
for T$e$V$e\,$S in Sec.~\ref{sec:Mondish}.

$\hskip -5pt\star $\quad \textit{Agreement with phenomenology of
gravitational lenses}: The theory should predict correctly the
lensing of electromagnetic radiation by extragalactic structures
which is responsible for gravitational lenses and arcs.  In
particular, it should give predictions similar to those of GR within
the dark matter paradigm.  This point is established for T$e$V$e\,$S
in Sec.~\ref{sec:lensing}.

$\hskip -5pt\star $\quad \textit{Concordance with the solar
system}: The theory should make predictions in  agreement with the
various solar system tests of relativity~\cite{Will}: deflection of
light rays, time delay of radar signals, precessions of the
perihelia of the inner planets,  the absence of the Nordtvedt effect
in the lunar orbit, the nullness of aether drift, etc.  T$e$V$e\,$S
is confronted with the first three tests in Sec.~\ref{sec:postN}.

$\hskip -5pt\star $\quad \textit{ Concordance with binary pulsar
tests} : The theory should make predictions in harmony with the
observed pulse times of arrival from the various binary pulsars. 
These contain information about relativistic time delay, periastron
precession and the orbit's decay due to gravitational radiation. 
They thus constitute a test of the strong
\textit{potential} limit of the theory.

$\hskip -5pt\star $\quad \textit{Harmony with cosmological facts}:
 The theory should give a picture of cosmology in harmony with  basic
empirical facts such as the Hubble expansion, its timescales for
various eras, existence of the microwave background,  light element
abundances from primordial nucleosynthesis, etc.  The similarity of
cosmological evolution in GR and in T$e$V$e\,$S is established in
Sec.~\ref{sec:models}, though the problem of how to eliminate
cosmological dark matter with T$e$V$e\,$S is left open.

\subsection{Some antecedent relativistic
theories\label{sec:previous}}

 It is now in order to briefly review \textit{some} of the previous
attempts to give a relativistic theory of MOND.  This will introduce
the concepts to be borrowed by T$e$V$e\,$S, and help to establish the
notation and conventions that we shall follow. A metric signature
$+2$, and units with
$c=1$ are used throughout this paper.  Greek indeces run over four
coordinates while Latin ones run over the spatial coordinates alone.

\subsubsection{\label{sec:rAQUAL}Relativistic AQUAL}

 It is well known that theories constructed, for example, by using a
local function of the scalar curvature as Lagrangian density, have a
purely Newtonian limit for weak potentials.  So if we steer away from
nonlocal actions, then AQUAL behavior cannot arise from merely
modifying the gravitational action.  The theory one seeks has to
involve degrees of freedom other than the metric.  

In the first relativistic theory with MOND aspirations, relativistic
AQUAL~\cite{BekMilg},  the physical metric $\tilde g_{\alpha\beta}$
was taken as conformal to a primitive (Einstein) metric
$g_{\alpha\beta}$, i.e., $\tilde g_{\alpha\beta}=e^{2\psi}
g_{\alpha\beta}$ with $\psi$ a real scalar field.   In order not to
break  violently with GR, which is well  tested in the solar system
(and to some extent in cosmology), the gravitational action was taken
as the Einstein-Hilbert's one built out of
$g_{\alpha\beta}$.  The MOND phenomenology was implanted by taking
for the Lagrangian density for $\psi$ 
\begin{equation}
\mathcal{L}_{\psi}=-{1\over 8\pi G L^2}
\tilde f\left(L^2 g^{\alpha\beta}\psi,_\alpha \psi,_\beta\right), 
\label{rAQUAL}
\end{equation} where $\tilde f$ is some function (not known
\textit{a priori}), and 
$L$ is a constant with dimensions of length  introduced for
dimensional consistency. Note that when $\tilde f(y)=y$, 
$\mathcal{L}_\psi$ is just the lagrangian density for a linear scalar
field,  but in general $\mathcal{L}_\psi$ is aquadratic. 

To implement the universality of free fall, one must write all
lagrangians of matter fields using a single metric, which is taken as
$\tilde g_{\alpha\beta}$ (not
$g_{\alpha\beta}$ which choice would make the theory GR).  Thus for
example, the action for a particle of mass $m$ is taken as
\begin{equation} S_m=-m\int e^{\psi}\,(-g_{\alpha\beta} dx^\alpha
dx^\beta)^{1/2}. 
\label{Lm}
\end{equation}  Hence test particle motion is nongeodesic w.r.t.
$g_{\alpha\beta}$ but, of course, geodesic w.r.t. $\tilde
g_{\alpha\beta}$.   Evidently this last is the metric measured by
clocks and rods, hence the physical metric.   Addition of a constant
to
$\psi$ merely multiplies all masses  by a constant (irrelevant global
redefinition of units), so that the theory is insensitive to the
choice of zero of
$\psi$. 

For slow motion in a quasistatic situation with nearly flat metric
$g_{\alpha\beta}$, and in a weak field $\psi$,
$e^{\psi}(-g_{\alpha\beta} dx^\alpha dx^\beta)^{1/2}\approx 
(1+\Phi_{\rm N}+\psi-\mathbf{v}^2/2)dt$, were 
$\Phi_{\rm N}=-(g_{tt}+1)/2$ is the Newtonian potential determined
by the mass density $\rho$ through the linearized Einstein equations
for
$g_{\alpha\beta}$, and $\mathbf{v}$ is the velocity defined w.r.t.
the Minkowski metric which is close to $g_{\alpha\beta}$. Thus the
particle's lagrangian is
$m(\mathbf{v}^2/2-\Phi_{\rm N}-\psi)$; this leads to the equation of
motion
\begin{equation}
\mathbf{a}\approx -\bm{\nabla}(\Phi_{\rm N}+\psi).
\label{acceleration}
\end{equation}

How is $\psi$ determined ?  For stationary weak fields the Lagrangian
\textit{density} for $\psi$, including a point source of physical
mass
$M$ at 
$\mathbf{r}=0$, is from the above discussion and
Eqs.~(\ref{rAQUAL})-(\ref{Lm}),
\begin{equation}
\mathcal{L}_{\psi}=-{1\over 8\pi G L^2}\tilde f\left(L^2
(\bm{\nabla}\psi)^2\right)-\psi M\delta(\mathbf{r}).
  \label{effAQUAL}
\end{equation} Comparing Eqs.~(\ref{effAQUAL}) and~(\ref{AQUAL})
we  conclude that $\psi$ here corresponds to 
$\Phi$ of mass $M$ as computed from AQUAL's Eq.~(\ref{AQUALeq}),
provided we take $\tilde f=f$ and
$L=1/\mathfrak{a}_0$.  Whenever $|\bm{\nabla}\psi|\gg
|\bm{\nabla}\Phi_{\rm N}|$ ($\Phi_{\rm N}$ is the Newtonian
potential  of the same mass distribution),  the equation of motion
(\ref{acceleration}) reduces to (\ref{Newton}), and we  obtain MOND
like dynamics.  For the choice of MOND function (\ref{cases}) the
said strong inequality is automatic in the deep MOND regime,
$|\bm{\nabla}{\psi}|\ll \mathfrak{a}_0$, because $\tilde\mu\ll 1$
there. 

In the regime $|\bm{\nabla}\psi|\gg \mathfrak{a}_0$,
$\tilde\mu\approx 1$ and
$f(y)
\approx y$  so that $\psi$ reduces to $\Phi_{\rm N}$.  It would seem
from Eq.~(\ref{acceleration}) that a particle's acceleration is then
twice the correct Newtonian value.  However, this just means that the
measurable Newton's constant $G_N$ is twice the bare $G$ appearing in
$\mathcal{L}_\psi$ or in Einstein's equations. It is thus clear,
regarding dynamics, that the relativistic AQUAL theory has the
appropriate MOND and Newtonian limits depending on the strength of
$\bm{\nabla}\psi$.   

But  relativistic AQUAL has problems. Early
on~\cite{BekMilg,Can2,Rosen} it was realized that $\psi$ waves can
propagate faster than light.   This acausal behavior can be traced to
the aquadratic form of  the lagrangian, as  explained in
Appendix~\ref{sec:A}.   A second problem ~\cite{Kyoto,Romatka} 
issues from the conformal relation $\tilde 
g_{\alpha\beta}=e^{2\psi}g_{\alpha\beta}$.   Light  propagates on
the null cones of the physical metric; by the conformal relation
these coincide with the lightcones of the Einstein metric.  This
last is calculated from Einstein's equations with the visible
matter and
$\psi$ field as sources.  Thus so long as the $\psi$ field
contributes comparatively little to the energy momentum tensor, it
cannot affect light deflection, which will thus be that due to the
visible matter alone.  But in reality galaxies and clusters of
galaxies are observed to deflect light stronger than the visible
mass in them would suggest. Thus relativistic AQUAL fails to
accurately describe light deflection in situations in which GR
requires dark matter.  It is thus empirically falsified.

Relativistic AQUAL bequeaths to T$e$V$e\,$S the use of a scalar
field to connect Einstein and physical metrics, a field which
satisfies an equation reminiscent of the nonrelativistic AQUAL
Eq.~(\ref {AQUALeq}).

\subsubsection{\label{PCG}Phase coupling gravitation} 

The Phase Coupled Gravity (PCG)  theory was
proposed~\cite{Can2,phase,Rosen} in order to resolve relativistic
AQUAL's acausality problem.  It retains the two metrics related by
$\tilde g_{\alpha\beta}=e^{2\psi} g_{\alpha\beta}$, but envisages
$\psi$ as one of a pair of mutually coupled real scalar fields with
the Lagrangian density (our definitions here differ slightly from
those in Ref.~\onlinecite{Can2})
\begin{equation}
\mathcal{L}_{\psi,A}=-{\scriptstyle 1\over \scriptstyle 
2}\left[g^{\alpha\beta} (A_{,\alpha} A_{,\beta} + \eta^{-2} A^2
\psi_{,\alpha} \psi_{,\beta})+\mathcal{V}(A^2)\right]
\label{PCGL}
\end{equation} Here $\eta$ is a real parameter and $\mathcal{V}$ a
real valued function.  The coupling between $A$ and $\psi$ is
designed to bring about AQUAL-like features for small $|\eta|$. The
theory receives its name because matter is coupled to $\psi$, which
is proportional to the phase of the self-interacting  complex field
$\chi=Ae^{\imath
\psi/\eta}$.  

Variation of $\mathcal{L}_{\psi,A}$ w.r.t. $A$ leads to (all
covariant derivatives and index raising w.r.t.
$g_{\alpha\beta}$)
\begin{equation}
A^{,\alpha}{}_{;\alpha}-\eta^{-2}A\,\psi_{,\alpha}\psi^{,\alpha}
-A\,\mathcal{V}'(A^2)=0
\label{Aequation}
\end{equation} In the variation w.r.t. $\psi$ we must include the
Lagrangian density of a source, say a point mass $M$ at rest at
$\mathbf{r}=0$ [c.f. 
$S_m$ in Eq.~(\ref{Lm})]:
\begin{equation} 
\left(A^2 g^{\alpha\beta}\psi_{,\beta}\right)_{;\alpha}=\eta^2
 e^{\psi} M\delta(\mathbf{r})
\label{phiequation}
\end{equation} The connection with AQUAL is now clear.  For
sufficiently small
$|\eta|$ the $A^{,\alpha}{}_{;\alpha}$ term in
Eq.~(\ref{Aequation}) becomes negligible, and the other two
establish an algebraic relation between
$\psi_{,\alpha}\psi^{,\alpha}$ and $A^2$.  Substituting this in
Eq.~(\ref{phiequation}) gives the AQUAL type of equation for $\psi$
that would derive from $\mathcal{L}_\psi$ in Eq.~(\ref{rAQUAL}). 

The PCG Lagrangian's advantage  over that of the relativistic
AQUAL's is precisely in that it involves first derivatives only in
quadratic form.  This would seem to rule out the superluminality
generating 
$X^\alpha$ dependent terms discussed in Appendix~\ref{sec:A}.  In
practice things are more complicated.  A detailed local analysis
employing the eikonal approximation~\cite{Rosen} shows that there
are superluminal
$\psi$ perturbations, for example when $\mathcal{V}''<0$.  However,
the same analysis shows that such superluminality occurs only when
the background solution is itself locally unstable.  This makes the
said causality violation moot. 

One way to obtain MOND phenomenology from PCG is to choose
$\mathcal{V}(A^2)=-{\scriptscriptstyle 1\over \scriptscriptstyle
3}\varepsilon^{-2} A^6$ with $\varepsilon$ a constant with dimension
of energy.  Although with this choice  $\mathcal{V}''<0$ which makes
for unstable backgrounds, we only need this form for small
$A$; $\mathcal{V}$ can take different form for large argument.  
Then in a static situation with nearly flat
$g_{\alpha\beta}$ and  weak $\psi$, 
Eqs.~(\ref{Aequation})-(\ref{phiequation}) reduce to
\begin{eqnarray}
\nabla^2 A-\eta^{-2}A(\bm{\nabla}\psi)^2+\varepsilon^{-2} A^5=0,
\label{phieq1}
\\
\bm{\nabla}\cdot(A^2\bm{\nabla}\psi)=\eta^2 M 
\delta(\mathbf{r}).
\label{phieq2}
\end{eqnarray}  The spherically symmetric solution of
Eqs.~(\ref{phieq1})-(\ref{phieq2}) is
\begin{eqnarray}  A&=&(\kappa\varepsilon/r)^{1/2};\quad d\psi/dr=
(\eta\varpi/4\kappa r)
\\
\varpi&\equiv& (\eta M/\pi\varepsilon);\quad\kappa\equiv 2^{-3/2}
\left(1+\sqrt{1+4\varpi^2}\right)^{1/2}
\end{eqnarray}

One may evidently still use Eq.~(\ref{acceleration}):
\begin{equation} a_r = -GM/r^2 - (\eta^2M/4\pi\varepsilon
\kappa r)
\label{modified}
\end{equation} Thus a $1/r$ force competes with the Newtonian one. 
For small
$M$ it starts to dominate at a fixed radius scale $r_c$, just as in
Tohline's~\cite{Tohline} and Kuhn-Kruglyak's~\cite{KuhnKruglyak}
non-Newtonian gravity theories.  Here  $r_c=2\pi
G\varepsilon/\eta^2$.  By contrast for $M\gg M_c\equiv
{\scriptscriptstyle 1\over\scriptscriptstyle 2}\pi\varepsilon/\eta$,
$\kappa\approx  {\scriptscriptstyle 1\over\scriptscriptstyle
2}\surd \varpi$ and the
$1/r$ force scales as $M^{1/2}$ and begins to dominate when the
Newtonian acceleration drops below the fixed acceleration scale
\begin{equation}  \mathfrak{a}_0\equiv \eta^3/(4\pi G\varepsilon).
\end{equation}

For $a_r\ll \mathfrak{a}_0$ the circular velocity whose centripetal
acceleration balances the
$1/r$ force is $v_c=(G\mathfrak{a}_0M)^{1/4}$, precisely as in
MOND.  Thus
$\mathfrak{a}_0$ here is to be identified with Milgrom's constant
$\mathfrak{a}_0$.  We conclude that, with a suitable choice of
parameters, PCG with a sextic potential recovers the main features of
MOND: asymptotically flat rotation curves and the TFL
for disk galaxies. Specifically, the choice $\eta=10^{-8}$ and
$\varepsilon=10^{53}\,
\textrm{erg}\ $ gives
$\mathfrak{a}_0=8.7\times 10^{-9} \textrm{cm s}^{-2}$, 
$M_c=8.7\times 10^{6} M_\odot$ and $r_c=5.2\times 10^{19}$ cm.  Now
since $r_c$ is larger than the Hubble scale, the
Tohline-Kuhn-Kruglyak $ 1/r$ force is comparatively unimportant. 
Hence for $M\gg 10^{7} M_\odot$ we should have MOND, and for low
masses almost Newtonian behavior.  This is about right: globular
star clusters at $10^4-10^5 M_\odot$ show no missing mass problem.
  
However, the above parameters are bad from the point of view of the
solar system tests of gravity, as summarized in
Appendix~\ref{sec:B}.  But the gravest problem with PCG is that it,
just as AQUAL,  provides insufficient light
deflection~\cite{Kyoto}.  Here again, the conformal relation
between Einstein and physical metric is to blame.  T$e$V$e\,$S
incorporates PCG's Lagrangian density (\ref{PCGL})  in the limit of
small
$\eta$ in which $A$ becomes nondynamical.

\subsubsection{\label{disf}Theories with disformally related
metrics}

The light deflection problem can be solved only by giving up the
relation $\tilde g_{\alpha\beta}=e^{-2\psi} g_{\alpha\beta}$.  It
was thus suggested~\cite{Kyoto} to replace this conformal relation
by a disformal one, namely 
\begin{equation}
\tilde g_{\alpha\beta}=e^{-2\psi}( \mathcal{A} g_{\alpha\beta}+
\mathcal{B} L^2\psi_{,\alpha}\psi_{,\beta}),
\end{equation} with $ \mathcal{A}$ and $ \mathcal{B}$ functions of
the invariant
$g^{\mu\nu}\psi_{,\mu}\psi_{,\nu}$ and $L$ a constant length
unrelated, of course, to that in Eq.~(\ref{rAQUAL}).  This relation
already allows
$\psi$ to deflect light via the
$\psi_{,\alpha}\psi_{,\beta}$ term in the physical metric. However,
it was found~\cite{BekSan} that if one insists on causal propagation
of both light and gravitational waves w.r.t. the light cones of the
physical metric, then the sign required of $ \mathcal{B}$ is opposite
that required to enhance the light deflection coming from the metric
$g_{\alpha\beta}$ alone. Thus the cited disformal relation between
metrics, if respecting causality, will give weaker light deflection
than would
$g_{\alpha\beta}$ were it the physical metric.  

This last observation of Ref.~\onlinecite{BekSan} has given rise to
a folk belief that relativistic gravity theories which attempt to
supplant dark matter's dynamical effects necessarily reduce light
deflection rather than enhancing
it~\cite{Edery,Aguirre,Soussa1,Soussa2}.  However, as remarked by
Sanders, the mentioned problem disappears if the term 
$\psi_{,\alpha}\psi_{,\beta}$ is replaced by
$\mathfrak{U}_\alpha \mathfrak{U}_\beta$, where
$\mathfrak{U}_\alpha$  is a constant 4-vector which, at least in the
solar system and within galaxies, points in the time
direction~\cite{Sandersstratified}.  Specifically Sanders takes
$\tilde g_{\alpha\beta} = e^{-2\psi} g_{\alpha\beta}
-2\mathfrak{U}_\alpha
\mathfrak{U}_\beta \sinh (2\psi)$.  

This ``stratified'' gravitation theory is reported to do well in the
confrontation with the solar system tests, and to possess the right
properties to explain the coincidence between $\mathfrak{a}_0$ of
MOND and the Hubble scale~\cite{M1}.  But its vector
$\mathfrak{U}_\alpha$ is an \textit{a priori} nondynamical element
whose direction is selected in an unspecified way by the cosmological
background.   This means the theory is a preferred frame theory
(although it is reported to be protected on this account against
falsification in the solar system and other strong acceleration
systems by its AQUAL behavior~\cite{Sandersstratified}).   This is
obviously a conceptual shortcoming which T$e$V$e\,$S removes, but the
latter's debt to the stratified theory should be underlined.

\section{Fundamentals of T$e$V$e\,$S\label{sec:fundamentals}}

\subsection{Fields and actions\label{sec:actions}}

T$e$V$e\,$S is based on three dynamical gravitational fields: an
Einstein metric
$g_{\mu\nu}$ with a well defined inverse
$g^{\mu\nu}$, a timelike 4-vector field $\mathfrak{U}_\mu$  such that
\begin{equation} g^{\alpha\beta} \mathfrak{U}_\alpha
\mathfrak{U}_\beta=-1,
\label{norm}
\end{equation} and a scalar field $\phi$; there is also a
nondynamical scalar field  $\sigma$ (the acronym T$e$V$e\,$S recalls
the theory's Tensor-Vector-Scalar content). The physical metric in
T$e$V$e\,$S, just as in Sanders' stratified theory, is obtained by
stretching the Einstein metric in the spacetime directions orthogonal
to
$\mathfrak{U}^\alpha\equiv g^{\alpha\beta}\mathfrak{U}_\beta$ by a
factor
$e^{-2\phi}$, while shrinking it by the same factor in the direction
parallel to
$\mathfrak{U}^\alpha$:
\begin{eqnarray}
\tilde g_{\alpha\beta} &=& e^{-2\phi}
(g_{\alpha\beta}+\mathfrak{U}_\alpha
\mathfrak{U}_\beta) - e^{2\phi}\mathfrak{U}_\alpha
\mathfrak{U}_\beta
\\
 &=&  e^{-2\phi} g_{\alpha\beta} -2\mathfrak{U}_\alpha
\mathfrak{U}_\beta
\sinh (2\phi)  
\label{physg}
\end{eqnarray}     It is easy to verify that the inverse physical
metric is
\begin{equation}
\tilde g^{\alpha\beta} = e^{2\phi} g^{\alpha\beta} +
2\mathfrak{U}^\alpha
\mathfrak{U}^\beta \sinh (2\phi)  
\label{inverse}
\end{equation} where $\mathfrak{U}^\alpha$ will {\it always} mean
$g^{\alpha\beta}\mathfrak{U}_\beta$.

The geometric part of the action, $S_g$, is formed from the Ricci
tensor
$R_{\alpha\beta}$ of $g_{\mu\nu}$ just as in GR:
\begin{equation} S_g=(16\pi G)^{-1}\int g^{\alpha\beta}
R_{\alpha\beta} (-g)^{1/2} d^4 x.
\label{EH}
\end{equation} Here $g$ means the determinant of metric
$g_{\alpha\beta}$. This choice is made in order to keep  T$e$V$e\,$S
close to  GR in some sense to be clarified below.

In terms of two constant positive parameters, $k$ and $\ell$, the
action for the pair of scalar fields is taken to be of roughly PCG
form,
\begin{equation} S_s =-{\scriptstyle 1\over\scriptstyle
2}\int\big[\sigma^2
h^{\alpha\beta}\phi_{,\alpha}\phi_{,\beta}+{\scriptstyle
1\over\scriptstyle 2}G
\ell^{-2}\sigma^4 F(kG\sigma^2) \big](-g)^{1/2} d^4 x,
\label{scalar}
\end{equation}   where $h^{\alpha\beta}\equiv g^{\alpha\beta}
-\mathfrak{U}^\alpha \mathfrak{U}^\beta$ and $F$ is a free
dimensionless function (it is related to PCG's potential
$\mathcal{V}$).  No overall coefficient is required for the kinetic
term; were it included, it could be absorbed into a redefinition of
$\sigma$ and thereby in $k$ and
$\ell$.  Because $\phi$ is obviously dimensionless, the dimensions of
$\sigma^2$ are those of $G^{-1}$.  Thus $k$ is a dimensionless
constant (it could be absorbed into the definition of $F$, but we
choose to exhibit it), while $\ell$ is a constant length. 

Because no kinetic $\sigma$ terms appear, the ``equation of motion''
of
$\sigma$ takes the form of an algebraic relation between it and the
invariant $h^{\alpha\beta}\phi_{,\alpha}\phi_{,\beta}$, and when
this is substituted for
$\sigma$ in $S_s$, the phenomenologically successful AQUAL type
action for $\phi$ appears.  We could, of course, have written this
last action directly.  The present route is more suggestive of the
possible origin of the action; for example, $S_s$ resembles the
action for a complex self-interacting scalar field $\eta\sigma
\exp(\imath\phi/\eta)$ in the limit of small $\eta$.    The term
$-\sigma^2 \mathfrak{U}^\alpha
\mathfrak{U}^\beta\phi_{,\alpha}\phi_{,\beta}$ here included in the
scalar's action is new; its role is to eliminate superluminal
propagation of the $\phi$ field, a recalcitrant problem in AQUAL
type theories. 

The action of the vector $\mathfrak{U}^\alpha$ is taken to have the
form
\begin{equation} S_v =-{K\over 32\pi G}\int
\big[g^{\alpha\beta}g^{\mu\nu} 
\mathfrak{U}_{[\alpha,\mu]} \mathfrak{U}_{[\beta,\nu]}
-2(\lambda/K)(g^{\mu\nu}\mathfrak{U}_\mu
\mathfrak{U}_\nu +1)\big](-g)^{1/2} d^4 x,
\label{vector}
\end{equation}  where antisymmetrization in a pair of indeces is
indicated by surrounding them by square brackets, e.g. $A_{[\mu}
B_{\nu]}=A_\mu B_\nu-A_\nu B_\mu$. In Eq.~(\ref{vector}) 
$\lambda$ is a spacetime dependent Lagrange multiplier enforcing the
normalization Eq.~(\ref{norm}) (we shall calculate
$\lambda$ later), while $K$ is a dimensionless constant since
$\mathfrak{U}^\alpha$ is dimensionless.  Thus T$e$V$e\,$S has two
dimensionless parameters, $k$ and $K$, in addition to the
dimensional constants $G$ and
$\ell$. The kinetic terms in Eq.~(\ref{vector}) are chosen
antisymmetric not because of any desire for gauge symmetry, which is
broken by the form of the physical metric anyway, but because this
choice precludes appearance of second derivatives of
$\mathfrak{U}_\alpha$ in the energy--momentum tensor of T$e$V$e\,$S
(see next subsection).  The action $S_v$ is a special case of that
in Jacobson and Mattingly's generalization of GR with a preferred
frame~\cite{Jac_Mat}.   

In accordance with the equivalence principle, the matter action in
T$e$V$e\,$S is obtained by transcribing the flat spacetime lagrangian
$\mathcal{L}(\eta_{\mu\nu}, f^\alpha, \partial_{\mu}f^\alpha,\,
\cdots)$ for fields written schematically
$f^\alpha$ as
\begin{equation} S_m=\int \mathcal{L}(\tilde g_{\mu\nu}, f^\alpha,
f^\alpha{}_{|\mu},
\, \cdots)(-\tilde g)^{1/2} d^4x,
\label{matter}
\end{equation} where the covariant derivatives denoted by $_|$ are
taken w.r.t.  $ \tilde g_{\mu\nu}$.  This has the effect that the
spacetime delineated by matter dynamics has the  metric $\tilde
g_{\mu\nu}$.  The appearance of
$(-\tilde g)^{1/2}$ here requires us to specify its relation to
$(-g)^{1/2}$.  In Appendix~\ref{sec:C} we show that
\begin{equation} (-\tilde g)^{1/2}= e^{-2\phi}(-g)^{1/2} 
\label{twog's}
\end{equation} 
By coupling to matter only through $\tilde g_{\alpha\beta}$, the
field $\mathfrak{U}_\alpha$ is totally different from the Lee-Yang
4-vector field with gravitation strength
interaction~\cite{LeeYang}, whose existence is ruled out by the
equivalence principle tests as well as by cosmological symmetry
arguments~\cite{LeeYang,Dicke}.

\subsection{Basic equations\label{sec:eqns}}

We shall obtain the basic equations by varying the total action
$S=S_g+S_s+S_v+S_m$ wth respect to the basic fields
$g^{\alpha\beta}$,
$\phi$, $\sigma$ and $\mathfrak{U}_\alpha$.  To this end we must be
explicit about how
$\tilde g_{\alpha\beta}$, which enters into $S_m$, varies with the
basic fields. Taking increments of Eq.~(\ref{inverse}) we get
\begin{widetext}
\begin{equation}
\delta\tilde g^{\alpha\beta} = e^{2\phi} \delta g^{\alpha\beta} + 2
\sinh (2\phi)  \mathfrak{U}_\mu\delta
g^{\mu(\alpha}\mathfrak{U}^{\beta)} +
 2 \big[e^{2\phi} g^{\alpha\beta} + 2\mathfrak{U}^\alpha
\mathfrak{U}^\beta \cosh (2\phi)\big]\delta\phi+ 2\sinh(2\phi)
\mathfrak{U}^{(\alpha} g^{\beta)\mu} \delta
\mathfrak{U}_\mu
\label{dg}
\end{equation}
\end{widetext} where symmetrization in a pair of indeces is
indicated by surrounding them by round brackets, e.g. $A^{(\mu}
B^{\nu)}=A^\mu B^\nu+A^\nu B^\mu$.

\subsubsection{Equations for the metric\label{sec:geqns}}

When varying $S$ w.r.t. $g^{\alpha\beta}$ we recall that 
$\delta S_g = (16\pi G)^{-1} G_{\alpha\beta} (-g)^{1/2} \delta
g^{\alpha\beta}$ ($G_{\alpha\beta}$ denotes the Einstein tensor of
$g_{\alpha\beta}$) while
\begin{equation}
\delta S_m=-{\scriptstyle 1\over\scriptstyle 2}
 \tilde T_{\alpha\beta}(-\tilde g)^{1/2}\delta \tilde
g^{\alpha\beta} +
\dots
\label{T}
\end{equation}
 where the ellipsis denotes variations of the
$f^\alpha$ fields, and $\tilde T_{\alpha\beta}$ stands for the
physical energy--momentum tensor defined with the metric
$\tilde g_{\alpha\beta}$.  We get
\begin{equation} G_{\alpha\beta} = 8\pi G\Big[\tilde
T_{\alpha\beta} +(1-e^{-4\phi}) \mathfrak{U}^\mu
\tilde T_{\mu(\alpha} \mathfrak{U}_{\beta)}
+\tau_{\alpha\beta}\Big]+
\Theta_{\alpha\beta}
\label{gravitationeq}
\end{equation} where
\begin{eqnarray}
\tau_{\alpha\beta}&\equiv&
\sigma^2\Big[\phi_{,\alpha}\phi_{,\beta}-{\scriptstyle 1\over
\scriptstyle 2}g^{\mu\nu}\phi_{,\mu}\phi_{,\nu}\,g_{\alpha\beta}-
\mathfrak{U}^\mu\phi_{,\mu}\big(\mathfrak{U}_{(\alpha}\phi_{,\beta)}-
{\scriptstyle 1\over \scriptstyle
2}\mathfrak{U}^\nu\phi_{,\nu}\,g_{\alpha\beta}\big)\Big]
\label{tau}
\nonumber
\\ &-&{\scriptstyle 1\over\scriptstyle 4}G
\ell^{-2}\sigma^4 F(kG\sigma^2)  g_{\alpha\beta}
\end{eqnarray}
\begin{equation}
\Theta_{\alpha\beta}\equiv
K\Big(g^{\mu\nu}\mathfrak{U}_{[\mu,\alpha]}
\mathfrak{U}_{[\nu,\beta]} -{\scriptstyle 1\over \scriptstyle 4}
g^{\sigma\tau}g^{\mu\nu}\mathfrak{U}_{[\sigma,\mu]}
\mathfrak{U}_{[\tau,\nu]}\,g_{\alpha\beta}\Big)-
\lambda \mathfrak{U}_\alpha\mathfrak{U}_\beta
\label{Theta}
\end{equation} When varying $g^{\alpha\beta}$ in $S_v$ we have used
Eq.~(\ref{norm}) to drop a term proportional to
$g_{\alpha\beta}$.

\subsubsection{Scalar equation\label{sec:seqns}}

Variation of $\sigma$ in $S_s$ gives the relation between $\sigma$
and
$\phi_{,\alpha}$ ($F'\equiv dF(\mu)/d\mu$),
\begin{equation}
 -kG\sigma^2 F-{\scriptstyle 1/\scriptstyle 2 }\,(kG\sigma^2)^2
F'=k\ell^2 h^{\alpha\beta}\phi_{,\alpha}\phi_{,\beta}
\label{sigma}
\end{equation}  In carrying out the variation w.r.t. $\phi$ it must
be remembered that this quantity enters in $S_m$ exclusively through
$\tilde g^{\alpha\beta}$, so that use must be made of
Eqs.~(\ref{dg})-(\ref{T}):
\begin{equation}
\big[\sigma^2 h^{\alpha\beta}\phi_{,\alpha}\big]_{;\beta}=  \big[
g^{\alpha\beta} + (1+e^{-4\phi})\mathfrak{U}^\alpha
\mathfrak{U}^\beta
\big]
\tilde T_{\alpha\beta}
\label{phi} 
\end{equation} In view of Eq.~(\ref{sigma}) this is an equation for
$\phi$ only, with $\tilde T_{\alpha\beta}$ as source.  

Suppose we define a function $\mu(y)$ by
\begin{equation} 
 -\mu F(\mu ) -{\scriptscriptstyle 1\over \scriptscriptstyle 2}\,
\mu ^2F'(\mu ) = y.
\label{F}
\end{equation} 
 so that
$kG\sigma^2=\mu
(k\ell^2h^{\alpha\beta}\phi_{,\alpha}\phi_{,\beta})$.   We may now
recast Eq.~(\ref{phi}) as
\begin{equation}
\left[\mu\left(k\ell^2 h^{\mu\nu}\phi_{,\mu}\phi_{,\nu}\right)
h^{\alpha\beta}\phi_{,\alpha} \right]_{;\beta}= 
kG\big[g^{\alpha\beta} + (1+e^{-4\phi}) \mathfrak{U}^\alpha
\mathfrak{U}^\beta\big] \tilde T_{\alpha\beta}.
\label{s_equation}
\end{equation}  This equation is reminiscent of the relativistic
AQUAL scalar equation [see Appendix~\ref{sec:A},
Eq.~(\ref{rAQUALeq})], albeit with the replacement
$g^{\alpha\beta}\mapsto h^{\alpha\beta}$ in the l.h.s.  In
quasistatic situations we may replace
$h^{\alpha\beta}$ by $g^{\alpha\beta}$ so that
Eq.~(\ref{s_equation}) has the same structure as the AQUAL equation.

\subsubsection{Vector equation\label{sec:veqns}}

Variation of $S$ w.r.t. $\mathfrak{U}_\alpha$ and use of
Eq.~(\ref{dg}) gives the vector equation
\begin{equation} K
\mathfrak{U}^{[\alpha;\beta]}{}_{;\beta}+\lambda
\mathfrak{U}^\alpha+8\pi G\sigma^2
\mathfrak{U}^\beta\phi_{,\beta}g^{\alpha\gamma}\phi_{,\gamma} =
8\pi G (1-e^{-4\phi}) g^{\alpha\mu} \mathfrak{U}^\beta 
\tilde T_{\mu\beta}
\label{vectoreq}
\end{equation} As mentioned, $\lambda$ here is a Lagrange
multiplier.  It can be solved for by contracting the previous
equation with
$\mathfrak{U}_\alpha$.  Substituting it back gives
\begin{eqnarray} &K&
\left(\mathfrak{U}^{[\alpha;\beta]}{}_{;\beta}+\mathfrak{U}^\alpha
\mathfrak{U}_\gamma 
\mathfrak{U}^{[\gamma;\beta]}{}_{;\beta}\right)+ 8\pi G\sigma^2
\big[\mathfrak{U}^\beta\phi_{,\beta}\,g^{\alpha\gamma}\phi_{,\gamma}+
\mathfrak{U}^\alpha (\mathfrak{U}^\beta\phi_{,\beta})^2\big]
\nonumber
 \\ &=&8\pi G (1-e^{-4\phi})
\big[g^{\alpha\mu} \mathfrak{U}^\beta  \tilde T_{\mu\beta} + 
\mathfrak{U}^\alpha
\mathfrak{U}^\beta
\mathfrak{U}^\gamma \tilde T_{\gamma\beta}\big] 
\label{vectoreq2}
\end{eqnarray} This equation has only three independent components
since both sides of it are orthogonal to
$\mathfrak{U}_\alpha$.  It thus determines three components of
$\mathfrak{U}^\alpha$ with the fourth being determined by the
normalization (\ref{norm}).  Like any other partial differential
equation, the vector equation does not by itself determine
$\mathfrak{U}_\alpha$ uniquely. 

\subsection{General relativity limit\label{sec:GRlimit}}

T$e$V$e\,$S has three parameters: $k,\ell$ and $K$.  Here we show
first that in several familiar contexts the limit $k\rightarrow 0$,
$\ell\propto k^{-3/2}$, $K\propto k$ of it corresponds to standard
GR for any form of the function $F$.   Many of the intermediate
results will beuseful in Sec.~\ref{sec:postN} and \ref{sec:models}. 
We then expand on a remark by Milgrom that the GR limit actually
follows under more general circumstances: $K\rightarrow 0$ and
$\ell\rightarrow\infty$.  

Whenever a specific matter content is needed, we shall assume the
matter to be an ideal fluid.  Its energy-momentum tensor has the
familiar form
\begin{equation}
\tilde T_{\alpha\beta}=\tilde\rho \tilde u_\alpha\tilde u_\beta
+\tilde p(\tilde g_{\alpha\beta}+\tilde u_\alpha \tilde u_\beta)
\label{oldT},
\end{equation} where 
$\tilde\rho$ is the proper energy density, $\tilde p$ the pressure
and
$\tilde u_\alpha$ the 4-velocity, all three expressed in the physical
metric.  We may profitably simplify Eq.~(\ref{s_equation}) in any
case when for symmetry reasons $\tilde u_\alpha$ is collinear with
$\mathfrak{U}_\alpha$.  In order that the velocity be normalized
w.r.t.
$\tilde g_{\alpha\beta}$, we must take in that case
$\tilde u_\alpha=e^{\phi}\,\mathfrak{U}_\alpha$ from which follows
\begin{equation}
\tilde g_{\alpha\beta}+\tilde u_\alpha \tilde u_\beta =
e^{-2\phi}(g_{\alpha\beta}+\mathfrak{U}_\alpha\mathfrak{U}_\beta).
\end{equation} Substituting this in $\tilde T_{\alpha\beta}$ allows
us to rewrite Eq.~(\ref{s_equation}) as
\begin{equation}
\left[\mu\left(k\ell^2 h^{\mu\nu}\phi_{,\mu}\phi_{,\nu}\right)
h^{\alpha\beta}\phi_{,\alpha} \right]_{;\beta}= 
kG(\tilde\rho+3\tilde p)\,e^{-2\phi}.
\label{s_equation2}
\end{equation}  This form is suitable for the analysis of cosmology
as well as static systems.

\subsubsection{Cosmology\label{sec:cosmology}}

Not only important in itself, cosmology is relevant for setting
boundary conditions in the study of T$e$V$e\,$S in the solar system
and other localized weak gravity situations.  We shall confine our
remarks to Friedmann-Robertson--Walker (FRW) cosmologies, for which
the metric can be given the form
\begin{equation}
\label{RW}  g_{\alpha\beta}\, dx^\alpha
dx^\beta=-dt^2+a(t)^2[d\chi^2+f(\chi)^2(d\theta^2+\sin^2\theta\,
d\varphi^2)].
\end{equation}  Here $f(\chi)\equiv \sin\chi, \chi, \sinh\chi$ for
closed, flat and open spaces, respectively.

In applying Eq.~(\ref{s_equation}) we shall assume that the fields
$\phi$, $\sigma$ and $\mathfrak{U}^\alpha$ partake of the symmetries
of the FRW spacetime.  Thus we take these fields to depend solely on
$t$. Also since there are no preferred spatial directions,
$\mathfrak{U}^\alpha$ must point in the cosmological time
direction: $\mathfrak{U}^\alpha=\delta_t{}^\alpha$ (that this is
possible distinguishes $\mathfrak{U}^\alpha$ from the Lee-Yang field
which is ruled out in FRW cosmology~\cite{Dicke}). Obviously this is
a case where
$\tilde u_\alpha=e^{\phi}\,\mathfrak{U}_\alpha$; the scalar
equation  then takes  the form
\begin{equation} a^{-3}\partial_t[a^3
\mu(-2k\ell^2 \dot\phi^2)\dot\phi]=-{\scriptscriptstyle 1\over 
\scriptscriptstyle 2} kG(\tilde
\rho+3\tilde p)e^{-2\phi}
\label{scalar_RW},
\end{equation}   where an overdot signifies $\partial/\partial t$. 
The first integral is
\begin{equation}
 \mu(-2k\ell^2\dot\phi^2)\dot\phi={-k\over 2a^3}\int_0^t G(\tilde
\rho+3\tilde p)e^{-2\phi}a^3 dt.
\label{first}
\end{equation}   As is customary in scalar--tensor theories, we have
dropped  an additive  integration constant; this has the effect of
ameliorating any divergence of
$\dot\phi$ as $a\rightarrow 0$.  In fact we can see that the r.h.s.
of the equation behaves there as $k(\tilde
\rho+3\tilde p)e^{-2\phi}t$.  We observe that as
$k\rightarrow 0$ with $\ell\propto k^{-3/2}$, $\dot\phi$ will behave
as $k$ with the argument of $\mu$ staying constant.  Thus regardless
of the form of
$\mu$, we have $\dot\phi\sim k$.  It is thus consistent  to assume
that
$\phi$ itself is of $\mathcal{O}(k)$ throughout cosmological
history.  This despite the possible divergence of $\dot\phi$ at the
cosmological singularity, since the rate of that divergence is also
proportional to
$k$, as we have just seen. Recalling that $kG\sigma^2=\mu$, we
conclude that
$\sigma^2$ is of $\mathcal{O}(k^{-1})$ in the cosmological solutions
(otherwise
$\mu$ would vary with $k$ whereas its argument stayed constant).

Let us check whether our assumption that
$\mathfrak{U}^\alpha=\delta_t{}^\alpha$ is consistent with the
vector equation (\ref{vectoreq}).  The choice
$\mathfrak{U}^\alpha=\delta_t{}^\alpha$ makes
$\mathfrak{U}^{[\alpha;\beta]}=0$.  For a comoving  ideal fluid
$\mathfrak{U}^\beta\tilde T_{\alpha\beta}=-e^{2\phi}\tilde\rho\,
\mathfrak{U}_\alpha$.  Thus the spatial components of the vector
equation (\ref{vectoreq}) vanish identically, while the temporal one
informs us that
\begin{equation}
\lambda= 8\pi G\big[\sigma^2 \dot\phi^2-2\tilde\rho
\sinh(2\phi)\big].
\label{lambda}
\end{equation} Our previous comments make it clear that
$\lambda$ is of $\mathcal{O}(k)$.

Turning to the gravitational equations
(\ref{gravitationeq})-(\ref{Theta}) we first note that in the limit
$\{k\rightarrow 0$, $\ell\propto k^{-3/2}$,
$K\propto k\}$, $\tau_{\alpha\beta}$ and
$\Theta_{\alpha\beta}$ are both $\mathcal{O}(k)$. It follows that
$G_{\alpha\beta}=8\pi G\tilde T_{\alpha\beta}+\mathcal{O}(k)$. 
Since the difference between $\tilde g_{\alpha\beta}$ and
$g_{\alpha\beta}$ is also of
$\mathcal{O}(k)$, it is obvious that $\tilde G_{\alpha\beta}=8\pi
G\tilde T_{\alpha\beta}+\mathcal{O}(k)$ so that any cosmological
model based on T$e$V$e\,$S differs from the corresponding one in GR
only by terms of $\mathcal{O}(k)$.   In FRW cosmology  T$e$V$e\,$S
has GR as its limit when $k\to 0$ with $\ell\propto k^{-3/2}$ and
$K\propto k$.

\subsubsection{Quasistatic localized
system\label{sec:quasistationary}}

We now turn to systems such as the solar system, or a neutron star,
which may be thought of as quasistatic situations in asymptotically
flat spacetime (at least up to sub--cosmological distances).  We
shall idealize them as truly static systems with time independent
metrics of the form
\begin{equation} g_{\alpha\beta}\, dx^\alpha
dx^\beta=g_{tt}(x^k)\,dt^2 + g_{ij}(x^k)\,dx^i dx^j
\label{staticmetric}
\end{equation}  and no energy flow.  The scalar and vector equations
have a variety of joint solutions.  We shall single out the physical
one by requiring the boundary condition that $\phi\rightarrow $
const. at spatial infinity, the constant being just the value of
$\phi$ from the cosmological model in which our localized system is
embedded.  Likewise, we shall require that
$\mathfrak{U}^\alpha\rightarrow \delta_t{}^\alpha$ so that the
vector field matches the cosmological field at ``spatial infinity''.

We first show that $\mathfrak{U}^\alpha=N\,\xi^\alpha$,  with
$\xi^\alpha=\delta_t{}^\alpha$ the Killing vector associated with
the static character of the spacetime, is an acceptable solution
(with
$N\equiv (-g_{\alpha\beta}\xi^\alpha\xi^\beta)^{-1/2}$, 
$\mathfrak{U}^\alpha$ is properly normalized).  Let us consider the
expression
$g^{\alpha\mu} \mathfrak{U}^\beta  \tilde T_{\mu\beta} + 
\mathfrak{U}^\alpha
\mathfrak{U}^\beta \mathfrak{U}^\gamma
\tilde T_{\gamma\beta}$ appearing in the source of the vector
equation (\ref{vectoreq2}) for this choice of $\mathfrak{U}^\alpha$.
Its
$\alpha=t$ component is
$N\left(\tilde T^t{}_t +\mathfrak{U}_t \mathfrak{U}^t\tilde
T^t{}_t\right) =0$, while the
$\alpha=i$ component is $N\left(g^{ij}\tilde T_{jt}+\mathfrak{U}^i
(\mathfrak{U}^t)^2\tilde T_{tt}\right)$ which also vanishes because 
$\tilde T_{jt}=0$ (no energy flow).  Turn now to the l.h.s. of
Eq.~(\ref{vectoreq2}).  Because $\mathfrak{U}^\alpha$ has only a
(time--independent) temporal component,
$\mathfrak{U}^\alpha\phi_{,\alpha}=0$, and the only nonvanishing
components of
$\mathfrak{U}^{[\alpha,\beta]}$ are the $jt$ ones, and they depend
only on the
$x^j$.  Hence $\mathfrak{U}^{[i,\beta]}{}_{;\beta}=0$ so that the
$\alpha=i$ components of the l.h.s. of the equation vanish.  What is
left of the
$\alpha=t$ component is
$K(\mathfrak{U}^{[t,\beta]}{}_{;\beta}+\mathfrak{U}^t
\mathfrak{U}_t
\mathfrak{U}^{[t,\beta]}{}_{;\beta})$ which vanishes by the
normalization of
$\mathfrak{U}^\alpha$.  Hence
$\mathfrak{U}^\alpha=N\,\xi^\alpha$ satisfies the vector equation
for any
$k$ and $K$.  We have not succeeded in proving that this is the
unique solution, but this seems to be a reasonable supposition.
 
Now, as $k\rightarrow 0$, the scalar equation (\ref{s_equation})
reduces to $(\mu h^{\alpha\beta}\phi_{,\alpha})_{;\beta}= 0$. 
Multiplying this by  $\phi(-g)^{1/2}$, discarding all time
derivatives, and integrating over space gives, after an integration
by parts and application of the boundary condition at infinity, that
$\int
\mu\ g^{\alpha\beta}\phi_{,\alpha}\phi_{,\beta} (-g)^{1/2} d^3x =
0$. Because for any static metric, $g^{ij}$ is positive definite and,
when defined, $\mu >0$, this equation is satisfied only by
$\phi=$ const. throughout.  But for $k\rightarrow 0$, the
cosmological model has
$\phi\rightarrow 0$.  Hence as $k\rightarrow 0$,
$\phi\rightarrow 0$ in all the space.  

Returning to the full scalar equation (\ref{s_equation}) and
recalling that
$\ell\propto k^{-3/2}$, it is easy to see that for small but finite
$k$ the
\textit{gradient} of $\phi$ scales as $k$.  From the last paragraph
it then follows that $\phi=\mathcal{O}(k)$.    These last
conclusions are actually independent of the form of
$\mu$ because its argument goes to a nonzero constant in the limit
$k\rightarrow 0$. We recall [see Eq.~(\ref{sigma})] that as
$k\rightarrow 0$, 
$\sigma^2\propto k^{-1}$.  Thus the scalars' energy-momentum tensor
$\tau_{\alpha\beta}$ is of $\mathcal{O}(k)$ (recall
$\ell\propto k^{-3/2}$). From the $\alpha=t$ component of
Eq.~(\ref{vectoreq}) we see that $\lambda=
\mathcal{O}(k)+\mathcal{O}(K)$. Hence
$\Theta_{\alpha\beta}=\mathcal{O}(k)+\mathcal{O}(K)$.  In addition,
the term in the gravitational equations (\ref{gravitationeq})
proportional to
$1-e^{-4\phi}$ is itself of
$\mathcal{O}(k)$; hence we have  $G_{\alpha\beta}=8\pi G\tilde
T_{\alpha\beta}+\mathcal{O}(k)+\mathcal{O}(K)$.  Since the
difference between
$\tilde g_{\alpha\beta}$ and $g_{\alpha\beta}$ is of
$\mathcal{O}(\phi)$, namely 
$\mathcal{O}(k)$, it is obvious that
$\tilde G_{\alpha\beta}=8\pi G\tilde
T_{\alpha\beta}+\mathcal{O}(k)+\mathcal{O}(K)$.  Thus for
quasistatic situations also, T$e$V$e\,$S has GR as its limit when
$k\to 0$ with $\ell\propto k^{-3/2}$ and $K\propto k$.

In conclusion, the limit $\{k\rightarrow 0$, $\ell\propto k^{-3/2}$,
$K\propto k\}$ of T$e$V$e\,$S is GR, both in cosmology and in
quasistatic localized systems.  

\subsection{\label{sec:milg}Generic general relativity limit}

Milgrom (private communication) has remarked that GR actually 
follows from T$e$V$e\,$S in the more general limit $K\rightarrow 0$
and $\ell\rightarrow\infty$ with $k$ arbitrary.   This is easily seen
after the change of variables $\phi\mapsto \phi_*\equiv \ell\phi$,
$\sigma\mapsto\sigma_*\equiv \surd k\sigma$, whereby only
$\tilde g_{\alpha\beta}$ and $S_s$ are changed:
\begin{eqnarray} 
\tilde g_{\alpha\beta} &=&  e^{-2\phi_*/\ell} g_{\alpha\beta}
-2\mathfrak{U}_\alpha\mathfrak{U}_\beta\sinh (2\phi_*/\ell) 
\\  S_s &=&-{1\over 2 k^2 \ell^2}\int\big[k\sigma_*{}^2
h^{\alpha\beta}\phi_{*,\alpha}\phi_{*,\beta}+{\scriptstyle
1\over\scriptstyle 2}G
\sigma_*{}^4 F(G\sigma_*{}^2) \big](-g)^{1/2} d^4 x,
\label{scalar*}
\end{eqnarray}   Thus as $\ell\rightarrow \infty$ the scalar action
disappears and $\phi_*$ decouples from the theory.  In addition,
with $K\rightarrow 0$, the vector's action $S_v$ disappears apart
from the term with $\lambda$.  All this means that the r.h.s. of the
Einstein equations (\ref{gravitationeq}) retains only the $\tilde
T_{\alpha\beta}$ and
$\lambda\mathfrak{U}_\alpha\mathfrak{U}_\beta$ terms.  But
according to the vector equation (\ref{vectoreq}), from which the
terms with differentiated
$\phi_*$ and $\mathfrak{U}_\alpha$ have dropped out,
$\lambda\rightarrow 0$ because $(1-e^{-4\phi_*/\ell})\rightarrow
0$.  Accordingly, we get the usual Einstein equations.  Since  
$g_{\alpha\beta}$ and $\tilde g_{\alpha\beta}$ coincide as
$\ell\rightarrow\infty$, we get exact GR. 

In this paper we shall assume that $k\ll 1$ and $K\ll 1$ without
restricting
$\ell$.  Empirical bounds on $k$ and $K$ are discussed in
Secs.~\ref{sec:Newtonian} and~\ref{sec:postN}.

\subsection{\label{sec:choice}The choice of $F$}

\begin{figure}[htbp]
\begin{center}
\includegraphics[width=3.0in]{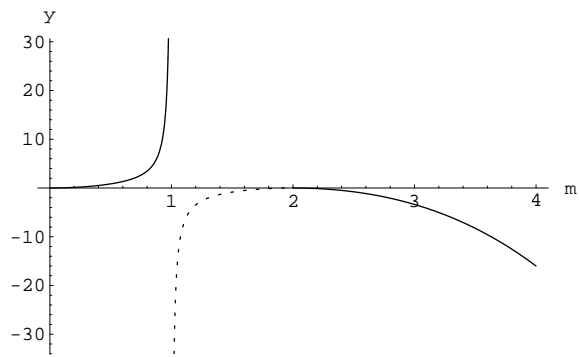}
\vspace{0.0in}
\caption{{\bf The function $y(\mu)$ as relevant for quasistationary
systems,  $0<\mu<1 $, and for cosmology,
$2<\mu<\infty $.}}
\label{fig:y}
\end{center}
\end{figure}

Because we have no theory for the functions $F(\mu)$ or $y(\mu)$,
there is great freedom in choosing them.  In this paper we shall
adopt, as an example, the form
\begin{equation} y={3\over 4}{ \mu^2(\mu-2)^2\over 1-\mu}
\label{y}
\end{equation} plotted in Fig.~\ref{fig:y}. As $y$ ranges from 0 to
$\infty$,
$\mu(y)$ increases monotonically from 0 to unity; for small $y$,
$\mu(y)\approx \sqrt{y/3}$.  For negative $y$ the function
$\mu(y)$ is double-valued.   As $y$ decreases from $0$, one branch
decreases monotonically from $\mu=2$ and tends to unity as
$y\rightarrow -\infty$, while the second increases monotonically from
$\mu=2$ and diverges as
$y\rightarrow -\infty$.  We adopt the second (far right) branch as
the physical one.
 \vspace{0.0in}

What features of the above $y(\mu)$ are essential for the following
sections ?  The denominator in Eq.~(\ref{y}) is included so that
$\mu$ shall asymptote to unity for $y\rightarrow\infty$ (the
Newtonian limit, c.f. Sec.~\ref{sec:Newtonian}).     The factor
$\mu^2$ ensures that  the MOND limit is contained in the theory (see
Sec.~\ref{sec:Mondish}), while the factor $(\mu- 2)^2$ ensures 
there exists a monotonically decreasing branch of $\mu(y)$ which
covers the whole of the range
$y\in \left[0,-\infty\right)$ (relevant to cosmology, c.f.
Sec.~\ref{sec:models}) and only it. 

\begin{figure}[htbp]
\begin{center}
\includegraphics[width=3.0in]{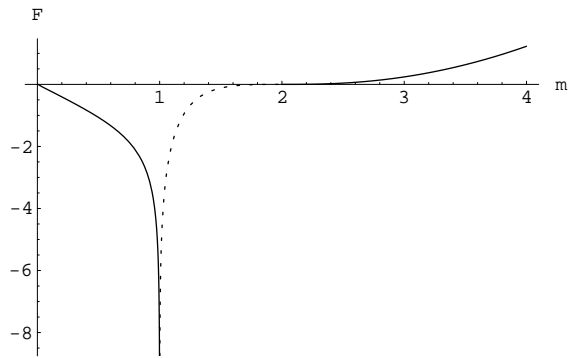}
\vspace{0.0in}
\caption{{\bf The function $F(\mu)$ as relevant for quasistationary
systems,  $0<\mu<1 $, and for cosmology,
$2<\mu<\infty $.}}
\label{fig:F}
\end{center}
\end{figure}

Integrating Eq.~(\ref{F}) with $y(\mu)$ we obtain (see 
Fig.~\ref{fig:F})
\begin{equation} F={3\over 8}\,\frac{ \mu \,\left( 4 + 2\,\mu  -
4\,{\mu }^2 + {\mu }^3 \right)  + 2\,\ln [{\left( 1 - \mu  \right)
}^2]  }{{\mu }^2},
\label{def_F}
\end{equation} where we ignore a possible integration constant
(which will, however, be useful in Sec.~\ref{sec:acc} below).  
Obviously $F<0$ in the range  $\mu\in(0,1)$ (relevant for
quasistationary systems) but
$F>0$ for $\mu>2$ (the cosmological range).    Where negative, $F$
contributes negative energy density in the energy momentum tensor
(\ref{tau}).  Despite this there seems to be no collision with the
requirement of positive overall energy density  (see
Secs.~\ref{sec:postN} and~\ref{sec:gene}).

 \section{\label{sec:nonrel}Nonrelativistic limit of T$e$V$e\,$S}
 
Sec.~\ref{sec:quasistationary} shows that  in quasistatic systems 
T$e$V$e\,$S approaches GR in the limit 
$\{k\rightarrow 0$, $\ell\sim k^{-3/2}$, $K\sim k\}$.  But in what
limit  do we recover standard Newtonian gravity ? And where is MOND,
which is antagonistic to Newtonian gravity, in all this  ?  This
section shows that with our choice of $F$, both Newtonian and MOND
limits emerge from T$e$V$e\,$S for small gravitational potentials,
but that MOND requires in addition small gravitational
\textit{fields}, just as expected from Milgrom's original scheme.  
\subsection{\label{sec:quasi}Quasistatic systems}
 
We are here concerned with a quasistatic, {\it weak} potential and
{\it slow} motion situation, such as a galaxy or  the solar
system.    As in Sec.~\ref{sec:quasistationary}, quasistatic means
we can neglect time derivatives in comparison with spatial ones. 
Let us assume that the metric
$g_{\alpha\beta}$ is nearly flat and that
$|\phi|\ll 1$.  Then linearization  of Eq.~(\ref{gravitationeq}) in
terms of the Newtonian potential $V$ generated by the energy content
on its r.h.s. gives
$g_{tt}=-(1+2 V)+\mathcal{O}( V^2)$.    From the prescription given
in Sec.~\ref{sec:quasistationary},
$\mathfrak{U}_\alpha=-[1+V+\mathcal{O}(V^2)]\delta_{t\alpha}$. It
follows from Eq.~(\ref{physg})  that to
$\mathcal{O}(\phi)$ and $\mathcal{O}(V)$,  $\tilde g_{tt}=-(1+2
V+2\phi)$.  Thus in T$e$V$e\,$S the total potential governing all
nonrelativistic motion is
$ \Phi= V+\phi$.  We should remark that if asymptotically
$\phi\rightarrow
\phi_c\neq 0$, the $\tilde g_{tt}$ does not there correspond to a
Minkowski metric.  This is remedied by rescaling the time
 (or spatial) coordinates by factors 
$e^{\phi_c}$  ( or $e^{-\phi_c}$ ).  With respect to the new
coordinates the metric is then asymptotically Minkowskian.   In this
paper we assume throughout that $|\phi_c|\ll 1$;
Sec.~\ref{sec:models} shows this is consistent  with cosmological
evolution of $\phi$.

How is $\Phi$ related to  $ \Phi_N$, the  Newtonian gravitational
potential generated by the mass density $\tilde\rho$ according to
Poisson's equation with gravitational constant $G$ ?   To relate
$\phi$ to
$\Phi_N$ we first set temporal derivatives in Eq.~(\ref{s_equation2})
to zero which means replacing
$h^{\alpha\beta}\phi_{,\alpha}\rightarrow
g^{\alpha\beta}\phi_{,\alpha}$:  
\begin{equation}
\left[\mu\left(k\ell^2 g^{\mu\nu}\phi_{,\mu}\phi_{,\nu}\right)
g^{\alpha\beta}\phi_{,\alpha} \right]_{;\beta}= 
kG(\tilde\rho+3\tilde p)\,e^{-2\phi}.
\end{equation}  This equation is still exact.  Next we replace
$g^{\alpha\beta}\to
\eta^{\alpha\beta}$ as well as
$e^{-2\phi}\to 1$.  This is the nonrelativistic approximation. 
Further, to be consistent we must neglect $\tilde p$ compared to
$\tilde\rho$; keeping the former would be tantamount to accepting
that $V$ is not small.  Thus
\begin{equation}
\bm\nabla\cdot\Big[\mu\Big(k\ell^2 (\bm\nabla\phi)^2\Big)
\bm\nabla\phi\Big]=kG\tilde\rho.
\label{AQUAL2}
\end{equation}  This is just the AQUAL equation (\ref{AQUALeq})
with a suitable reinterpretation of the function $\mu$.   Now
comparing Eq.~(\ref{AQUAL2}) with Poisson's equation we see that
\begin{equation} k^{-1}\mu|\bm\nabla\phi|=\mathcal{O}(|\bm\nabla 
\Phi_N|)
\label{potratio}
\end{equation}  This will be made more precise below in situations
with symmetry.

We now show that it is consistent to take $V=C\Phi_N$, with $C$ a
constant close to unity (to be determined).  The starting point are
the modified Einstein equations (\ref{gravitationeq}).     With $F$
as in (\ref{def_F}),
$F<0$ simultaneously with $F'<0$ for $0<\mu<1$; it follows from
Eq.~(\ref{F}) that
$\mu|F|<y$.   Now the $F$ term on the r.h.s. of
Eq.~(\ref{gravitationeq}) is
$-2\pi G^2
\ell^{-2}\sigma^4 F(kG\sigma^2)  g_{\alpha\beta}=-2\pi 
k^{-2}\ell^{-2}\mu^2 F(\mu)  g_{\alpha\beta}$.  Similarly, since
$\phi_{,t}=0$ here, the terms on the r.h.s. involving
$\phi_{,\alpha}$ are of order
$8\pi G\sigma^2 h^{\gamma\delta}\phi_{,\gamma}\phi_{,\delta}\,
g_{\alpha\beta}=8\pi k^{-2}\ell^{-2}\mu y(\mu) g_{\alpha\beta}$.  
Thus by our earlier remark the $\phi$ derivative terms in
$\tau_{\alpha\beta}$ dominate the
$F$ term, and by Eq.~(\ref{potratio}) they are of order $8\pi
k\mu^{-1}(\bm\nabla 
\Phi_N)^2$.  But $(\bm\nabla 
\Phi_N)^2$  is precisely the type of source (Newtonian gravitational
energy or stress density) needed to compute  the first nonlinear or
$\mathcal{O}(\Phi_N{}^2)$ contributions to the metric.   As we shall
see in Sec.~\ref{sec:models}, we  need $k\sim 10^{-2}$, so that if
all we desire is to compute the metric to
$\mathcal{O}(\Phi_N)$, and $\mu$ is not very small, then all of
$\tau_{\alpha\beta}$ may be neglected.  

Further, since
$\mathfrak{U}_\alpha=-[1+V+\mathcal{O}(V^2)]\delta_{t\alpha}$, the
$\mathfrak{U}_{[\alpha,\beta]}{}^2$ terms in
$\Theta_{\alpha\beta}$ have the form
$(C\bm\nabla\Phi_N)^2$; we drop them for the same reason that we
dropped the
$\mathcal{O}(\Phi_N{}^2)$ term in
$\tau_{\alpha\beta}$.   It follows that in the weak potential
approximation the spatio-temporal and spatial-spatial components of
Einstein's equations are exactly the same as in GR because the term
proportional to
$1-e^{-4\phi}$ can be dropped by virtue of the slow motion condition
which suppresses the spatio-temporal components of
$T_{\alpha\beta}$.    The temporal-temporal component of Einstein's
equations depends on $\lambda$, and is thus another story.   From
Eqs.~(\ref{vectoreq})  and (\ref{oldT}) and the observation that
$\mathfrak{U}^\alpha\phi_{,\alpha}=0$, 
\begin{equation}
\lambda=K\mathfrak{U}_\alpha\mathfrak{U}^{[\alpha;\beta]}{}_{;\beta}-16\pi
G \tilde\rho\sinh(2\phi).
\end{equation}
 With our $\mathfrak{U}_\alpha$ the first term is
$K\mathfrak{U}_t
\mathfrak{U}^{[t;\beta]}{}_{;\beta}=-KC\nabla^2
\Phi_N+KC^2\mathcal{O}(\bm\nabla\Phi_N{}^2)$, where by Poisson's
equation 
$\nabla^2 \Phi_N =4\pi G\tilde\rho$.  Further, as we shall see in
Sec.~\ref{sec:postN}, $\phi$ is always very close to its
aforementioned asymptotic value $\phi_c$ (which is just $\phi$'s
very slowly varying cosmological value).  Dropping the
$C^2\mathcal{O}(\bm\nabla\Phi_N{}^2)$ contribution for the same
reason as above gives
\begin{equation}
\lambda\approx -8\pi G[ KC/2 +2\sinh(2\phi_c)]   
\tilde\rho.
\label{lambdaexpl}
\end{equation} 

Substituting this in Eq.~(\ref{Theta}) and combining the result with
the $(1-e^{-4\phi_c})$ term in the $G_{tt}$ equation
Eq.~(\ref{gravitationeq}), we see that
$(e^{-2\phi_c}+KC/2)\tilde\rho$ replaces the source 
$\tilde\rho$  appropriate in the weak potential approximation to
GR.  By linearizing the 
$G_{tt}$ equation as done in GR, we conclude that
\begin{equation} 
V=(e^{-2\phi_c}+KC/2)\Phi_N
\end{equation} which verifies the claim that $V$ is proportional to
$\Phi_N$.  Indeed, since the proportionality constant here must be
identical with $C$, we have $C=(1-K/2)^{-1}e^{-2\phi_c}$.  Since we
shall show in Sec.~\ref{sec:models} that it is consistent to assume 
$|\phi_c|\ll 1$, and assume that $K\ll1$, we shall replace $C$ everywhere by $ \Xi\equiv
1+K/2-2\phi_c$.  In particular
\begin{equation}
\Phi=\Xi\Phi_N+\phi.
\label{corrV}
\end{equation}

In summary, Eq.~(\ref{corrV}), which is subject to corrections of
$\mathcal{O}(\Phi_N{}^2)$, quantifies the difference at the
nonrelativistic level between T$e$V$e\,$S and GR, a difference
which  is in harmony with our conclusion in
Sec.~\ref{sec:quasistationary}.  We shall use it until  we turn to
post-Newtonian corrections.  The condition ``$\mu$ is not very
small'' which we imposed above to be able to neglect the
$\tau_{\alpha\beta}$ contribution to the gravitational equations is 
not restrictive.  For the Newtonian limit we shall see that
$\mu\approx 1$.   And when $\mu\ll 1$ (extreme MOND limit relevant
for extragalactic phenomena),  the  consequent corrections of
$\mathcal{O}( \Phi_N{}^2)$ (with large coefficient) to $V$ are
entirely ignorable because this potential  is then dominated by
$\phi$ in the expression for $\Phi$, c.f. Eq.~(\ref{2phis}).

\subsection{\label{sec:Mondish}The MOND limit: spherical symmetry}
First for orientation we assume a spherically symmetric situation. 
Then from Eq.~(\ref{AQUAL2}) together with Gauss' theorem we infer
that
\begin{equation}
\bm\nabla\phi=(k/4\pi\mu)\bm\nabla  \Phi_N.
\label{2phis}
\end{equation}  In view of Eq.~(\ref{corrV}) we have
\begin{equation}
\tilde\mu\bm\nabla \Phi=\bm\nabla  \Phi_N.
\label{gradphi}
\end{equation}  with
\begin{equation}
\tilde\mu\equiv (\Xi+k/4\pi\mu)^{-1}.
\label{tildemu}
\end{equation} 

Consider the case  $\mu\ll1$ for which
$\mu\big( k\ell^2 (|\bm\nabla\phi|)^2\big)
\approx (k/3)^{1/2}\ell|\bm\nabla\phi|$ (see
Sec.~\ref{sec:choice}).  Eliminating $\bm\nabla\Phi_N$ between
Eqs.~(\ref{2phis}) and (\ref{gradphi}) and defining
\begin{equation} \mathfrak{a}_0\equiv {(3 k)^{1/2}\over 4\pi\Xi\ell}
\label{a_0}
\end{equation}  we obtain a quadratic equation for $\mu$ with
positive root
\begin{equation}
\mu=(k/8\pi\Xi)\big(-1+\sqrt{1+4|\bm\nabla\Phi|/\mathfrak{a}_0}\,\big)
\label{plainmu}
\end{equation} This is obviously valid only when
$|\bm\nabla\Phi|\ll (4\pi/k)^2\mathfrak{a}_0$ since otherwise
$\mu$ is not small.   From Eq.~(\ref{tildemu}) we now deduce the
MOND function
\begin{equation}
\tilde\mu={1\over \Xi}\,{-1+\sqrt{1+4|\bm\nabla\Phi|/
\mathfrak{a}_0} 
\over 1+\sqrt{1+4|\bm\nabla\Phi|/ \mathfrak{a}_0}  }
\label{tildemu2}
\end{equation} For $|\bm\nabla\Phi|\ll \mathfrak{a}_0$  (which is
consistent with the above restriction since $k\ll 1$) this equation
gives to lowest order in $K$ and $\phi_c$
\begin{equation}
\tilde\mu\approx |\bm\nabla\Phi|/\mathfrak{a}_0.
\end{equation}  Thus if we identify our $\mathfrak{a}_0$ with
Milgrom's constant, Eq.~(\ref{gradphi}) with this $\tilde\mu$
coincides with the MOND formula~(\ref{MOND}) in the extreme low
acceleration regime.    Therefore, T$e$V$e\,$S recovers MOND's
successes in regard to low surface brightness disk galaxies, dwarf
spheroidal galaxies, and the outer regions of spiral galaxies.  For
all these the low acceleration limit of Eq.~(\ref{MOND}) is  known to
summarize the phenomenology correctly.

Now suppose $|\bm\nabla\Phi|$ varies from an order below
$\mathfrak{a}_0$ up to a couple of orders above it.
 This  respects the condition  $|\bm\nabla\Phi|\ll
(4\pi/k)^2\mathfrak{a}_0$.   Then Eq.~(\ref{tildemu2}) shows
$\tilde\mu$ to grow monotonically from about $0.1$ to $0.9$.  Then
Eq.~(\ref{gradphi}) is essentially formula~(\ref{MOND}) in the
intermediate MOND regime.  This regime is relevant for the disks of
massive spiral galaxies well outside the central bulges but not quite
in their outer reaches.  It is known that the precise form of
$\tilde\mu$ makes little difference for the task of predicting
detailed rotation curves from surface photometry. 

We see that T$e$V$e\,$S reproduces the MOND paradigm encapsulated in
Eq.~(\ref{MOND})  for not too large values of
$|\bm\nabla\Phi|/\mathfrak{a}_0$. What happens for very large
$|\bm\nabla\Phi|/\mathfrak{a}_0$ ?

\subsection{\label{sec:Newtonian} The Newtonian limit: spherical
symmetry}

According to  our choice of $y(\mu)$, Eq.~(\ref{y}), the limit
$\mu\rightarrow 1$ corresponds to
$y\rightarrow\infty$, that is to say
 $|\bm\nabla\phi|\rightarrow\infty$.    By
Eqs.~(\ref{2phis})--(\ref{tildemu})  we simultaneously  have
$|\bm\nabla\Phi|\rightarrow\infty$ and 
$\tilde\mu\rightarrow(\Xi+k/4\pi)^{-1}$.    Defining the Newtonian 
gravitational constant by
\begin{equation}
 G_N=(\Xi+k/4\pi)G,
\label{GNewton}
\end{equation}  we see from Eq.~(\ref{gradphi}) that 
$\bm\nabla \Phi$ is obtained from $\bm\nabla\Phi_N$ by just replacing
$G\rightarrow G_N$ in it.  In other words, in the nonrelativistic and
arbitrarily large $|\bm\nabla\Phi|$ regime, T$e$V$e\,$S   is
equivalent to  Newtonian gravity, but with a ``renormalized'' value
of the gravitational constant.  Now $\Xi$ is really a surrogate of
$C=(1-K/2)^{-1}e^{-2\phi_c}$; hence for $K<2$, $G_N$ is positive.  As mentioned, we here assume $K\ll 1$.

But how close are dynamics to Newtonian for \textit{large but finite}
$|\bm\nabla\Phi|/\mathfrak{a}_0$ ?   Expanding the r.h.s. of
Eq.~(\ref{y}) in the neighborhood of $\mu=1$  gives
\begin{equation} y={3/4 \over 1-\mu} +
\mathcal{O}(1-\mu).
\end{equation}  We also have by  Eqs.~(\ref{2phis}) and
(\ref{gradphi}) that
$y\equiv k\ell^2|\bm\nabla\phi|^2\approx
(k^3\ell^2/16\pi^2)|\bm\nabla\Phi|^2$ where we have dropped
corrections of higher order in $(k/4\pi)$. Dropping the
$\mathcal{O}(1-\mu)$ term in $y(\mu)$ and eliminating $\ell$ in
favor of $\mathfrak{a}_0$ (with $\Xi= 1$) we get
\begin{equation}
\mu \approx 1-{64\pi^4\over k^4}{\mathfrak{a}_0{}^2\over
|\nabla\Phi|^2}
\label{muapprox}
\end{equation}

Thus to trust the approximation $\mu\approx 1$ we must have
$|\bm\nabla\Phi|/\mathfrak{a}_0\gg 8\pi^2 k^{-2} $.   Using
Eqs.~(\ref{muapprox}) and (\ref{tildemu}) we obtain, again after
dropping higher order terms in $k$, that
\begin{equation}
\tilde\mu\approx {G\over G_N}
\left(1-{16\pi^3\over  k^3}{\mathfrak{a}_0{}^2\over
|\nabla\Phi|^2}\right).
\label{mutilde}
\end{equation}  Here the factor $(G/G_N)$ just reflects the
mentioned ``renormalization'' of the gravitational constant; it is
the next factor which interests us as a measure of departures from
strict Newtonian behavior.    For example, if
$k=0.03$ there is  a  $5.3\times 10^{-9}$ fractional enhancement of
the sun's Newtonian field at Earth's orbit where 
$|\bm\nabla\Phi|=0.59
\,\textrm{cm s}^{-2}$.  This is probably unobservable today.   At
Saturn's orbit where $|\bm\nabla\Phi|=0.0065\,\textrm{cm s}^{-2}$
the fractional correction is  $4.3\times 10^{-5}$, corresponding to
an excess acceleration
$2.8\times 10^{-7} \,\textrm{cm s}^{-2}$ (at this point $\mu$ departs
from unity by only $0.018$ so that Eq.~(\ref{muapprox})  is still
reliable).  Although this departure from Newtonian predictions
seems serious, it  should be remembered that navigational data from
the Pioneer 10 and 11  spacecrafts seem to disclose a constant
acceleration in excess of Newtonian of about  
$8\times 10^{-8} \,\textrm{cm s}^{-2}$ between Uranus' orbit and the
trans-Plutonian region~\cite{nieto}.   It is, however, unclear
whether the correction in Eq.~(\ref{mutilde}), sensitive as it is to
the choice of
$F$, has anything to do with the ``Pioneer anomaly''.

\subsection{\label{sec:nonspherical}Nonspherical systems}

We now consider \textit{generically} asymmetric systems.  Since any
system has a region where
$\mu$ differs from unity and is variable,  Eq.~(\ref{2phis}) is not
the general solution of  Eq.~(\ref{AQUAL2}) and must be replaced by
\begin{equation}
\bm\nabla\phi=(k/4\pi\mu)(\bm\nabla 
\Phi_N+\bm\nabla\times\bm h),
\label{correct}
\end{equation} where $\bm h$ is some regular vector field which is
determined up to a gradient by the condition that the curl of the
r.h.s. of Eq.~(\ref{correct}) vanish. 

The freedom inherent in  $\bm h$ allows it to be made
divergenceless.   Then by Gauss' theorem $\bm h$ must fall off faster
than
$1/r^2$ and $\bm\nabla\times\bm h$ faster than
$1/r^3$ at large distances.  On physical grounds
$|\bm\nabla\times\bm h|$ is expected to be of the same order as
$|\bm\nabla  \Phi_N|$ well inside the matter.  But since the latter
quantity falls off as $1/r^2$ well outside the matter, the curl term
in Eq.~(\ref{correct}) must rapidly become negligible well outside
the system.  We thus expect the discussion in Sec.~\ref{sec:Mondish}
to apply well outside any nonspherical galaxy  just as it applies
anywhere inside a spherical one.  The interior and near exterior of
such a galaxy, where
$\bm\nabla\times\bm h$ is still important, must be treated by
numerical methods which would be no different than those developed
by Milgrom within the old AQUAL theory~\cite{Mnumer}.

Needless to say, an asymmetric system so dense that the Newtonian
regime ($\mu$ approximately constant)  obtains in its interior, e.g.
an oblate globular cluster like $\omega$ Centauri, can be described
everywhere without an
$\bm h$.  For in the interior $\bm h$ is not needed since even in its
absence the curl of the r.h.s. of Eq.~(\ref{correct}) vanishes
(approximately).  And
$\mu$ begins to differ substantially from unity only well outside the
system where we know from our previous argument that any $\bm h$ is
becoming negligible.  Hence both Newtonian and MOND regimes of the
system may be described as in Secs.~\ref{sec:Mondish}
and~\ref{sec:Newtonian}. 

In summary, we see that the extragalactic predictions of the MOND
equation (\ref{MOND}) are recovered from T$e$V$e\,$S; at the same
time T$e$V$e\,$S hints at non-Newtonian behavior in the reaches of
the solar system, though the effect is sensitive to the choice of
$F$ in the theory.

\section{\label{sec:postN}The post-Newtonian corrections}

The upshot of the discussion at the end of
Sec.~\ref{sec:quasistationary} is that in the solar system
(regarded as a static system---with rotation neglected---embedded in
a FRW cosmological background),  $\tilde G_{\alpha\beta}=8\pi G\tilde
T_{\alpha\beta}+\mathcal{O}(k)+\mathcal{O}(K)$.   Here we compute
the consequent $\mathcal{O}(k)+\mathcal{O}(K)$ corrections to the
Schwarzschild metric 
\begin{equation}  g_{\alpha\beta}\, dx^\alpha
dx^\beta=-{(1-Gm/2\varrho)^2\over
(1+Gm/2\varrho)^2}dt^2+(1+Gm/2\varrho)^4[d\varrho^2+
\varrho^2(d\theta^2+\sin^2\theta d\varphi^2)]
\label{Schwarzschild2}
\end{equation}     that describes the exterior of a spherically mass
$m$, and determine the post-Newtonian parameters of T$e$V$e\,$S 
which we compare with those of GR.

Rather than just extending the Newtonian limit calculation of
Sec.~\ref{sec:Newtonian}, we start from scratch.  First we write the
spherically symmetric and static metric of the sun (inside and
outside it) as
\begin{equation}  g_{\alpha\beta}\, dx^\alpha dx^\beta=-e^\nu
dt^2+e^\varsigma [d\varrho^2+
\varrho^2(d\theta^2+\sin^2\theta d\varphi^2)] 
\label{spherical}
\end{equation} with $\nu=\nu(\varrho)$ and
$\varsigma=\varsigma(\varrho)$.   Just as for metric
(\ref{Schwarzschild2}), outside the sun these functions should admit
the expansions ($\alpha_i$ and
$\beta_i$ are dimensionless constants)
\begin{eqnarray} e^\nu=1-r_g/\varrho+\alpha_2 (r_g/\varrho)^2
+\cdots
\label{expansion1}
\\ e^\varsigma=1+\beta_1 r_g/\varrho+\beta_2 (r_g/\varrho)^2
+\cdots\, ,
\label{varsigma}
\label{expansion2}
\end{eqnarray}  where $r_g$ is a lengthscale to be determined (see
Appendix~\ref{sec:D}).  The magnitude of the coefficient of the 
$r_g/\varrho$ term in Eq.~(\ref{expansion1}) has been absorbed into
$r_g$; its sign must be negative, as shown, because gravity is
attractive.  From the fact that T$e$V$e\,$S approaches GR for small
$k$ and $K$, we may infer that $r_g$ is close to $2G$ times the
system's Newtonian mass.  This is made precise below.
 
Taking  $\phi=\phi(\varrho)$ and $\tilde T_{\alpha\beta}$ from
Eq.~(\ref{oldT}), we may  write the scalar equation
(\ref{s_equation2}) as  
\begin{equation}
\varrho^{-2} e^{-(\nu+3\varsigma)/2}[\mu e^{(\nu+\varsigma)/2}
\varrho^2 \phi']'=kGe^{-2\phi}(\tilde\rho+3\tilde p).
\label{newscalar}
\end{equation}Here $'$ stands for $d/d\varrho$.    The first integral
of Eq.~(\ref{newscalar})  is
\begin{equation}
\phi'={kG e^{-(\nu+\varsigma)/2}\over  \mu\varrho^2 }\int_0^\varrho
(\tilde\rho+3\tilde p) e^{\nu/2+3\varsigma/2-2\phi}
\varrho^2 d\varrho,
\label{phitag}
\end{equation} where the integration constant has been chosen so that
$\phi$ is regular at $\varrho=0$. 

Supposing the matter's boundary is at
$\varrho=R$, we define the (positive) ``scalar mass''
\begin{equation} m_s\equiv 4\pi \int_0^R (\tilde\rho+3\tilde p)
e^{\nu/2+3\varsigma/2-2\phi} \varrho^2 d\varrho.
\end{equation}     Because for a nonrelativistic fluid $\tilde p\ll
\tilde\rho$,
$m_s$ must be close to the Newtonian mass.  In fact, as shown in
Appendix~\ref{sec:D},
$m_s$ and an appropriately defined gravitational mass $m_g$  differ
only by a fraction of $\mathcal{O}(Gm_g/R)$ which amounts to
$10^{-5}$ for the inner solar system.  For $\varrho>R$ we may expand
$\phi'$ as
\begin{equation}
\phi'={kGm_s\over 4\pi\mu}\Big[{1\over
\rho^2}+{(1-\beta_1)r_g\over 2\varrho^3}+\mathcal{O}({
\varrho^{-4}})\Big].
\label{scalar'}
\end{equation}  It is obvious from this that $\phi$ {\em decreases
inward}. Its asymptotic value, as will be explained in
Sec.~\ref{sec:models},  is positive and of $\mathcal{O}(k)$.  The
decrement in $\phi$ down to ``radius''
$\varrho$ is, according to Eq.~(\ref{phitag}), or its integral
Eq.~(\ref{intscalar}) below,   of
$\mathcal{O}(kGm_s/4\pi\varrho)$.  In  any weakly gravitating
system,
$Gm_s/\varrho\ll 1$ and for strongly gravitating systems like a
neutron star,
$Gm_s/\varrho$ is still well below unity  (black holes require a
special discussion which we defer to another occasion).  Thus $\phi$
remains positive and small throughout space for all systems, and for
the solar system in particular.  This will have repercussions for the
causality question examined in Sec.~\ref{sec:causality}.   

Since we are not here interested in purely MOND corrections, we shall
take
$\mu=1$ in Eq.~(\ref{scalar'}) as well as in the terms in
$\tau_{\alpha\beta}$, Eq.~(\ref{tau}), which explicitly involve
$\phi$ derivatives.  The $\mu$ in the $F$ term of
$\tau_{\alpha\beta}$ is not so easily disposed of because with our
choice of $F$, and indeed with any viable one,
$F$ must be singular at $\mu=1$.   If neglecting the $F$ term in
$8\pi G\tau_{\alpha\beta}$  can be justified, then using
Eq.~(\ref{scalar'})  we may compute from Eq.~(\ref{tau}) that for
$\varrho>R$
\begin{equation} 8\pi G\tau_{tt}=8\pi
G\tau_{\varrho\varrho}={kG^2m_s^2\over 4\pi\varrho^4}+\mathcal{O}({
\varrho^{-5}}).
\label{tautt}
\end{equation}  Now by the approximation (\ref{muapprox}) the ratio
of the
$F$ term in $8\pi G\tau_{\alpha\beta}$ to these last terms is 
\begin{equation} {8\pi^2\mu^2 |F(\mu)|\varrho^4\over
k^3\ell^2G^2m_s{}^2}={128\pi^4
\mathfrak{a}_0{}^2 \mu^2 |F(\mu)|\over 3
k^4|\bm\nabla\Phi_N|^2}\approx {2\over 3}(1-\mu)|F(\mu)| 
\end{equation} which numerically does not exceed 0.04 for
$\mu>0.99$.  This justifies Eq.~(\ref{tautt}) in any region where
MOND effects are totally negligible.  However, as pointed out in
Sec.~\ref{sec:Newtonian},  at Saturn's orbit $\mu$ already departs
from unity by two percent.  In such cases the contribution of the $F$
term to $\tau_{\alpha\beta}$ must be taken into account, and its
post-Newtonian effects compared with the MOND departure from strict
Newtonian behavior calculated in Sec.~\ref{sec:Newtonian}.  Here we
shall only be concerned with inner solar system dynamics where
$\mu$ is very close to unity.  Because
$\tau_{tt}$ is dominated by the derivative terms, the energy density
contributed by the scalar fields is evidently positive.

Clearly  in our situation (see Sec.~\ref{sec:quasistationary})
\begin{equation}
\mathfrak{U}^\alpha=\{e^{-\nu/2},0,0,0\}.
\label{U*}
\end{equation} Using this  in Eqs.~(\ref{Theta}) and
(\ref{vectoreq})  we find for $\varrho>R$ that 
 \begin{eqnarray}
\lambda&=&{K(2+\beta_1-4\alpha_2)r_g{}^2\over
4\varrho^4}+\mathcal{O}({
\varrho^{-5}})
\\
\Theta_{tt}&=&{K(-2\beta_1-3+8\alpha_2)r_g{}^2\over
8\varrho^4}+\mathcal{O}({
\varrho^{-5}})
\label{Thetatt}
\\
\Theta_{\varrho\varrho}&=&-{K r_g{}^2\over 8\varrho^4}+\mathcal{O}({
\varrho^{-5}})
\end{eqnarray}

With this we now turn to  Einstein's equations (\ref{gravitationeq})
for all
$\varrho$.   By virtue of $\mathfrak{U}^\alpha$'s form here, the $tt$
and
$\varrho\varrho$ components simplify to
\begin{eqnarray} -e^{\nu-\varsigma}\Big(\varsigma''+{\scriptstyle
1\over
\scriptstyle 4}\varsigma'^2+2\varsigma'/
\varrho\Big)&=&8\pi G\big[(2e^{-4\phi}-1)
\tilde T_{tt}  +\tau_{tt}\big]+\Theta_{tt}
\label{ttEq}
\\ {\scriptstyle 1\over \scriptstyle 4}\varsigma'^2+{\scriptstyle
1\over
\scriptstyle 2}\varsigma'\nu'+(\varsigma'+\nu')/\varrho &=&8\pi
G\big[\tilde T_{\varrho\varrho}+\tau_{\varrho\varrho}\big]+
\Theta_{\varrho\varrho}
\label{rhorhoEq}
\end{eqnarray} First we solve these for $\varrho>R$  where $\tilde
T_{\alpha\beta}=0$.  From Eqs.~(\ref{expansion1}) and
(\ref{expansion2}) it follows that
\begin{eqnarray}
\nu'&=&r_g/\varrho^2+(1-2\alpha_2)r_g{}^2/\varrho^3 +\cdots
\label{nutag}
\\
\varsigma'&=&-\beta_1r_g/\varrho^2+(\beta_1{}^2-2\beta_2)r_g{}^2/\varrho^3
+\cdots
\label{zetatag}
\end{eqnarray}  Substituting these together with
Eqs.~(\ref{expansion1}), (\ref{expansion2}), (\ref{tautt}) and
(\ref{Thetatt}) in Eqs.~(\ref{ttEq})-(\ref{rhorhoEq}), matching
coefficients of like powers of
$1/\varrho$, and solving the three resulting algebraic conditions
gives to lowest order in $k$ and $K$
\begin{eqnarray}
\beta_1&=&1
\label{beta_1}
\\
\alpha_2&=&{\scriptstyle 1\over \scriptstyle 2}
\\
\beta_2&=&{\scriptstyle 3\over \scriptstyle 8}+{\scriptstyle 1\over
\scriptstyle 16}K-{kG^2m_s{}^2\over 8\pi r_g{}^2}
\label{beta_2}
\end{eqnarray} Using these results we show in Appendix~\ref{sec:D} 
that $r_g=2Gm_g[1+\mathcal{O}(k G m_g/R)+\mathcal{O}(K G m_g/R)]$
with $m_g$, the gravitational mass, defined by Eq.~(\ref{massg}). 
The relative correction here is
$\ll 10^{-5}$ for the inner solar system.   We also remark that with
the values (\ref{beta_1})-(\ref{beta_2}) the energy density
contributed by
$\Theta_{tt}$  is positive (see Eq.~(\ref{Thetatt})).

For solar system tests of T$e$V$e\,$S we must know the physical
metric
$\tilde g_{\mu\nu}$.  According to Eqs.~(\ref{physg}) and (\ref{U*}),
$\tilde g_{tt}=-e^{2\phi+\nu}$, $\tilde g_{\varrho\varrho}=\tilde
g_{\theta\theta}/\varrho^2=g_{\varphi\varphi}/\varrho^2
\sin^2\theta =e^{-2\phi+\varsigma}$, so we need $\phi$.  
Integration of  Eq.~(\ref{scalar'}) in light of Eq.~(\ref{beta_1})
gives
\begin{equation}
\phi(\varrho)=\phi_c-{kGm_s\over 4\pi
\varrho}+\mathcal{O}({\varrho^{-3}}),
\label{intscalar}
\end{equation} whereupon
\begin{equation} e^{\pm2\phi}=e^{\pm2\phi_c}\Big(1\mp {kGm_s\over
2\pi
\varrho}+{k^2 G^2 m_s^2\over 8\pi^2
\varrho^2}+\mathcal{O}({\varrho^{-3}})\Big).
\end{equation}   The integration constant $\phi_c$ is evidently the
cosmological value of
$\phi$ at the epoch in question.  This value changes slowly over
solar system timescales, so we can ignore its drift for most
purposes.  Thus by taking the advantage of the isotropic form of the
metric (\ref{spherical}), and rescaling the
$t$ and $\varrho$ coordinates appropriately, we absorb the factors
$e^{2\phi_c}$ and  $e^{-2\phi_c}$ that would otherwise appear in
$\tilde g_{\mu\nu}$ so that it can asymptote to Minkowskian form as
expected.    With this precaution one can calculate as if $\phi_c$
vanished.  It must be stressed that this strategy works at a
particular cosmological era.

Accordingly
\begin{eqnarray}
\tilde g_{tt}&=&-1+2G_N\, m\, \varrho^{-1}-2\beta G_N{}^2 m^2
\varrho^{-2}+\mathcal{O}({\varrho^{-3}})
\\
\tilde g_{\varrho\varrho}&=&1+2\gamma G_N\, m\,
\varrho^{-1}+\mathcal{O}({\varrho^{-2}})
\\ G_N\,m &\equiv&  r_g/2+(kG m_s/4\pi)
\\
\beta &=& 1
\\
\gamma&=& 1
\end{eqnarray} As previously,  $G_N$ is defined by
Eq.~(\ref{GNewton}).  Recalling the relations between $r_g$, $m_g$
and
$m_s$ (Appendix~\ref{sec:D}), we find that
$m=m_g[1+\mathcal{O}(k G m_g/R)]+\mathcal{O}(K G m_g/R)]$, i.e., in
the inner solar system $m$ and $m_g$ differ fractionally by $\ll
10^{-5}$.   Setting $r_g=2Gm_g=2Gm$ gives the second form of
$\beta$.  Our results for $\beta$ and $\gamma$ are consistent with
those obtained by Eiling and Jacobson~\cite{Eil_Jac} for the
relevant case of the Jacobson-Mattingly theory.

The $\beta$ and $\gamma$ are the standard post-Newtonian coefficients measurable by the classical tests of gravity theory~\cite{Will}.  They are both unity in T$e$V$e$S, exactly as in GR (for $\beta$ this was first noticed by Giannios).  Consequently the classical tests (perihelion precession, light deflection and radar time delay) cannot distinguish between the two theories with present experimental precision.

The $\beta$ and $\gamma$ are \textit{not} the only PPN
coefficients.  Future work should look at those coefficients
having to do with preferred frame effects, as well as at the
Nortvedt effect, which should not be null in T$e$V$e\,$S.

\section{\label{sec:lensing}Gravitational Lensing in T$e$V$e\,$S}

In Sec.~\ref{sec:postN} we touched upon gravitational lensing in the
Newtonian regime.  Here we show that  in the low acceleration
regime, T$e$V$e\,$S predicts gravitational lensing of the correct
magnitude to explain the observations of intergalactic lensing
without any dark matter. First by following the essentially exact
method of Ref.~\onlinecite{BekSan}, we show this for a spherically
symmetric structure; in nature many elliptical galaxies and galaxy
clusters are well modelled as spherically symmetric.  We then use
linearized theory to give a short proof of the same result for
asymmetric systems.  Our discussion refers to lensing of both  rays
that pass through the system and those that skirt it, and is thus a
generalization of the implicit result about light deflection in
Sec.~\ref{sec:postN}  in more than one way.

\subsection{\label{sec:lensespherical}Spherically symmetric systems}

We adopt the Einstein metric (\ref{spherical}); the physical metric
is obtained by replacing $e^\nu\rightarrow e^{\nu+2\phi}$ and
$e^\varsigma\rightarrow e^{\varsigma-2\phi}$ in it.  Consider a
light ray which propagates in the equatorial plane of the metric
(which may, of course, be chosen to suit any light ray).  The
4-velocity
$\dot x^\alpha$ of the ray (derivative taken with respect to some
suitable parameter)  must satisfy
\begin{equation} -e^{\nu+2\phi}\, \dot t^2+
e^{\varsigma-2\phi}(\dot\varrho^2+\varrho^2\dot
\varphi^2)=0.
\label{ray}
\end{equation} From the metric's stationarity follows the
conservation law
$e^{\nu+2\phi} \dot t=E$ where
$E$ is a constant characteristic of the ray.  From spherical
symmetry it follows that
$e^{\varsigma-2\phi}\varrho^2\dot\varphi=L$ where
$L$ is another constant property of the ray.   Let us write
$\dot\varrho=(d\varrho/d\varphi) \dot\varphi $. Now eliminating
$\dot t$ and $\dot\varphi$ from Eq.~(\ref{ray}) in favor of $E$ and
$L$, and dividing by $E^2$ yields
\begin{equation} -e^{-\nu-2\phi}+(b/\varrho)^2
e^{-\varsigma+2\phi}[\varrho^{-2}(d\varrho/d\varphi)^2+1]=0,
\end{equation} where $b\equiv L/E$.   By going to infinity where the
metric factors approach unity one sees that $b$ is just the ray's
impact parameter with respect to the matter distribution's center at
$\varrho=0$.   This last equation has the quadrature
\begin{equation}
\varphi= \int^\varrho
\Big[e^{\varsigma-\nu-4\varphi}\Big({\varrho\over
b}\Big)^2-1\Big]^{-1/2} {d\varrho\over \varrho}.
\end{equation} Were the physical metric exactly flat, this relation
would describe a line with $\varphi$ varying from $0$ to $\pi$ as
$\varrho$ decreased from infinity to its value $\varrho_{turn}$ at
the turning point, and then returned to infinity.  Hence the
deflection of the ray due to gravity is
\begin{equation}
\Delta\varphi=2\int^\infty_{\varrho_{turn}}
\Big[e^{\varsigma-\nu-4\varphi}\Big({\varrho\over
b}\Big)^2-1\Big]^{-1/2}{d\varrho\over
\varrho}-\pi.
\end{equation} 

This last integral is difficult.  So let us take advantage of the
weakness of extragalactic fields which allow us to assume that $\nu$,
$\varsigma$ and $\phi$ are all small compared to unity.  Then the
above result is closely approximated by         
\begin{equation}
\Delta\varphi=-4{\partial\over\partial\alpha}\int^\infty_{\varrho_{turn}}
\Big[(1+\varsigma-\nu-4\varphi)\Big({\varrho\over
b}\Big)^2-\alpha\Big]^{1/2} {d\varrho\over
\varrho}\Big|_{\alpha=1}-\pi.
\end{equation} The rewriting in terms of an $\alpha$ derivative
allows us to Taylor expand the radical in the small quantity
$\varsigma-\nu-4\varphi$ without incurring a divergence of the
integral at its lower limit.  The zeroth order of the expansion 
yields a well known integral which cancels the $\pi$.  Thus, to
first order in small quantities
\begin{equation}
\Delta\varphi=-{2\over
b}{\partial\over\partial\alpha}\int^\infty_{b\surd\alpha}
{(\varsigma-\nu-4\phi)\varrho d\varrho\over(\varrho^2-\alpha
b^2)^{1/2}}\Big|_{\alpha=1}.
\end{equation} 

At this point it pays to integrate by parts:
\begin{equation}
\Delta\varphi=-{2\over
b}{\partial\over\partial\alpha}\Big[\lim_{\varrho\rightarrow
\infty} (\varsigma-\nu-4\phi)(\varrho^2-\alpha
b^2)^{1/2}-\int_{b\surd
\alpha}^\infty (\varsigma'-\nu'-4\phi')(\varrho^2-\alpha b^2)^{1/2}
d\varrho 
\Big]\Big|_{\alpha=1}
\label{firstint}
\end{equation} Since $\nu$, $\varsigma$ and $\phi$ all decrease
asymptotically as 
$\varrho^{-1}$, the integrated term, being $\alpha$ independent,
contributes nothing.  Carrying out the $\alpha$ derivative, and
introducing the usual Cartesian $x$ coordinate along the initial ray
by
$x\equiv \pm (\varrho^2-b^2)^{1/2}$, we have
\begin{equation}
\Delta\varphi={b\over 2}\int_{-\infty}^\infty
{\nu'-\varsigma'+4\phi'\over\varrho}\, dx.
\label{bending}
\end{equation}  A factor $1/2$ appears  because we have included the
integral in Eq.~(\ref{firstint}) twice, once with $\varrho$
decreasing to, and once with $\varrho$ increasing from $b$.  The
integral is now performed over an infinite straight line following
the original ray.

The difference between GR with dark matter and T$e$V$e\,$S in this
respect is that with dark matter one would have $\phi=0$ and would
compute
$\nu$ and $\varsigma$ from Einstein's equations including dark
matter as source, whereas in T$e$V$e\,$S one has a nontrivial $\phi$,
and computes $\nu$ and $\varsigma$ on the basis of the visible
matter alone. 

We may simplify the above result by means of Einstein's
equation~(\ref{rhorhoEq}).  We shall neglect the $\varsigma'^2$ and
$\varsigma'\nu'$ terms because they are of second order, and thus
smaller than $\nu'/\varrho$ by a factor $G\cdot {\rm mass}/\varrho$
which amounts to $v^2$, with $v$ the typical orbital velocity in the
system.  Using the residual terms we eliminate
$\varsigma'$ from Eq.~(\ref{bending}):
\begin{equation}
\Delta\varphi=b\int_{-\infty}^\infty {\nu'+2\phi'\over\varrho}\, dx
-4\pi Gb\int_{-\infty}^\infty \big(\tilde
T_{\varrho\varrho}+\tau_{\varrho\varrho}+\Theta_{\varrho\varrho}/8\pi
G\big)\, dx.
\end{equation}  Now by Sec.~\ref{sec:quasi},
$\nu=2V+\mathcal{O}(V^2)$ and
$\Phi=V+\phi$.  Hence with fractional corrections of
$\mathcal{O}(V^2)$,
\begin{equation}
\Delta\varphi=2b\int_{-\infty}^\infty {\Phi'\over\varrho}\, dx
-4\pi Gb\int_{-\infty}^\infty \big(\tilde
T_{\varrho\varrho}+\tau_{\varrho\varrho}+\Theta_{\varrho\varrho}/8\pi
G\big)\, dx.
\label{lastbend}
\end{equation} The first integral here depends exclusively on the
potential 
$\Phi$ which determines nonrelativistic motion.  That is, the
observed stellar or galactic dynamics will uniquely fix this part of
$\Delta\varphi$.  For this reason the first term makes the same
predictions for lensing by nonrelativistic systems in T$e$V$e\,$S as
in GR (where $\Phi=\Phi_N$, the last calculated assuming dark
matter).   We next show that for nonrelativistic systems the second
integral is negligible.

In astrophysical matter the radial pressure $\tilde
T_{\varrho\varrho}$ is of order
$\tilde\rho$ times the local squared random velocity of the matter
particles (stars, gas clouds, galaxies).  Thus $\int \tilde
T_{\varrho\varrho}\, dx=\langle v^2\rangle 
\int\tilde\rho \, dx$ with $\langle v^2\rangle$ a suitably averaged
$v^2$.  But by Poisson's equation $4\pi
G\tilde\rho=\bm\nabla\cdot\bm\nabla\Phi_N\sim
\Phi_N{}'/\varrho=\tilde \mu \Phi'/\varrho$ where we have also used
Eq.~(\ref{gradphi}).  Thus the term with the integral over
$\tilde T_{\varrho\varrho}$ is smaller than the first term in
Eq.~(\ref{lastbend}) by a factor of $\mathcal{O}(\tilde\mu
\langle v^2\rangle )$.  In GR (for which effectively $\tilde\mu=1$)
this factor is no larger than $10^{-5}$ for all extragalactic systems
which have a missing mass problem; in T$e$V$e\,$S it is even smaller
because typically
$\tilde
\mu\ll 1$ for such systems.

Turning now to $\tau_{\varrho\varrho}$  we recall from
Sec.~\ref{sec:quasi} that in the quasistatic situation in question,
the
$F$ part is dominated by the term quadratic in
$\phi$ derivatives.  Using Eqs.~(\ref{2phis})-(\ref{gradphi})  we
work out that $4\pi G\tau_{\varrho\varrho}\approx (k\tilde\mu/8\pi
\mu)
\Phi'\Phi_N{}' $.  Evidently
$\Phi'\sim \Phi/\varrho$, and since $\Phi=\mathcal{O}(v^2)$ and
$(k\tilde\mu/8\pi\mu)<{\scriptstyle 1\over \scriptstyle 2}$, the
contribution of
$\tau_{\varrho\varrho}$ to the second term of Eq.~(\ref{lastbend})
is no larger than that coming from $\tilde T_{\varrho\varrho}$.

Finally we note that the $\lambda$ term in
$\Theta_{\varrho\varrho}$ vanishes in a quasistatic situation
because then
$\mathfrak{U}_\alpha\approx -(1+\Phi_N)\delta_{\alpha t}$.  And
from this last formula we estimate
$|\Theta_{\varrho\varrho}|\approx {\scriptstyle 1\over
\scriptstyle 2}K(\Phi_N{}')^2\sim 
K\tilde\mu^2|\Phi\Phi'|/\varrho$.  Since $\tilde\mu<1$ and by 
Sec.~\ref{sec:postN} we must take
$K<10^{-2}$, it is clear that the contribution of
$\Theta_{\varrho\varrho}$ is much smaller than that coming from
$\tilde T_{\varrho\varrho}$.  From all the above the light ray
deflection  in T$e$V$e\,$S is
\begin{equation}
\Delta\varphi=2b \big[1+\mathcal{O}(\tilde\mu
v^2)\big]\int_{-\infty}^\infty {\Phi'\over\varrho} dx.
\end{equation} In GR with dark matter the same formula is valid with
$\mathcal{O}(\tilde\mu v^2)$ replaced by $\mathcal{O}( v^2)$.  Since
these corrections are beyond foreseeable accuracy of extragalactic
astronomy,  it is clear that for given dynamics (given $\Phi$), both
theories predict identical lensing.  We shall elaborate on this
statement shortly.

\subsection{\label{sec:asymmetric}Asymmetric systems}

We now turn to systems with no particular spatial symmetry.  The
weakness of the gravitational potentials typical of nonrelativistic
systems entitles us to use linearized theory~\cite{MTW} in which the
metric is viewed as a perturbed Lorentz metric:
\begin{equation} g_{\alpha\beta}=\eta_{\alpha\beta}+\bar
h_{\alpha\beta}-{\scriptscriptstyle 1\over \scriptscriptstyle 2}\,
\eta_{\alpha\beta} \,
\eta^{\gamma\delta} \bar h_{\gamma\delta}
\label{metricpert} 
\end{equation} with $|\bar h_{\alpha\beta}|\ll 1$.  By small
coordinate transformations one enforces the gauge conditions
$ \eta^{\beta\delta}\bar h_{\gamma\delta}{}_{,\beta} =0$; as a
consequence to first order in the
$\bar h$ fields
\begin{equation} G_{\alpha\beta}=-{\scriptscriptstyle 1\over
\scriptscriptstyle 2}\,
\eta^{\gamma\delta}\partial_\gamma\partial_\delta
\,\bar h_{\alpha\beta},
\end{equation} so that Einstein's equations take the form of wave
equations in flat spacetime with the r.h.s. of
Eq.~(\ref{gravitationeq}) as sources.  Of course there are motions
and changes in galaxies and clusters of galaxies, but the associated
changes in the metric are mostly very slow.   Thus we confine
ourselves to quasistationary situations where we can drop time
derivatives (but not yet the
$g_{ti}$ components since galaxies do rotate).  This tells us that 
\begin{equation} G_{tt}=-{\scriptscriptstyle 1\over
\scriptscriptstyle 2}\,
\nabla^2 \,\bar h_{tt}  = 8\pi G\Big[\tilde T_{tt} +2(1-e^{-4\phi})
\mathfrak{U}^\mu
\tilde T_{\mu t} \mathfrak{U}_{t} +\tau_{tt}\Big]+ \Theta_{tt}.
\end{equation} The various parts of the source here were explored in
Sec.~\ref{sec:quasi}; from that discussion it follows that
\begin{equation}
\bar h_{tt}=-4V= -4\Xi\Phi_N.
\label{htt}
\end{equation}

In regard to the spatio-temporal source components of
Eq.~(\ref{gravitationeq}), we observe that the $\tilde T_{it}$ is an
$\mathcal{O}(v)$ below $\tilde T_{tt}$ (momentum density is velocity
times mass density).  Further, the dominant contributions to
$\tau_{ti}$ are
$\bar h_{ti}$ multiplied by $\sigma^2\eta^{jk}\phi_{,j}\phi_{,k}$ 
and by
$(G/\ell^2)\sigma^4 F$.   Of these the first dominates (see
Sec.~\ref{sec:quasi}),  and it is small on the scale of $\tilde\rho$
both because it is of second order (c.f. Sec.~\ref{sec:postN}), and
because
$|\bar h_{ti}|\ll 1$.  We can guess that $\mathfrak{U}_i$ is at most
of order
$\bar h_{ti}$ (it would vanish in a truly static situation), and
since by Eq.~(\ref{lambdaexpl})
$\lambda$ is below $8\pi G\tilde\rho$ by factors of
$\mathcal{O}(K)$ and
$\mathcal{O}(\phi_c)$, we see that the $\lambda \mathfrak{U}_t
\mathfrak{U}_i$ term contribution to $\Theta_{ti}$ is small compared
to
$8\pi G\tilde\rho$.  Similarly, the
$Kg^{\mu\nu}\mathfrak{U}_{[\mu,t]}\mathfrak{U}_{[\nu,i]}$
contribution to $\Theta_{ti}$, being of second order in $V_{,i}$ and
first order in $\bar h_{ti}$, or first order in $V_{,i}$ and first
order in
$\bar h_{ti,j}$ (aside of carrying the small coefficient
$K$), must be very small.  We conclude that the source of the
spatio-temporal Einstein equation can be neglected, so that to the
accuracy of Eq.~(\ref{htt}),
$\bar h_{ti}\approx 0$.

Things are similar for the spatial-spatial components.  We have
already remarked that $\tilde T_{ij}$ is an $\mathcal{O}(v^2)$ below
$\tilde T_{tt}$.  The $\tau_{ij}$ consists of a term quadratic in
$\phi_{,i}$ and one with a $F$ factor which has been argued to be
smaller.  Hence $\tau_{ij}$ is small.  Again the $K
g^{\mu\nu}\mathfrak{U}_{[\mu,i]}
\mathfrak{U}_{[\nu,j]}$ contributions to $\Theta_{ij}$ are quadratic
in
$V_{,i}$ and suppressed by the $K$ coefficient, so they are also
small.  And the $\lambda$, which we remarked above to be small, is
multiplied by two factors $\bar h_{ti}$, and so is also small.  So by
the same logic as above we neglect the sources of the spatial-spatial
components $\bar h_{ij}$ and conclude that $\bar h_{ij}\approx 0$.

Substituting all these results in Eq.~(\ref{metricpert}) we obtain
\begin{equation}
g_{\alpha\beta}=(1-2V)\eta_{\alpha\beta}-4V\delta_{\alpha t}
\delta_{\beta t}.
\end{equation}  The absence of $g_{ti}$ in this approximation makes
the situation truly static (rather than just stationary); hence
$\mathfrak{U}^\alpha=\delta_t^\alpha$.  Calculating the physical
metric from Eq.~(\ref{physg}) with $e^{\pm2\phi}\approx 1\pm 2\phi$
we have
\begin{equation}
\tilde g_{\alpha\beta}=(1-2V-2\phi)\eta_{\alpha\beta}-4(V+\phi)
\delta_{\alpha t} \delta_{\beta t}
\end{equation} which is equivalent to
\begin{equation} \tilde g_{\alpha\beta} dx^\alpha dx^\beta
=-(1+2\Phi)dt^2+(1-2\Phi)\delta_{ij} dx^i dx^j
\label{weakmetric}
\end{equation}  with $\Phi=V+\phi$ as in Sec.~\ref{sec:quasi}.   

Metric (\ref{weakmetric}) has the same form as the GR metric for
weak gravity~\cite{MTW}.   Thus in T$e$V$e\,$S just as in GR the
same potential governs dynamics and gravitational lensing.   This
accords with the conclusion of Sec.~\ref{sec:lensespherical} for the
spherically symmetry case.  What does this mean in practice ? In GR
$\Phi$'s role is played by the Newtonian potential due to the
visible matter together with the putative dark matter; in
T$e$V$e\,$S $\Phi$ is the sum of the scalar field and the
renormalized Newtonian potential generated by the visible matter
alone.  These two prescriptions for $\Phi$ need not agree
\textit{a priori}, but as we argued in Sec.~\ref{sec:Mondish},
nonrelativistic dynamics in T$e$V$e\,$S are approximately of MOND
form, and MOND's predictions have been found to agree with much of
galaxy dynamics phenomenology.  We thus expect T$e$V$e\,$S's
predictions for gravitational lensing by galaxies and some clusters
of galaxies to be as good as those of dark halo models within
GR.  But, of course, the early MOND formula (\ref{MOND}), and
T$e$V$e\,$S with our choice (\ref{def_F}) for
$F(\mu)$ both claim that asymptotically the potential $\Phi$ of an isolated  galaxy grows logarithmically with distance indefinitely.  Dark halo models do not.  So T$e$V$e\,$S for a specific choice of $F$ is in principle falsifiable.  Dark matter is less falsifiable because of the essentially unlimited choice of halo models and choices of their free parameters.  One should also remember that gravitational
lensing affords the opportunity to map the $\Phi$ to greater
distances than can dynamics; for unlike the latter, lensing can be
measured outside the gas or galaxy distribution.  Using this $\Phi$ both GR and T$e$V$e\,$S would predict the same dynamics for stars or galaxies, while disagreeing on the implied distribution of mass.

\section{\label{sec:models}Cosmological evolution of $\phi$}

\subsection{\label{sec:gene}Persistence of cosmological expansion}

This section (where we write $\phi$ rather than $\phi_c$) shows that
for a range of initial conditions, FRW cosmological models with flat
spaces in T$e$V$e\,$S  expand forever, have
$0\leq
\phi\ll1$ throughout, and their law of expansion is very similar to
that in GR.    The second point is crucial for our discussion of
causality in Sec.~\ref{sec:causality}.

First using Eq.~(\ref{physg}) we transform metric (\ref{RW}) to the
physical metric
\begin{eqnarray} \tilde g_{\alpha\beta}\, dx^\alpha dx^\beta
&=&-d\tilde t^2+\tilde a(\tilde t)^2
\big[d\chi^2+f(\chi)^2(d\theta^2+\sin^2\theta\, d\varphi^2)\big],
\label{FRW}
\\ d\tilde t&=&e^\phi  dt; \qquad\tilde a=e^{-\phi}\, a. 
\label{transf}   
\end{eqnarray}
 In what follows we take the initial moment, conventionally written
as
$\tilde t=0$, at the end of the quantum era with $\tilde a(0)$ a very
small scale; furthermore we take the zero of $t$ to coincide with
$\tilde t=0$.  For illustration we assume  the initial conditions
$\dot\phi(0)= 0$ (an overdot always denotes $\partial/\partial t$) 
and
$0<\phi_0\equiv \phi(0)\ll 1$.    Hence $a$ also starts off from a
very small scale, $a_0$, and can only increase initially.

 We now show that the spatially flat ($f(\chi)\equiv
\chi$) FRW models in T$e$V$e\,$S persist and cannot recollapse, i.e.
$\tilde a$ has no finite maximum.     As in Sec.~\ref{sec:cosmology}
we have
$\mathfrak{U}^\alpha=
\delta_t{}^\alpha$ which causes $\mathfrak{U}^{[\alpha;\beta]}$ to
vanish. As a consequence $\Theta_{\alpha\beta}=-\lambda
\delta_\alpha^t\delta_\beta^t$ with $\lambda$ given by
Eq.~(\ref{lambda}). Since $\phi=\phi(t)$, Eq.~(\ref{tau}) gives
$\tau_{tt} =2\sigma^2\dot\phi^2+G(4\ell^2)^{-1}\sigma^4 F(\mu)$. 
As mentioned in Sec.~\ref{sec:cosmology}, 
$\mathfrak{U}^\beta\tilde T_{\alpha\beta}=-\tilde\rho e^{2\phi}
\mathfrak{U}_\alpha$.  Using
$g^{\alpha\beta}\mathfrak{U}_\alpha\mathfrak{U}_\beta=-1$ gives us
$\tilde T_{tt}+(1-e^{-4\phi})\mathfrak{U}^\alpha\tilde
T_{\alpha(t}\mathfrak{U}_{t)}=(2e^{-4\phi}-1)\tilde \rho
e^{2\phi}$.  Substituting all the above in the $tt$ component of
Eq.~(\ref{gravitationeq}), we get the following analog of Friedmann's
equation:
\begin{eqnarray}  {\dot a^2\over a^{2}}&=&{8\pi G\over 3}\tilde\rho
e^{-2\phi}+{8\pi G\sigma^2 \dot\phi^2\over 3}+ {2\pi\over
3k^2\ell^2}\mu^2 F(\mu)
\nonumber
\\ &=&{8\pi G\over 3}\tilde\rho e^{-2\phi}+{4\pi\over
3k^2\ell^2}\big[-\mu y(\mu)+{\scriptstyle 1\over \scriptstyle
2}\mu^2 F(\mu) \big]
\label{Friedmann}
\end{eqnarray}

With the choice (\ref{y}) for $y(\mu)$ we have  $\mu>0$,
$y(\mu)<0$ and
$F> 0$ in the cosmological domain.  Thus the scalar fields contribute
positive energy density and the r.h.s. of Eq.~(\ref{Friedmann}) is
positive definite ($\tilde\rho<0$ is physically unacceptable).  It
follows that $\dot a$ cannot vanish  for any $t$, so that by our
earlier remark it must always be positive.  Now the relations
(\ref{transf})  imply that
\begin{equation} d\tilde a/d\tilde t=e^{-2\phi}(\dot a-a\dot \phi).
\label{a_a}
\end{equation}     We shall show in the sequel that although
$\dot\phi$ can be positive, it is always the case that $|\dot\phi|\ll
\dot a/a$.  As a consequence $d\tilde a/d\tilde t$ is always strictly
positive: in T$e$V$e\,$S a FRW model with flat spaces cannot
recollapse. 

The fact that $\dot\phi$ is given by an integral over time [see
Eq.~(\ref{first})] means that in a cosmological phase transition,
where
$\tilde\rho$ may change suddenly, $\dot\phi$ (and of course
$\phi$) will nevertheless evolve continuously in time.  It follows
that 
$F$ will also evolve continuously in time [see Eq.~(\ref{F})].  A
consequence of Eq.~(\ref{Friedmann}) is that any jump in
$\tilde\rho$ will be reflected in a similar jump in  $(\dot a/a)^2$
or in the square of the Hubble function $\tilde H\equiv \tilde
a^{-1}\,d\tilde a/d\tilde t$.

\subsection{\label{sec:proto}The proto-radiation era}

Contemporary  cosmology regards the inflationary era as preceded by
a brief radiation dominated era, the proto-radiation era, in which
the physical scale factor $\tilde a$ expands  by just a few orders
following the quantum gravity regime.  As in any radiation dominated
regime, here the equation of state is
$\tilde\rho=3\tilde p$ with both $\tilde p$ and $\tilde\rho$ varying
as
$\tilde a^{-4}$.  It follows from Eq.~(\ref{first}) that throughout
the era
\begin{equation}
\mu\dot\phi=-{k\over  a^3}\int_{0}^t  G\tilde\rho e^{-2\phi} a^3 dt
,
\end{equation}  Because in the cosmological regime
$\mu> 2$, we have $\dot\phi<0$ throughout this era.  Thus as promised
$d\tilde a/d\tilde t$ in Eq.~(\ref{a_a}) is positive.   Using the
constancy of
$(G\tilde\rho)^{1/2}a^2 e^{-2\phi}$ we can  now write 
\begin{equation}
\mu\dot\phi=-{k(G\tilde\rho)^{1/2}e^{-2\phi}\over  a}\int_0^t 
(G\tilde\rho)^{1/2} a dt .
\end{equation}  Tentatively assuming that $|\phi|\ll 1$ throughout
the era we may, according to Eq.~(\ref{Friedmann}), bound both
instances of
$(G\tilde\rho)^{1/2}$ from above by
$(3/8\pi)^{1/2}\dot a/a$.  The consequent integral is then trivial,
and since
$a_0$  is essentially zero we get
\begin{equation}
\mu|\dot\phi|<(3k/8\pi)(\dot a/ a).
\label{phidotpr}
\end{equation}   Thus $|\dot\phi|<  (3k/16\pi)(\dot a/a)$;  since
$k\ll 1$, we have by Eq.~(\ref{a_a}) that $d\tilde a/d\tilde
t\approx \dot a$.  

We can now show that the cosmological evolution during the
proto-radiation era is very similar to that within GR.  For the
choice (\ref{def_F})  both $F$ and
$F'$ are positive in the cosmological domain (see
Fig.~\ref{fig:F}).  It follows from Eq.~(\ref{F})  that $\mu^2
F<-\mu y$ (recall that
$y<0$), so the last term on  the r.h.s. of the Friedmann equation is
less than half the second.   Next we use
$y=-2k\ell^2\dot\phi^2$ to infer from Eq.~(\ref{phidotpr}) that 
\begin{equation} {4\pi\over 3k^2\ell^2}\mu |y|<{3k\over
8\pi\mu}\big({\dot a\over a}\big)^2
\label{ymu}
\end{equation} But this means that the scalar field contributions to
the Friedmann equation are small compared to its l.h.s. 
Specifically, to within a fractional correction of
$\mathcal{O}(k/16)$ (actually smaller than this because $\mu$ will
turn out to be large),  the relation between $\tilde H$ and
$\tilde\rho$ is the same as in GR.

The fact that the scalar field contributions to the Friedmann
equation are small compared to its l.h.s. also means that inequality
(\ref{phidotpr}) is nearly saturated, as must be its kin
(\ref{ymu}).  Then 
\begin{equation}
\mu^2 |y(\mu)|\approx {\scriptstyle 1\over
\scriptstyle 6}(3k/4\pi)^4(\dot a/ a)^2\,
\mathfrak{a}_0{}^{-2}.
\label{mu2y}
\end{equation}
 But $a/\dot a$ is a very short scale (in standard cosmological
models
$\tilde H^{-1}
\sim 10^{-35}\, {\rm s}$ in the proto-radiation era) while
$\mathfrak{a}_0{}^{-1}\sim 3\times 10^{18}\, {\rm s}$.   
  Thus $\mu^2 y(\mu)\gg 1$.  Since by Eq.~(\ref{y}) this is possible
only for
$\mu\gg 1$, we can sharpen our earlier conclusion from
Eq.~(\ref{phidotpr}):
$|\dot\phi|\ll (3k/8\pi)\dot a/a$.  Now it is even clearer that $a$
and $\tilde a$ (as well as $t$ and $\tilde t$) are essentially equal,
so that the expansion in this era proceeds just as in GR.  Further,
integrating this last inequality gives
\begin{equation} |\phi_{pr}-\phi_0|\ll (3k/8\pi) \ln(a_{pr}/a_0),
\label{diffphi}
\end{equation} where the subscript ``$pr$'' stands for the end of the
proto-radiation era.  Since this era spans just a few $e$-foldings of
the scale $a$,  the logarithm here is of order unity.  Hence $\phi$
is almost frozen at its initial value $\phi_0$, provided this last
is not extremely small.  By choosing as initial condition 
$0<\phi_0\ll 1$, as we proposed, but avoiding extremely small
$\phi_0$,  we get  $0<\phi\ll 1$ throughout the proto-radiation era,
as assumed earlier.  Thus our assumption was consistent.

\subsection{\label{sec:inflation}The inflationary era}

The equation of state during inflation is $\tilde p=-\tilde\rho=$
const.  Then (\ref{first}) tells us that
\begin{equation}
\mu\dot\phi={k\over  a^3}\int_{t_{pr}}^t  G\tilde\rho e^{-2\phi}
a^3 dt+
\mu_{pr}\dot\phi_{pr} \Big({a_{pr}\over a} \Big)^3.
\label{scalar_infl}
\end{equation}  The integration constant prefacing the last term is
fixed by the condition that
$\mu$ and $\dot\phi$ be continuous through the proto-radiation
inflation divide.  It is clear that after rapid expansion has
suppressed the last (negative) term here, $\dot\phi$ becomes
positive. Because
$\tilde\rho$ is constant, we may pull a factor
$(G\tilde\rho)^{1/2}$ out of the integral.     Then by
Eq.~(\ref{Friedmann}) and assuming everywhere that
$e^{-\phi}\approx 1$  (which we verify below), we have 
$ (G\tilde\rho)^{1/2} e^{-2\phi}<  (3/8\pi)^{1/2}
\dot a/a$ both in and outside the integral.  Thus
\begin{eqnarray}
\mu\dot\phi\  &<&{3 k\dot a\over 8\pi a^4}\int_{t_{pr}}^t  a^2
\dot a\, dt+\mu_{pr}\dot\phi_{pr} \Big({a_{pr}\over a} \Big)^3
\\ &=&{ k\dot a
\over 8\pi a}\Big(1-{a_{pr}{}^3\over a^3}\Big) -{3k\over
8\pi}\Big({\dot a\over a}\Big)_{pr}\Big({a_{pr}\over a}
\Big)^3.
\label{phirate}
\end{eqnarray}   where we have used Eq.~(\ref{phidotpr}) as an
equality as the end of the proto-radiation era. Thus during inflation
\begin{equation} -(3k/8\pi)(\dot
a/a)_{pr}<\mu\dot\phi<(k/8\pi)(\dot a/a).
\label{ineq1}
\end{equation}  The l.h.s. here comes from the last term in
Eq.~(\ref{scalar_infl}) in light of inequality~(\ref{phidotpr}). 
In the passage from the proto-radiation era, which involves a phase
transition,   $\tilde\rho$ can change by a factor of order unity, but
then settles down to a constant.  Thus by Eq.~(\ref{Friedmann}) 
$\dot a/a$ remains \textit{at least} of the same order of magnitude
as
$(\dot a/a)_{pr}$.   Hence inequality (\ref{ineq1}) translates into
one of the same form as (\ref{phidotpr}) but valid during
inflation.  As in Sec.~\ref{sec:proto}, this tells us that $d\tilde
a/d\tilde t\approx
\dot a$ also during inflation.  And the argument following
inequality  (\ref{phidotpr}) can now be repeated to show that the
$-\mu y$ and
$\mu^2 F$ terms in Friedmann's equation amount to relative
corrections of $\mathcal{O}(k/16)$ (actually smaller), so that
inflation in T$e$V$e\,$S  proceeds very much like in GR.  

Repeating the argument  leading to Eq.~(\ref{phirate}) in light of
this last conclusion and the added realization that the $a^{-3}$
terms disappear very rapidly, we conclude that during the
$\dot\phi>0$ part of inflation, inequality~(\ref{phidotpr}) is very
nearly saturated. One can then rederive Eq.~(\ref{mu2y}) as in
Sec.~\ref{sec:proto}.   Because the inflation timescale is again very
short compared to
$\mathfrak{a}_0{}^{-1}$, the argument yielding Eq.~(\ref{diffphi})
can be repeated with slight modifications to show that during
inflation
$\mu\gg 1$, and consequently
\begin{equation}
\phi_{i}-\phi_{pr}\ll (3k/8\pi) \ln(a_{i}/a_{pr}),
\label{diffphi2}
\end{equation} where a subscript ``$i$'' stands for the end of
inflation.    Thus, although in standard models inflation can span up
to 70 e-foldings of $a$, the r.h.s. of this inequality is very small
compared to unity.  We conclude that inflation manages to raise
$\phi$ above its value at the end of the proto-radiation era by a
very small fraction of unity.  This justifies our replacement of 
$e^{-\phi}$ by unity in deriving Eq.~(\ref{phirate}). 

In what follows we shall denote by $\tilde H_i$, $\mu_i$, $\phi_{i}$
and
$\dot\phi_{i}$ the values of the Hubble parameter,
$\mu(-2k\ell^2\dot\phi^2)$, 
$\phi$ and $\dot\phi$, respectively, at the end of inflation,
$t=t_i$, where
$a=a_i$.
 
\subsection{\label{sec:rad}The radiation era}

In the ensuing radiation era  the equation of state switches back to
$3\tilde p=\tilde\rho$ with both $\tilde p$ and $\tilde\rho$ varying
as
$\tilde a^{-4}$.   Thus  the integral in  Eq.~(\ref{first})  is
\begin{equation}
\mu\dot\phi=-{k\over  a^3}\int_{t_{i}}^t  G\tilde\rho e^{-2\phi}
a^3 dt +\mu_i\dot\phi_i \Big({a_i\over  a}\Big)^3,
\label{dotphir0} 
\end{equation}  with the integration constant $\mu_i\dot\phi_i$ set
so 
$\mu\dot\phi$ at the radiation's era outset equals that at
inflation's end.    Although initially
$\dot\phi>0$, clearly the integral will eventually dominate the last
term making
$\dot\phi$  negative thereafter.

Now according to Eq.~(\ref{phirate}), 
$\mu_i\dot\phi_i <(k/8\pi)  (\dot a/a)_i$.  Due to the approximate
continuity of
$(\dot a/a)$ across the inflation-radiation eras divide [which
itself follows from the approximate continuity of $\tilde\rho$ and
Eq.~(\ref{Friedmann})], and from the fact that $(\dot a/a)$  falls
off no faster than $(a_i/a)^2$ in the radiation era, 
Eq.~(\ref{dotphir0}) gives
\begin{equation}
\mu\dot\phi<(k/8\pi)  (\dot a/a)_i\, (a_i/a)^3<(k/8\pi) (\dot a/a).
\label{dotphir3}
\end{equation}

On the other hand, from  $\tilde\rho \tilde a^4=$ const. we can move
a factor
$(G\tilde\rho)^{1/2} a^2 e^{-2\phi}$ out of the integral in
Eq.~(\ref{dotphir0}).  Using again $ (G\tilde\rho)^{1/2}< 
(3/8\pi)^{1/2}
\dot a/a$ from Eq.~(\ref{Friedmann})  (if we assume provisionally
that
$e^{-\phi}\approx 1$) both in and outside the integral, we have
\begin{equation}
\mu\dot\phi>-\Big({3k\dot a\over 8\pi a^2}\Big)\int_{t_{i}}^t  
(\dot a/a)a\, dt +\mu_i\dot\phi_i \Big({a_i\over a} \Big)^3.
\label{dotphir1} 
\end{equation}
   The integral is
$a(t)-a_i$.  Hence
\begin{equation}
\mu\dot\phi>(-3k/8\pi)(1-a_i/a) (\dot
a/a)+\mu_i\dot\phi_i(a_i/a)^3>-(3k/8\pi) (\dot a/a)
\label{dotphir2}
\end{equation}  In view of Eqs.~(\ref{dotphir3}) and
(\ref{dotphir2}), inequality (\ref{phidotpr}) is again valid here. 
Because $\mu>2$ we get again from Eq.~(\ref{a_a}) that $d\tilde
a/d\tilde a\approx\dot a/a$.    We may now reproduce inequality
(\ref{ymu}) and show as in Sec.~\ref{sec:proto} that to within a
fractional correction of
$\mathcal{O}(k/16)$,  the relation between $\tilde H$ and
$\tilde\rho$ is the same as in GR. 

Because of this last result, Eq.~(\ref{dotphir3})  and the rapid
decay of
$a_i/a$  in Eq.~(\ref{dotphir2}), we may conclude that when
$\dot\phi<0$, inequality~(\ref{phidotpr}) is nearly saturated.  We
may then rederive Eq.~(\ref{mu2y}) as before.  Now in conventional
cosmology at redshift
$z$ during the radiation era $\tilde H\sim 3\times 10^{-20}(1+z)^2\,
{\rm s}^{-1}$, which by previous inference  closely approximates 
$\dot a/a$ in our model.  We thus obtain
$\mu^2 |y(\mu)|\approx 5\times 10^{-6} k^4 (1+z)^4$.  Taking
$k\sim 0.03$ on the basis of Sec.~\ref{sec:Newtonian} we see that at
the end of the radiation era ($z\approx 10^4$),
$\mu^2|y(\mu)|\approx 4\times 10^4$ which corresponds to
$\mu\approx 10$.  For earlier times
$\mu\propto (1+z)^{4/5}$ so that it rises to $10^{19}$ at the
beginning of the era at
$z\approx 10^{27}$.  Going back to inequality (\ref{phidotpr}) we
see that  in the last three
$e$-foldings of the era
$\phi(t)-\phi_i>-8\times 10^{-4}$ with the previous 50 $e$-foldings
contributing an even  smaller decrease.  Our assumption that
$e^{-\phi}\approx 1$ was evidently justified if  $\phi_0$ is taken
small compared to unity, yet sufficiently positive to keep $\phi(t)$
positive throughout the era.

We shall denote by $\mu_r$, $\phi_{r}$ and $\dot\phi_{r}$ the values
of
$\mu(-2k\ell^2\dot\phi^2)$,  $\phi$ and $\dot\phi$, respectively, at
the end of the radiation era, $t=t_r$ where $a=a_r$.
 
\subsection{\label{sec:mat}The matter era}

In the matter era $\tilde p\approx 0$ and $\tilde\rho$ varies as
$\tilde a^{-3}$.  Integrating  Eq.~(\ref{first}) gives, c.f.
Eq.~(\ref{dotphir0})
\begin{equation}
\mu\dot\phi={-k\over 2a^3}\int_{t_r}^t G\tilde\rho e^{-2\phi} a^3
dt+\mu_r\dot\phi_r \Big({a_r\over a} \Big)^3.
\label{dotphim0}
\end{equation} It is clear that $\dot\phi$ continues to be negative
throughout the matter era.  Using $\tilde\rho a^3\,e^{-3\phi} =$
const. and setting henceforth $e^\phi= 1$ (whose consistency will be
checked below),  we explicitly perform the integral in
Eq.~(\ref{dotphim0}) from
$t_r$ to
$t$:  
\begin{equation}
\mu\dot\phi=-{\scriptstyle 1\over \scriptstyle 2}\,
kG\tilde\rho(t-t_r) +\mu_r\dot\phi_r (a_r/a)^3.
\label{dotphi4}
\end{equation} Integrating the inequality
$(G\tilde\rho a^3)^{1/2}<(3/8\pi)^{1/2} a^{1/2} \dot a$ coming from
Eq.~(\ref{Friedmann}) we get
$(G\tilde\rho)^{1/2}(t-t_r)<(2/3)(3/8\pi)^{1/2}$.  Both together
give
$G\tilde\rho(t-t_r)<(\dot a/4\pi a)$, which when substituted in
Eq.~(\ref{dotphi4}) finally gives
\begin{equation}
\mu\dot\phi>(-k/8\pi) (\dot a/a)+\mu_r\dot\phi_r (a_r/a)^3.
\label{dotphim2}
\end{equation}

Now according to Eq.~(\ref{dotphir2})
 $\mu_r\dot\phi_r>(-3k/8\pi)(\dot a/a)_r$.  Thus at the beginning of
the matter era, where $a=a_r$,  the lower bound on the second term
on the r.h.s. of inequality (\ref{dotphim2}) maybe as much as three
times larger in magnitude than the first term, yet it decays as
$a^{-3}$ while the first term cannot do so faster than $a^{-3/2}$
[see Friedmann's equation (\ref{Friedmann})].  Hence within about
one
$e$-folding of $a$, the first term comes to dominate the r.h.s., and
over most of the matter era
\begin{equation}
\mu|\dot\phi|<(k/8\pi) (\dot a/a).
\label{newonk}
\end{equation} From this follows a tighter version of bound
(\ref{ymu}) which again demonstrates that the scalar field terms in
Einstein's equations are rather negligible.  The fact that
(\ref{newonk}) may be exceeded by a factor of a few early in the
matter era is no reason to exclude that epoch from the just mentioned
conclusion: the rather large 
$\mu$ at the end of the radiation era ($\mu\sim 10$ )---and a bit
beyond---acts to suppress that factor.  Using by now well worn logic
we conclude that in the matter era as well, the relation between
$\tilde H$ and $\tilde\rho$ is almost the same as in GR.

Integrating inequality (\ref{newonk}) with the use of $\mu>2$ (the
first
$e$-folding's  relatively larger contribution is suppressed by the
larger
$\mu$ which holds sway then), we get
\begin{equation} \phi(t)-\phi_r> -(k  /16\pi)\ln(a/a_r).
\label{delphi5}
\end{equation} Because the matter era thus far has spanned nine
$e$-foldings,  $\phi$ has decreased by less than  $0.0054$ during
this era.  

Note that we have not addressed the cosmological matter problem. In
T$e$V$e\,$S the expansion is driven by just $\tilde\rho$, the visible
matter's density, whereas the observations require that the source of
Friedmann's equation which falls off like $\tilde a^{-3}$ should be
larger by a factor of perhaps 6.  There are at least two possible
avenues for dealing with this embarrassment.  First, we have stuck to
a particular $F(\mu)$; possibly a more realistic $F(\mu)$ would
change late cosmological evolution enough to resolve the problem. 
Second, we have insisted on $\phi$ being small.  This is a consistent
solution as we have shown, but it is perhaps not the unique
solution.  Plainly nonegligible values of $\phi$ can affect the
Friedmann equation significantly.   

\subsection{\label{sec:acc}The accelerating expansion}

Lately data from distant supernovae indicate that in recent times
($z<0.5$) the cosmological expansion has began to accelerate, namely,
that
$d\tilde H/d\tilde t>-\tilde H^2$.   The data are best interpreted in
GR by accepting the existence a positive cosmological constant
$\Lambda\approx 2\tilde H_{today}^2$~\cite{Riess}.  One can
incorporate such accelerating epoch in the T$e$V$e\,$S Einstein
equations (\ref{gravitationeq}) by adding to $\mu^2
F(\mu)$---purely phenomenologically---a constant
($\mu$-independent) term of magnitude $\approx \Lambda
k^2\ell^2/2\pi$.   Such constant part, which corresponds to the
integration constant involved in solving Eq.~(\ref{F}), leaves
$y(\mu)$ unchanged, merely shifting the curve for
$F(\mu)$ in Fig.~\ref{fig:F} up.   Furthermore, according to
Eq.~(\ref{a_0}) and the empirical connection
$\mathfrak{a}_0\sim H_0$~\cite{M1}, the added constant is
$\sim 3k^3/16\pi^2$, that is very small.  It cannot thus affect the
discussion in earlier sections, and in particular $F$ continues to
make a positive contribution to the energy both in static systems,
and in cosmology. 

The appearance of the cosmological constant in $F$ has almost no
effect on the value of $\phi$.  To see why note that $\Lambda$ does
not directly affect the scalar equation (\ref{s_equation2}), but
only the Friedmann equation (\ref{Friedmann}).    Hence
Eq.~(\ref{dotphi4}) is still valid.  As the expansion accelerates,
$a$ begins to grow exponentially with $t$.  Both terms on the r.h.s.
of Eq.~(\ref{dotphi4}) thus fall off drastically, and $\phi$ becomes
``stuck'' at the value it had soon after the onset of acceleration. 
Consolidating the results of Secs.~\ref{sec:proto}-\ref{sec:mat} with
our conclusion we see that the range of initial conditions
$0.007<\phi_0\ll 1$ insures that  $\phi>0$ and
$e^\phi\approx 1$ throughout cosmological evolution.

\section{\label{sec:causality}Causality in T$e$V$e\,$S}

T$e$V$e\,$S's predecessors, AQUAL and PCG, permitted superluminal
propagation of scalar waves on a static background.  In the case of
PCG with a convex potential this occurs hand in hand with an
instability of the background, so it is unclear if true causality
violation occurs.  How does T$e$V$e\,$S handle causality issues ?  

The question is complicated here by of the existence of two metrics,
$g_{\alpha\beta}$ and $\tilde g_{\alpha\beta}$, whose null cones do
not coincide (except where $\phi=0$).  Which of the two cones is the
relevant one for causal considerations ?   We shall take the view
that since common rods and clocks are material systems with
negligible self-gravity, the coordinates to which the Lorentz
transformations of special relativity refer are those of local
orthonormal frames of the physical metric
$\tilde g_{\alpha\beta}$ and not of
$g_{\alpha\beta}$.  It is by ascertaining that in no such physical
Lorentz frame can physical signals travel back in time that we shall
certify the causal behavior of T$e$V$e\,$S.  Now Lorentz
transformations involve a parameter, the critical speed ``$c$''.  We
shall identify this with the speed of electromagnetic disturbances so
that, as customary, the speed of light is the same in all Lorentz
frames.  Since we have built special relativity into T$e$V$e\,$S by
insisting that all nongravitational physics equations (including
Maxwell's equation) take their standard form when written with
$\tilde g_{\alpha\beta}$, this procedure is consistent.  In fact, all
signals associated with particles of all sorts are subluminal or
travel at light's speed with respect to $\tilde g_{\alpha\beta}$.

There remains the question of whether gravitational perturbations
(tensor, vector or scalar) can ever exit $\tilde g_{\alpha\beta}$'s
null cone.  The analysis given below is quite different for tensor
and vector perturbations on the one hand, and scalar perturbations
on the other.  One point in common, however, is that causality is
guaranteed only in spacetime regions for which $\phi> 0$.  As shown
in Sec.~\ref{sec:models}, there is gamut of reasonable cosmological
models for which $\phi$ is indeed positive throughout the expansion.

\subsection{\label{sec:tensor_p}Propagation of tensor and vector
disturbances is causal}

 The characteristics of both Einstein's equations
(\ref{gravitationeq}) and the vector equation (\ref{vectoreq}) lie
on the null cone of
$g_{\alpha\beta}$ because all terms in them with two derivatives are
the usual ones in Einstein's and gauge field's equations. 
Accordingly, we do not expect metric and vector perturbations to
travel outside the null cone of the Einstein metric
$g_{\alpha\beta}$.  However, the interesting question is rather what
is the speed of a wave of this class in terms of the physical metric
$\tilde g_{\alpha\beta}$ ?

In the eikonal approximation the wavevector $\kappa_\alpha$ of
metric perturbations, that is the 4-gradient of the characteristic
function, will satisfy
$g^{\alpha\beta}\kappa_\alpha
\kappa_\beta = 0$.    Hence Eq.~(\ref{inverse}) gives
\begin{equation}
\tilde g^{\alpha\beta}\kappa_\alpha \kappa_\beta -
2(\mathfrak{U}^\alpha
\kappa_\alpha)^2
\sinh(2\phi)=0.
\label{xx}
\end{equation} We consider a generic situation where
$\mathfrak{U}^\alpha$ may have both temporal and spatial
components.  The normalization (\ref{norm}) implies by
Eq.~(\ref{physg}) that $\tilde
g_{\alpha\beta}\mathfrak{U}^\alpha\mathfrak{U}^\beta=-e^{2\phi}$.
Thus in an appropriately oriented local Lorentz frame,
$\mathfrak{L}$, of the metric $\tilde g_{\alpha\beta}$ we may
parametrize
$\mathfrak{U}^\alpha$ by
\begin{equation}
\mathfrak{U}^{\alpha}=e^{\phi} (1-V^2)^{-1/2}\{1,-V,0,0\}
\label{U}
\end{equation} with  $-1<V<1$.  This $V$ is actually the ordinary
velocity (measured by the physical metric) of $\mathfrak{L}$ w.r.t.
the privileged frame in which the matter is at rest (whether in
cosmology or in a local static configuration), namely that in which 
$\mathfrak{U}^\alpha =\{e^{\phi},0,0,0\}$.   This is evident by
considering a Lorentz transformation from the matter rest frame  to
the coordinates appropriate to frame $\mathfrak{L}$.

In view of the above, Eq.~(\ref{xx}) reduces to
\begin{eqnarray} 0&=& A\omega^2+2B
\bm{\kappa}_\|\omega+D\bm{\kappa}_\|{}^2-(1-V^2)\bm{\kappa}_\bot{}^2 
\label{dispersion}
\\ A&\equiv& e^{4\phi}-V^2
\\ B&\equiv&V(e^{4\phi}-1)
\\ D&\equiv& -1+V^2e^{4\phi}
\end{eqnarray} with $\omega=-\kappa_t$ and $\bm{\kappa}_\|$ and
$\bm{\kappa}_\bot$ the spatial components of
$\kappa_\alpha$ collinear and normal to $\mathfrak{U}_i$ (the space
part of
$\mathfrak{U}_\alpha$), respectively.  For arbitrary $V$
(\ref{dispersion}) is an anisotropic inhomogeneous dispersion
relation ($\omega$ depends on position through $\phi$ as well as on
direction of the wavevector).  However, in the rest frame of the
matter ($V=0$), it is isotropic (though still position dependent
through
$\phi$) with group (or phase) speed equal to
\begin{equation} v_0=e^{-2\phi}.
\label{speed}
\end{equation} 

The condition for tensor and vector perturbations not to propagate
superluminally ($v_0\leq 1$ as judged in the physical metric) is thus
that $\phi> 0$, which as we saw, is satisfied in a wide range of
cosmological models (see Sec.~\ref{sec:models}) as well as
quasistatic systems embedded in them (Sec.~\ref{sec:postN}). 
Normally this conclusion could be carried over to all Lorentz frames
without further calculations.   But because T$e$V$e\,$S admits a
locally privileged frame, that in which
$\mathfrak{U}^\alpha =e^{\phi}\{1,0,0,0\}$, we investigate this
conclusion in more detail for any $V^2<1$. 

Solving Eq.~(\ref{dispersion}) for $\omega$ gives
\begin{eqnarray}
\omega &=&
 (-B\bm{\kappa}_\|\pm  S)A^{-1},
\label{omega}
\\ S&\equiv & (C\bm{\kappa}_\|^2+A(1-V^2)\bm{\kappa}_\perp^2)^{1/2},
\\ C&\equiv& B^2-AD=(1-V^2)^2e^{4\phi}.
\end{eqnarray}  The condition $\phi> 0$ makes $A$ here strictly
positive.  It is possible for the above expression for $\omega$ to
change sign, so for given
$\bm{\kappa} $ we must agree to always choose the branch of the
square root that makes $\omega$ positive (negative $\omega$ with
opposite sign
$\bm{\kappa}$ is the same mode, of course).   In what follows we
call the modes with upper (lower) signs of the square root
$+$~($-$)~modes. For the components of
\textit{group} velocity collinear and orthogonal to
$\mathfrak{U}_i$, respectively,  we derive
\begin{eqnarray}
\mathbf{v}_\|&=&\partial\omega/ \partial\bm{\kappa}_\| = (-B\pm C
S^{-1}\bm{\kappa}_\|)A^{-1},
\\
\mathbf{v}_\perp&=&{\partial\omega/
\partial\bm{\kappa}_\perp}=
\pm (1-V^2)S^{-1}\bm{\kappa}_\perp.
\label{vperp}
\end{eqnarray}  Since these expressions are homogeneous of degree
zero in
$\bm{\kappa}$, there is no dispersion, but for $V\neq 0$ the
propagation is  anisotropic.  For small $\phi$ one has analytically
\begin{equation} v=1-2(1\pm
V\cos\vartheta)^2\,(1-V^2)^{-1}\phi+\mathcal{O}(\phi^2)
\label{upsilon}
\end{equation} where  $v\equiv
(\mathbf{v}_\|^2+|\mathbf{v}_\perp|^2)^{1/2}$ and $\vartheta$ is
the angle between $\bm{\kappa}$ and $\mathfrak{U}_i$. Thus for
moderate $V$ the group speed $v$ is subluminal, but obviously
formula (\ref{upsilon}) becomes unreliable for $V$ close to unity. 

For arbitrary $V$ it is profitable, as remarked by Milgrom, to write
$v$ in terms of $\omega$.  In fact a straightforward calculation
gives
\begin{equation} 1-v^2=S^{-2} C
(\bm{\kappa}_\|^2+\bm{\kappa}_\perp^2-\omega^2),
\label{one-v2}
\end{equation} from which it is clear that $v$ can become
superluminal only if the (isotropic) phase speed
$\omega(\bm{\kappa}_\|^2+\bm{\kappa}_\perp^2)^{-1/2}$ does the same
simultaneously.  Since the latter was found subluminal at $V=0$, we
have only to ask if there is some $V<1$ for which
$\omega=(\bm{\kappa}_\|^2+\bm{\kappa}_\perp^2)^{1/2}$;  we might
then suspect there is  superluminal propagation for larger
$V$.  Suppose we substitute this last value of $\omega$ in
Eq.~(\ref{dispersion}) together with those of $A$, $B$ and $D$. 
Collecting terms one can put the condition for the transition to
superluminality in the form
\begin{equation}
(e^{4\phi}-1)\Big(V\bm{\kappa}_\|+\sqrt{\bm{\kappa}_\|^2+
\bm{\kappa}_\perp^2}\,\Big)^2=0.
\label{superl}
\end{equation}
 As we saw in Sec.~\ref{sec:models}, for a broad class of
cosmological models $\phi>0$ throughout the expansion, and as
Sec.~\ref{sec:postN} testifies,  variation of
$\phi$ in the vicinity of localized masses embedded in such a
cosmology is far short of what is required to turn the sign of
$\phi$. It is thus clear that even in the case $\bm{\kappa}_\|<0$, 
condition (\ref{superl}) cannot be satisfied for
$V<1$.  Hence superluminal propagation of vector and tensor
perturbations is forbidden.

How does $v$ vary with $V$ ?  When $\bm\kappa_\perp\neq 0$, we find
numerically the following behavior.  For the $+$~mode with
$\bm\kappa_\|\leq 0$, $\mathbf{v}_\|<0$ for all $V$, and after
experiencing a shallow maximum at modest $V$, $v$ reaches a minimum
at  $V$ very near unity, which is the deeper and farther from $V=1$
the larger $|\bm\kappa_\perp/\bm\kappa_\||$.  As
$V$ grows further, $v$ rises and approaches unity for $V\rightarrow
1$.   If $\bm\kappa_\|>0$,
$\mathbf{v}_\|$ starts positive for small $V$ but eventually turns
negative at a rather large $V$ which grows with
$|\bm\kappa_\perp/\bm\kappa_\||$.   As $V$ grows further,
$v$ reaches a minimum, which gets shallower with growing
$|\bm\kappa_\perp/\bm\kappa_\||$, and then begins to rise.   At a
critical
$V$ the positive $\bm\kappa_\|$ $+$~mode terminates.  However, the
$-$~mode with
\textit{negative} $\bm\kappa_\|$  takes over onward from the
critical
$V$; it features $\mathbf{v}_\|<0$, and for it $v$ rises with
$V$ and approaches unity as $V\rightarrow 1$.   The $-$~mode with
$\bm\kappa_\|>0$ is always unphysical. 

For $\bm \kappa_\perp=0$ and $\bm\kappa_\|< 0$ the $+$~mode has 
$\mathbf{v}_\|<0$ throughout, and $v$ rises monotonically with
$V$  approaching unity as $V\rightarrow 1$.   For  
$\bm\kappa_\|> 0$ that same mode has
$\mathbf{v}_\|>0$ and $v$ decreasing with increasing $V$ up to a
$V=V_c\approx e^{-2\phi}$  at which point both
$\mathbf{v}_\|$ and $v$ vanish.   The terminated sequence is
continued by the $-$~mode with $\bm\kappa_\|<0$  for which
$\mathbf{v}_\|<0$ and $v$ rises monotonically with $V$ from zero at
$V=V_c$ and approaches unity as $V\rightarrow 1$.   

\subsection{Propagation of scalar perturbations is also causal}

The terms with second derivatives in the scalar equation
(\ref{s_equation}) have a nonstandard form reminiscent of those in
relativistic AQUAL (see Appendix~\ref{sec:A}).  Do scalar
perturbations propagate across
$\tilde g_{\alpha\beta}$'s null cone, that is do they travel faster
than electromagnetic waves ?  We now show that the answer is
negative.  In the scalar equation (\ref{s_equation}) in free space
we break
$\phi$ into background and perturbation  $\phi=\phi_{\rm
B}+\delta\phi$, but ignore perturbations of
$g_{\alpha\beta}$ and $\mathfrak{U}_\alpha$.  For convenience we
shall call
$\phi_{\rm B}$ simply $\phi$.  To first order in
$\delta\phi$ we get [c.f. Eqs.~(\ref{pert})-(\ref{xi})]
\begin{eqnarray} 0&=&\left(h^{\alpha\beta}+2\xi H^\alpha
H^\beta\right)\delta\phi_{;\alpha\beta} + \cdots
\label{scalarpert}
\\ H^\alpha&\equiv& (h^{\mu\nu}\phi_{,\mu}
\phi_{,\nu})^{-1/2}h^{\alpha\beta}\phi_{,\beta}
\label{defU}
\\
\xi&\equiv& d\ln \mu(y)/d\ln y
\label{defxi}
\end{eqnarray} where the ellipsis denotes terms with $\delta\phi$
differentiated only once.   We have \textit{temporarily} assumed
that
$H^\alpha$ is spacelike.   Using Eq.~(\ref{inverse}) we reexpress
(\ref{scalarpert}) in terms of the physical metric:
\begin{equation} [e^{-2\phi}\tilde
g^{\alpha\beta}-(2-e^{-4\phi})\mathfrak{U}^\alpha\mathfrak{U}^\beta+2\xi
H^\alpha H^\beta]\delta\phi_{;\alpha\beta}+ \cdots=0
\label{modwaveeq}
\end{equation}

\subsubsection{Quasistatic background}

For a quasistatic background, e.g. a quiescent galaxy,  $H^\alpha$ is
indeed a  purely space vector in coordinates that reflect the time
symmetry.   By  (\ref{defU})  $H^\alpha$ is normalized to unity
w.r.t. metric
$g_{\alpha\beta}$ and to $e^{-2\phi}$ w.r.t. $\tilde
g_{\alpha\beta}$.  
   In a local Lorentz frame of $\tilde g_{\alpha\beta}$  at rest
w.r.t. to those coordinates and appropriately oriented,  a generic
$H^\alpha$  will have the form $e^{-\phi}\{0,s,0,\sqrt{1-s^2}\}$,
with
$s$ the cosine of the angle between $H_i$ and the positive $x$ axis.
Then in a Lorentz frame moving w.r.t. the first one at velocity $V$
in the positive
$x$ direction 
\begin{equation} H^{\alpha}=e^{-\phi}
(1-V^2)^{-1/2}\{-Vs,s,0,\sqrt{(1-s^2)(1-V^2)}\}
\end{equation} In this same frame $\mathfrak{U}^\alpha$ is given by
Eq.~(\ref{U}). 

In the eikonal approximation (c.f. Appendix~\ref{sec:A}) one
replaces in a Lorentz frame  
$\delta\phi_{;\alpha\beta}$
$\mapsto$ $-\kappa_\alpha\kappa_\beta\delta\phi$ and drops first
derivatives.  Again interpreting $-\kappa_t$ as $\omega$ this gives
a generalization of (\ref{dispersion}), namely
\begin{eqnarray} 0&=&\hat A\omega^2+2(B_\|
\bm{\kappa}_\|+B_\perp
\bm{\kappa}_\perp)\omega +\hat
D\bm{\kappa}_\|{}^2-(1-V^2)(\bm{\kappa}_\lfloor^2+E\bm\kappa_\perp^2)+2B_\perp
V^{-1} \bm{\kappa}_\|\bm{\kappa}_\perp
\\
\hat A &\equiv& 2 e^{4\phi}-(1+2\xi s^2)V^2
\\ B_\|&\equiv& V(2e^{4\phi}-1-2\xi s^2)
\\ B_\perp&\equiv& -2V\xi s\sqrt{(1-s^2)(1-V^2)}
\\
\hat D&\equiv& 2V^2 e^{4\phi}-(1+2\xi s^2)
\\ E&\equiv& 1+2\xi(1-s^2),
\label{dispersion2}
\end{eqnarray} where $\bm\kappa_\|$ is the component in the $x$
direction, $\bm\kappa_\perp$ is that in a direction orthogonal to
$x$  in the plane spanning the $x$ axis and $H_i$, and
$\bm\kappa_\lfloor$ is the component orthogonal to that plane (we
use vector symbols for components to keep with previous
notation).     

For $V=0$ (rest frame of matter) there is nothing to distinguish the
$x$ axis from
$H_i$'s direction, so without restricting generality we may set
$s=1$  and speak jointly of $\bm\kappa_\perp$ and
$\bm\kappa_\lfloor$ as a
\textit{vector}
$\bm\kappa_\perp$.  Then the group speed $v=|\partial
\omega/\partial
\bm{\kappa}|^{1/2}$  turns out to be
\begin{equation} v_0={e^{-2\phi}\over \surd 2}\left[{(1+2\xi)^2
\bm{\kappa}_\|^2+\bm{\kappa}_\perp^2
\over(1+2\xi)
\bm{\kappa}_\|^2+\bm{\kappa}_\perp^2}\right]^{1/2}.
\label{speed2}
\end{equation}  

\begin{figure}[htbp]
\begin{center}
\includegraphics[width=3.0in]{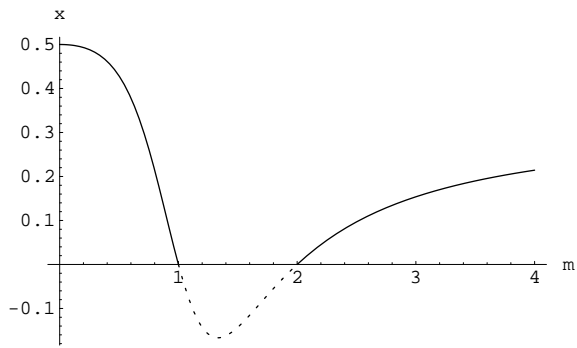}
\vspace{0.0in}
\caption{{\bf The logarithmic slope $\xi(\mu)$ as relevant for
quasistationary systems,  $0<\mu<1 $, and for cosmology,
$2<\mu<\infty $.}}
\label{fig:xi}
\end{center}
\end{figure}

From Sec.~\ref{sec:choice} we compute the logarithmic slope
\begin{equation}
\xi(\mu)=(\mu-1)(\mu-2)/(3\mu^2-6\mu+4)
\end{equation} whose graph is shown in Fig.~\ref{fig:xi}.
\vspace{0.0in}
 In particular,  $\xi\leq {\scriptstyle 1\over
\scriptstyle 2}$ in a quasistatic region.  In the deep MOND  regime
$\mu(y)\approx
\sqrt{y/3}$ so $\xi\approx {\scriptstyle 1\over \scriptstyle 2}$,
while in the high acceleration limit
$\mu(y)\approx 1$ so $\xi\approx 0$.   Consequently, in the deep
MOND regime,  $v_0\leq e^{-2\phi}$ with equality for
$\bm\kappa_\perp=0$.   In the Newtonian regime  $v_0=2^{-1/2}
e^{-2\phi}$ for all $\bm\kappa$.  Finally, in the intermediate
regime 
$2^{-1/2} e^{-2\phi}\leq v_0\leq   (1+2\xi)^{1/2}2^{-1/2}
e^{-2\phi}$, with lower and upper equality for
$\bm{\kappa}_\|=0$ and $\bm\kappa_\perp=0$, respectively.
Summarizing,  scalar waves  propagate subluminally in the frame in
which the matter is at rest, provided, of course, $\phi> 0$.

Since the vector $\mathfrak{U}^\alpha$ defines a privileged Lorentz
frame, the form of the wave equation (\ref{modwaveeq}) is different
in different frames.  Thus we must check explicitly that the
subluminal propagation of scalar waves remains valid in all Lorentz
frames.  Since the analytic expressions  for general
$\bm\kappa$ are cumbersome, we have done so numerically for small
positive $\phi$.  For small $V$ the group speed starts at the value
(\ref{speed2}).  If $\bm\kappa_\|<0$, $v$ for the $+$ mode rises
with increasing $V$ approaching unity as $V\rightarrow 1$.  By
contrast, if $\bm\kappa_\|>0$, $v$ at first decreases with
increasing
$V$ only to reach a minimum which can be quite narrow and deep for
$\bm\kappa_\|/|\bm\kappa|$ near unity.  Beyond the minimum is a
critical $V$ past which the  $+$ mode with positive
$\bm\kappa_\|$ is no longer possible.  It is replaced by the $-$
mode with opposite sign of
$\bm\kappa_\|$, whose  $v$ rises as $V$ rises beyond the critical
$V$, approaching unity for $V\rightarrow 1$.  

In summary, provided $\phi> 0$ as guaranteed (see
Sec.~\ref{sec:postN}) for the vicinity of masses embedded in the
cosmologies studied in Sec.~\ref{sec:models}, no case of
superluminal propagation is observed for scalar perturbations on a
quasistatic background.

\subsubsection{Cosmological background}

Consider now propagation of scalar perturbations in FRW cosmology. 
Here $\mathfrak{U}^\alpha$ remains pointed in the time direction,
and takes the form (\ref{U}) in a local Lorentz frame of the
physical metric which moves w.r.t. the matter at velocity $V$ in the
$x$ direction.   Since
$H^\alpha$ is now timelike,  one must change the sign of the
argument of the square root  in definition~(\ref{defU}). 
Definition~(\ref{defxi}) then requires a switch in sign of the
$\xi$ term in Eq.~(\ref{scalarpert}).  We may evidently write
$\phi_{,\alpha}=\zeta\mathfrak{U}_\alpha$ (with $\zeta$ spacetime
dependent).   It follows from definition (\ref{U}) that
$H^\alpha=\surd 2\,\mathfrak{U}^\alpha$ independent of
$\zeta$.  Using all this in the modified wave equation
(\ref{modwaveeq}), we obtain in the said Lorentz frame, after an
eikonal approximation, a dispersion relation of the form
(\ref{dispersion}) with the coefficients $A$, $B$ and $C$ modified
according to the rule
$e^{4\phi}\rightarrow (2+4\xi)e^{4\phi}$. Thus in the frame
$\mathcal{L}$ where the matter is at rest ($V=0$) we now find the
isotropic group speed, c.f. Eq.~(\ref{speed}),
\begin{equation}  
 v_0=(2+4\xi)^{-1/2} e^{-2\phi},
\end{equation}    so that according to Fig.~\ref{fig:xi}, for $\phi>
0$,
$v_0$ never exceeds $1/\surd 2$.

For $V> 0$ we use the analysis leading to
Eqs.~(\ref{one-v2})-(\ref{superl}) with the substitution
$e^{4\phi}\rightarrow (2+4\xi)e^{4\phi} $ to conclude that the
passage to superluminality is forbidden.  Numerical plots disclose a
behavior of $v(V)$ very similar to the one for tensor waves.  For
 $+$ type modes with $\bm\kappa_\|<0$, $v$ grows monotonically
approaching unity for $V\rightarrow 1$.   For $\bm\kappa_\|>0$ modes
there is a minimum of
$v$ at some high $V$, the narrower and deeper the larger
$\bm\kappa_\|/|\bm\kappa|$.  A mode of this type exists only up to a
critical $V$ beyond the minimum, and is thereafter taken over by the
$-$ type mode whose $\bm\kappa_\|$ is of opposite sign, and for
which $v$ approaches unity as $V\rightarrow 1$. 
 
\subsection{Caveats}
 
Summing up,  propagation of  weak perturbations of the  tensor,
vector or scalar gravitational fields of  T$e$V$e\,$S is always
subluminal with respect to the physical metric.  We have checked
this in detail only for waves propagating on pure cosmological
backgrounds or on quasistatic backgrounds.  Furthermore, the
analysis looked at perturbations of one field while keeping the
others ``frozen'' at their background values.  A more advanced
analysis would have examined propagation of joint
tensor-vector-scalar modes.   This said, no mechanism is evident for
the formation of causal loops.   This under the condition
$\phi> 0$ which, as we have seen, is widely obeyed in flat-space
cosmological models and quasistatic systems embedded therein.
 
\acknowledgements
I thank Mordehai Milgrom, Robert Sanders, Arthur Kosowsky for a number of critical comments and for pointing out algebraic errors as well as easier ways to do things.  Some results in Sec.V have been corrected for an error noticed by Dimitrios Giannios. 

\appendix
\section{\label{sec:A}Acausality in relativistic AQUAL}

  This comes about because the wave equation for free propagation of
$\psi$ deriving from the $\mathcal{L}_\psi$ in Eq.~(\ref{rAQUAL})
(covariant derivatives are w.r.t.
$g_{\alpha\beta}$),
\begin{equation} [\tilde f'(L^2 g^{\mu\nu}\psi,_\mu
\psi,_\nu)g^{\alpha\beta}
\psi_{,\beta}]_{;\alpha} =0,
\label{rAQUALeq}
\end{equation} leads to the following linear equation for propagation
of small perturbations
$\delta\psi$ on the background $\{g^{\alpha\beta}, \psi_{\rm B}\}$:
\begin{eqnarray} 0&=&\left(g^{\alpha\beta}+2\xi X^\alpha
X^\beta\right)\delta\psi_{;\alpha\beta} + \cdots
\label{pert}
\\ X^\alpha&\equiv& (g^{\mu\nu}\psi_{{\rm B},\mu} \psi_{{\rm
B},\nu})^{-1/2}g^{\alpha\beta}\psi_{{\rm B},\beta}
\\
\xi&\equiv& d\ln \tilde f'(y)/d\ln y
\label{xi}
\end{eqnarray} In Eq.~(\ref{pert}) the ellipsis stands for terms
where
$\delta\psi$ is differentiated only once.  

For a static background  $X^\alpha$ is a unit purely space vector
$\mathbf{X}$.  In  an appropriately oriented Cartesian coordinate
system in a local Lorentz frame, it will point in the $x$
direction.  In such frame Eq.~(\ref{pert}) takes the form
\begin{equation}
0=-\delta\psi_{,tt}+(1+2\xi)\delta\psi_{,xx}+\delta\psi_{,yy}+
\delta\psi_{,zz} + \cdots
\label{waveeq}
\end{equation} In the eikonal approximation appropriate for short
wavelengths, one sets
$\psi=Ae^{\imath\varphi}$ and neglects terms with derivatives of
$A$ or of $k_\alpha\equiv
\varphi_{,\alpha}$.   Then Eq.~(\ref{waveeq}) gives
\begin{equation}
\omega = -k_t =[(1+2\xi)k_x{}^2+k_y{}^2+k_z{}^2]^{1/2}
\end{equation} The group speed $v_{\rm g}=|\partial
\omega/\partial
\mathbf{k}|^{1/2}$ turns out to be
\begin{equation} v_{\rm g}=\left[{(1+2\xi)^2 k_x{}^2+k_y{}^2+k_z{}^2
\over (1+2\xi) k_x{}^2+k_y{}^2+k_z{}^2}\right]^{1/2}.
\end{equation}   In the deep MOND regime $[\tilde f(y)={\scriptstyle
2\over\scriptstyle 3}y^{3/2}$],
$2\xi=1$ while in the high acceleration limit [$\tilde f(y)\approx
y$],
$\xi\approx 0$.  Thus whatever the choice of $\tilde f$,  $0<\xi<1$
over some range of $y$ (acceleration).  There
$v_{\rm g}>1$ if $\mathbf{k}$ is not exactly orthogonal to
$\mathbf{X}$ (distances and times measured w.r.t. metric
$g_{\alpha\beta}$).  On the other hand, light waves travel on light
cones of
$\tilde g_{\alpha\beta}$ while metric waves do so on null cones of
$g_{\alpha\beta}$.  The two metrics are conformally related so their
null cones coincide: light and metric waves travel with unit speed. 
Thus most
$\psi$ waves are superluminal, in violation of the causality
principle [see Sec.~\ref{principles}]. 

\section{\label{sec:B}Problems for PCG in solar system}

The permissible ranges of $\eta$ and $\varepsilon$ are strongly
constrained by the solar system.  It can be shown~\cite{Can2} that
the
$1/r$ force in Eq.~(\ref{modified}) causes the Kepler ``constant'' of
planetary orbits with periods $P$ and semimajor axes $\tilde a$ to
vary slightly with $\tilde a$:
\begin{equation}  4\pi^2 \tilde
a^3/P^2=GM_\odot(1+\mathfrak{a}_0\tilde a/\kappa\eta).
\end{equation}  Assuming $M_\odot\ll M_c$, we get
$\kappa={\scriptscriptstyle 1\over\scriptscriptstyle 2}$ so that as
we pass from planet to planet, the ``constant'' varies by a fraction
$\sim 2\times 10^{-15}/\eta$.  The inner planet periods $P$ are
known to better than one part in $10^{8}$. Thus $\eta>2\times
10^{-7}$.   

A stronger constraint comes the perihelia precessions of the
planets.  The anomalous force in Eq.~(\ref{modified}) generates an
extra precession~\cite{Can2} which in Mercury's case (excentricity
$0.206$ and
$\tilde a=6\times 10^{12}$ cm) amounts to $3\times 10^{-8}\eta^{-1}$
arcsec$/$century.  With 
$\eta=2\times 10^{-7}$ this already amounts to $0.35\%$ of the
Einstein precession, which is measured to about that accuracy. 
Trying to assume
$M_\odot>M_c$ just aggravates the problem.  And we are not at
liberty to raise $\eta$ further because for fixed $\mathfrak{a}_0$,
$M_c$ scales as $\eta^2$.  Thus, for example, with $\eta=3\times
10^{-7}$,  the MOND limit of PCG would not apply to galaxies with
$M<8\times 10^{9}$, a range including many dwarf spirals with
missing mass problems !  Hence the perihelion precession marginally
rules out PCG with a sextic potential.   

\section{\label{sec:C}Relation between determinants $g$ and $\tilde
g$}

From Eqs.~(\ref{physg}-\ref{inverse}) it follows that
\begin{equation}
\tilde g^{\mu\nu} g_{\nu\alpha} = e^{2 \phi}
\big[\delta^\mu{}_\alpha+(1-e^{-4\phi})\mathfrak{U}^\mu
\mathfrak{U}_\alpha\big]
\end{equation} Viewing this as multiplication of two matrices, we
take the determinant:
\begin{equation}
\tilde g^{-1} g = e^{8\phi} {\rm Det}\, \mathcal{K}(\phi,
\mathcal{U});\qquad
\mathcal{K}(\phi, \mathcal{U})\equiv
\mathcal{I}+(1-e^{-4\phi})\mathcal{U} 
\label{Det}
\end{equation} where $\mathcal{I}$ is the unit matrix whose
components are
$\delta^\mu{}_\alpha$ while $\mathcal{U}$ is a matrix with components
$\mathfrak{U}^\mu \mathfrak{U}_\alpha$.  Now both $\tilde g$ and
$g$ are scalar densities, so that their ratio must be a true
scalar.  Hence
${\rm Det}\,
\mathcal{K}(\phi,\mathcal{U})$ is a scalar.   

In a local Lorentz frame in which the unit timelike vector
$\mathfrak{U}^\alpha$ has components $\{1,0,0,0\}$,
$\mathcal{U}$'s only nonvanishing component is
$\mathcal{U}^0{}_0=-1$.  Therefore, ${\rm Det}\,\mathcal{K} =
[1-(1-e^{-4\phi})]\times 1\times 1\times 1=e^{-4\phi}$.  
Substituting this in Eq.~(\ref{Det}) we recover Eq.~(\ref{twog's}).

\section{\label{sec:D}Relations between $m_s$, $m_g$ and $r_g$}

To determine $r_g$ one must delve into the region $\varrho<R$.
Assuming that the ideal fluid modeling the matter is at rest in the
global coordinates, we may write its 4-velocity as $\tilde
u_\alpha=e^{\phi}
\mathfrak{U}_\alpha=-e^{\phi+\nu/2}\delta_\alpha{}^t$ (see
Sec.\ref{sec:GRlimit}).  Let us return to Eq.~(\ref{ttEq}),
substitute
$\tilde T_{tt}$ from Eq.~(\ref{oldT}) and reorganize the left hand
side to obtain
\begin{eqnarray}
\varrho^{-2}e^{\nu-5\varsigma/4}(\varrho^2
\varsigma' e^{\varsigma/4})'=-8\pi G\mathfrak{P}
\\
 \mathfrak{P} \equiv  \tilde\rho
e^{\nu}(2e^{-2\phi}-e^{2\phi})+\tau_{tt}+\Theta_{tt}/8\pi G
\end{eqnarray} Integration gives for $\varrho > R$
\begin{eqnarray}
\varsigma' e^{\varsigma/4}&=& -{2Gm_g\over \varrho^2} -{1\over
\varrho^2}\int_R^\varrho (8\pi
G\tau_{tt}+\Theta_{tt})\,e^{5\varsigma/4-\nu}
\varrho^2 d\varrho
\label{intfors}
\\ m_g &\equiv& 4\pi\int_0^R
\mathfrak{P}\,e^{5\varsigma/4-\nu}
\varrho^2 d\varrho,
\label{massg}
\end{eqnarray} where the integral in Eq.~(\ref{intfors}) does not
contain
$\tilde\rho$ since it extends only outside the fluid.  

How much does the ``gravitational mass'' $m_g$ differ from the scalar
mass
$m_s$ ?  For a star the volume integral  of $\tilde p$ is of order
the random kinetic energy, which by the Newtonian virial theorem is
of order of the gravitational energy $\sim Gm_g/R$.   According to
Eqs.~(\ref{expansion1}), (\ref{expansion2}) and (\ref{intscalar})
this is also the order of the fractional correction to $m_s$ or to
$m_g$ coming from the metric factors and
$e^\phi$.   We have not worked out
$\tau_{tt}$ or
$\Theta_{tt}$ in the interior, but from Eqs~(\ref{tautt}) and
(\ref{Thetatt}) we may estimate that the
$\tau_{tt}$ and $\Theta_{tt}/8\pi G$ terms contribute  to $m_g$
terms of
$\mathcal{O}(kG m_s{}^2/R)$ and
$\mathcal{O}(Kr_g{}^2/GR)$, respectively.    Because we  assume
small
$k$ and
$K$, these last two terms are obviously subdominant contributions. 
We may conclude that
$m_g$ and
$m_s$ differ by a fraction of order $Gm_g/R$ which is $10^{-5}$ for
the solar system. 

Let us now calculate $\varsigma' e^{\varsigma/4}$ at
$\varrho=R$ using Eq.~(\ref{expansion2}), (\ref{beta_1}) and
(\ref{beta_2})
 and equate the result to $-2Gm_g/R^2$ as stipulated by
Eq.~(\ref{intfors}):
\begin{equation} r_g+{K r_g{}^2\over 8R}-{kG^2m_s{}^2\over 4\pi
R}+\mathcal{O}(r_g{}^3/R^2)=2Gm_g
\end{equation} For the sun $r_g/R\sim Gm_s/R\sim 10^{-5}$; we see
that 
$r_g\approx 2Gm_g$ with fractional accuracy much better than
$10^{-5}$.


\begin{thebibliography}{99}

\bibitem{Oort}{J. Oort}, \textit{ Bull. Astron. Soc. Neth.}
\textbf{6}, {249} (1932); \textbf{15}, 45 (1960). 

\bibitem{Zwicky}{F. Zwicky}, \textit{Helv. Phys. Acta\/}
\textbf{6}, {110} (1933); see also {S.   Smith}, Astrophys. Journ.
\textbf{83}, {23} (1936).

\bibitem{DMU}\textit{Dark Matter in the Universe}, G. R. Knapp and 
J. Kormendy, eds. (Reidel, Dordrecht 1987).

\bibitem{Can2}J. D. Bekenstein, in \textit{Second Canadian
Conference  on General Relativity and Relativistic Astrophysics}, A.
Coley, C. Dyer and T. Tupper, eds. (World Scientific, Singapore
1988), p. 68. 

\bibitem{Sweden} \textit{The Birth and Early Evolution of our
Universe:  Proceedings of Nobel Symposium \#79, Gr\"aft\aa vallen,
\"Ostersund, Sweden}, Physica Scripta \textbf{T36} (1991).

\bibitem{TurnerEllis} J. Ellis, Ref.~\onlinecite{Sweden};  M.
Turner, Ref.
\onlinecite{Sweden}; C. Munoz, ArXiv hep-ph/0309346.

\bibitem{M1}{M. Milgrom}, Astrophys. Journ. \textbf{270}, {365}
(1983).

\bibitem{M2}{M. Milgrom}, Astrophys. Journ. \textbf{270}, {371}
(1983).

\bibitem{M3}{M. Milgrom}, Astrophys. Journ. \textbf{270}, {384}
(1983).

\bibitem{Berendzen}{R. Berendzen, R. Hart, and D. Seeley},
\textit{Man  Discovers the Galaxies} (Columbia University Press,
New York 1987).

\bibitem{Jeans}{J. Jeans}, Month. Not. Roy. Astron. Soc.
\textbf{84}, 60 (1923).

\bibitem{Morpho}{F. Zwicky}, \textit{Morphological Astronomy}
(Springer, Berlin 1957).

\bibitem{Finzi}{A. Finzi}, Month. Not. Roy. Astron. Soc.
\textbf{127}, {21} (1963).

\bibitem{TullyFisher}{R. B. Tully and J. R. Fisher}, Astron.
Astrophys.
\textbf{54}, {661} (1977).

\bibitem{Aaronson}{M. Aaronson and J.  Mould}, Astrophys. Journ.
\textbf{265}, {1} (1983); {M. Aaronson, G. Bothun,  J.  Mould, J.
Huchra, R. A. Schommer and N. E. Cornell}, Astrophys. Journ.
\textbf{302}, {536} (1986); {W. L. Freedman}, Astrophys. Journ.
\textbf{355}, {L35} (1990), R. H Sanders and M. A. W. Verheijen,
Astrophys. Journ. \textbf{503},  97  (1998).

\bibitem{OstrikerPeeblesYahil}{J. P. Ostriker, P. J. E. Peebles and
A. Yahil}, Astrophys. Journ.  \textbf{193}, {L1} (1974).

\bibitem{Kalnajs}A. Kalnajs in \textit{The Internal Kinematics and
Dynamics of Galaxies}, E. Athanassoula, ed.  (Reidel, Dordrecht
1983), p. 87.

\bibitem{BahcallCassertano}J. N. Bahcall and S.  Cassertano,
Astrophys. Journ. \textbf{293}, {L7} (1985)

\bibitem{Sancisi}T. S. van  Albada and R. Sancisi, Phil. Trans. Roy.
Soc. London A \textbf{320}, {447} (1986). 

\bibitem{QuinnSalmonZurek}P. J. Quinn, J. K. Salmon and W. H. Zurek,
Nature \textbf{322}, {329} (1986); J. Silk in \textit{A Unified View
of the Micro- and Macro-Cosmos}, A. de Rujula, D. V. Nanopoulos and
P. A. Shaver, eds. (World Scientific, Singapore 1987). 

\bibitem{Barnes}G. Blumenthal, S. M. Faber, R. A.  Flores, and J. R.
Primack, Astyrophys. Journ. \textbf{301}, 27 (1986);  J. Barnes, in
\textit{Nearly Normal Galaxies}, S. M. Faber, ed. (Springer, New
York 1987), p. 154; B. S. Ryden and J. E. Gunn, Astrophys. Journ.
\textbf{318}, 15 (1987).

\bibitem{SReview}{R. H. Sanders}, Astron. Astrophys. Rev.
\textbf{2}, {1} (1990).

\bibitem{JobinCarignan}M. Jobin and C. Carignan, Astron. Journ.
\textbf{100}, {648} (1990); C. Carignan, R. Sancisi and T. S. van
Albada, Astron. Journ. \textbf{95}, {37}, (1988); C. Carignan and S.
Beaulieu, Astrophys. Journ.
\textbf{347}, {192}, (1989).

\bibitem{McGaugh}S. S. McGaugh and W. J. G. de Blok, Astrophys.
Journ.
\textbf{508}, 132 (1998). 

\bibitem{Kent}{S. M. Kent}, Astron. Journ. \textbf{93}, {816} (1987).

\bibitem{McKent}{M. Milgrom}, Astrophys. Journ. \textbf{333}, {689}
(1988). 

\bibitem{BBS}{K. G. Begeman, A. H. Broeils and R. H. Sanders},
Month. Not. Roy. Astron. Soc. \textbf{249}, {523} (1991). 

\bibitem{Gentile}G. Gentile, P. Salucci, U. Klein, D. Vergani  and P. Kalberla, ArXiv astro-ph/0403154.

\bibitem{MLSD}M. Milgrom and E. Braun, Astrophys. Journ.
\textbf{334}, {130} (1988).

\bibitem{MCGdB}S. S. McGaugh and E. de Blok, Astrophys. Journ. 
\textbf{499}, 66 (1998).

\bibitem{Begum}Begum, A., and Chengalur, J. N. , Astron. and Astrophys. \textbf{413}, 525 (2004).

\bibitem{SandersMcGaugh}R. H. Sanders and S. S. McGaugh, Ann.
Rev. Astron. Astrophys. \textbf{40},  263 (2002).

\bibitem{McGaugh2}S. S. McGaugh, ArXiv astro-ph/0403610.

\bibitem{Aguirre}A. Aguirre, {\it Proceedings of the IAU Symposium 220
``Dark matter in galaxies"\/}, eds. S. Ryder, D. J. Pisano, M. Walker
and K. Freeman (2003). 

\bibitem{Sanders}R.H.  Sanders,  Mon. Not. R. Ast. Soc. 342, 901 (2003).

\bibitem{more}M. Milgrom, \textbf{287}, 571, 1984; R. H. Sanders,
Astron. Astrophys. \textbf{284}, (1994); M. Milgrom and R. H. Sanders, Astrophys. J. \textbf{599}, L25 (2003);  R. Scarpa, G. Marconi and R. Gilmozzi, Astron. and Astrophys. \textbf{405}, L15  (2003). 

\bibitem{Felten}J. E. Felten, Astrophys. Journ. \textbf{286},  3
(1984).

\bibitem{SandersC}R. H. Sanders, Mon. Not. Roy. Astron. Soc.
\textbf{296}, 1009 (1998) .

\bibitem{BekMilg}J. D. Bekenstein and M.  Milgrom, Astrophys. Journ. 
\textbf{286}, 7 (1984).

\bibitem{phase}J. D. Bekenstein,  Physics Letters B\textbf{202}, 497
(1988).

\bibitem{SPCG}R. H. Sanders, Month. Not. Roy. Ast. Soc.
\textbf{235}, 105 (1988).

\bibitem{Rosen}J. D. Bekenstein in \textit{Developments in General
Relativity, Astrophysics and Quantum Theory}, F. I. Cooperstock, L.
P. Horwitz and J. Rosen, eds.  (IOP Publishing, Bristol 1990), p.
156.

\bibitem{Kyoto}J. D. Bekenstein, in \textit{Proceedings of the Sixth
Marcel Grossman Meeting on General Relativity}, H. Sato and T.
Nakamura, eds.  (World Scientific, Singapore 1992), p. 905.

\bibitem{BekSan}J. D. Bekenstein and R. H. Sanders, Astrophys. Journ.
\textbf{429}, 480 (1994).

\bibitem{Nester}V V. Zhytnikov and J. M. Nester, Phys. Rev. Lett. 
\textbf{73}, 2950 (1994). 

\bibitem{Edery}A. Edery, Phys. Rev. Lett.  \textbf{83}, 3990- (1999)
and rebuttal  J. D. Bekenstein, M. Milgrom and R. H. Sanders, Phys.
Rev. Lett.
\textbf{85}, 1346  (2000). 
 
\bibitem{Soussa1}M. E. Soussa  and R. P. Woodard, Class. Quant. Grav.
\textbf{20}, 2737  (2003).

\bibitem{Soussa2}M. E. Soussa  and R. P. Woodard, Phys. Lett. B
\textbf{578},  253 (2004).

\bibitem{Soussa3}M. E. Soussa, ArXiv astro-ph/0310531.

\bibitem{Sandersstratified}R. H. Sanders, Astrophys. Journ.
\textbf{480}, 492 (1997).

\bibitem{Mnumer}M. Milgrom, Astrophys. Journ. \textbf{302}, 617
(1986).

\bibitem{Smn86}R. H. Sanders,  Month. Not. Roy. Astron. Soc.
\textbf{223}, 559 (1986).

\bibitem{Romatka}R. Romatka, Dissertation (University of Munich,
1992).

\bibitem{Will}C. Will, {\it Theory and Experiment in Gravitational
Physics\/} (Cambridge University, Cambridge 1986).

\bibitem{VessotLevine}R. F. C. Vessot and M. A. Levine, Phys. Rev.
Letters
\textbf{45}, 2081 (1980).

\bibitem{Schiff}L. Schiff, Am. Journ. Phys. \textbf{28}, 340 (1960).

\bibitem{Witten}E. Witten, Commun. Math. Phys. \textbf{80}, 381
(1981).

\bibitem{Bmass}J. D. Bekenstein, Int. Journ. Theor. Phys. \textbf{
13}, 317 (1975).

\bibitem{Tohline} J. E. Tohline, in \textit{Internal Kinematics and
Dynamics of Galaxies}, A. Athanassoula, ed., (Reidel, Dordrecht
1982), p.205.

\bibitem{KuhnKruglyak}J. R. Kuhn and L. Kruglyak, Astrophys. Journ.
\textbf{313}, 1 (1987). 

\bibitem{Jac_Mat}T. Jacobson and D. Mattingly, Phys. Rev.
\textbf{D}64, 024028 (2001). 

\bibitem{Eil_Jac}C. Eiling and T. Jacobson, Phys. Rev.
\textbf{D}69, 064005 (2004). 

\bibitem{nieto}M. M. Nieto and S. G. Turyshev, ArXiv gr-qc/0308017.

\bibitem{MTW}C. Misner, K. S. Thorne and J. A. Wheeler, {\it
Gravitation} (Freeman, San Francisco 1973).

\bibitem{Riess}A. G. Riess, L. G. Strolger  et. al, ArXiv
astro-ph/0402512.

\bibitem{LeeYang}T. D. Lee and C. N. Yang, Phys. Rev. \textbf{98},
1501 (1955).

\bibitem{Dicke}R. H. Dicke, \textit{The Theoretical Significance of
Experimental Relativity} (Gordon and Breach, New York 1964).

\end{thebibliography}
\end{document}